\documentclass[longauth]{aa} 

\usepackage{graphicx}
\usepackage{xcolor}
\usepackage[varg]{txfonts}
\usepackage[breaklinks=true,colorlinks=true,urlcolor=blue,linkcolor=blue,citecolor=blue,bookmarks=true,bookmarksopen=true]{hyperref}
\usepackage{siunitx}
\usepackage{amsmath}
\DeclareMathAlphabet{\mathbbmsl}{U}{bbm}{m}{sl}

\newcommand\T{\rule{0pt}{2.6ex}}
\newcommand\B{\rule[-1.2ex]{0pt}{0pt}}

\usepackage{natbib,twoopt}
\bibpunct{(}{)}{;}{a}{}{,} 					
\makeatletter
\newcommandtwoopt{\citetds}[3][][]{\href{http://adsabs.harvard.edu/abs/#3}%
{\def\hyper@linkstart##1##2{}%
\let\hyper@linkend\@empty\citetlp[#1][#2]{#3}}}
\newcommandtwoopt{\citepads}[3][][]{\href{http://adsabs.harvard.edu/abs/#3}%
{\def\hyper@linkstart##1##2{}%
\let\hyper@linkend\@empty\citep[#1][#2]{#3}}}
\newcommandtwoopt{\citetads}[3][][]{\href{http://adsabs.harvard.edu/abs/#3}%
{\def\hyper@linkstart##1##2{}%
\let\hyper@linkend\@empty\citet[#1][#2]{#3}}}
\newcommandtwoopt{\citeyearads}[3][][]%
{\href{http://adsabs.harvard.edu/abs/#3}
{\def\hyper@linkstart##1##2{}%
\let\hyper@linkend\@empty\citeyear[#1][#2]{#3}}}
\makeatother

\begin{document} 

\title{The polarimetric imaging mode of VLT/SPHERE/IRDIS II:}

\subtitle{Characterization and correction of instrumental polarization effects\thanks{Based on observations made with ESO telescopes at the La Silla Paranal Observatory under program ID
60.A-9800(S), 60.A-9801(S) and 096.C-0248(C).}}

\titlerunning{The polarimetric imaging mode of VLT/SPHERE/IRDIS II}

\author{R.\,G. van Holstein\inst{\ref{inst:leiden},\ref{inst:esosantiago}}\fnmsep\thanks{E-mail corresponding author: vanholstein@strw.leidenuniv.nl} 
\and J.\,H. Girard \inst{\ref{inst:stsi},\ref{inst:ipag}}
\and J. de Boer \inst{\ref{inst:leiden}}
\and F. Snik \inst{\ref{inst:leiden}}
\and J. Milli \inst{\ref{inst:esosantiago},\ref{inst:ipag}}
\and D.\,M. Stam \inst{\ref{inst:delftae}}
\and C. Ginski \inst{\ref{inst:pannekoek},\ref{inst:leiden}}
\and D. Mouillet \inst{\ref{inst:ipag}}
\and Z. Wahhaj \inst{\ref{inst:esosantiago}}
\and H.\,M. Schmid \inst{\ref{inst:eth}}
\and C.\,U. Keller \inst{\ref{inst:leiden}}
\and M. Langlois \inst{\ref{inst:crallyon},\ref{inst:lam}}
\and K. Dohlen \inst{\ref{inst:lam}}
\and A. Vigan \inst{\ref{inst:lam}}
\and A. Pohl \inst{\ref{inst:mpia},\ref{inst:heidelberg}}
\and M. Carbillet\inst{\ref{inst:cotedazur}}
\and D. Fantinel\inst{\ref{inst:inaf}}
\and D. Maurel\inst{\ref{inst:ipag}}
\and A. Orign\'e\inst{\ref{inst:lam}}
\and C. Petit\inst{\ref{inst:onera}} 
\and J. Ramos\inst{\ref{inst:mpia}}
\and F. Rigal\inst{\ref{inst:pannekoek}}
\and A. Sevin\inst{\ref{inst:lesia}}
\and A. Boccaletti \inst{\ref{inst:lesia}}
\and H. Le Coroller \inst{\ref{inst:lam}}
\and C. Dominik \inst{\ref{inst:pannekoek}}
\and T. Henning \inst{\ref{inst:mpia}}
\and E. Lagadec \inst{\ref{inst:cotedazur}}
\and F. M\'enard \inst{\ref{inst:ipag}}
\and M. Turatto \inst{\ref{inst:inaf}}
\and S. Udry \inst{\ref{inst:geneva}}
\and G. Chauvin \inst{\ref{inst:ipag}} 
\and M. Feldt \inst{\ref{inst:mpia}}
\and J.-L. Beuzit \inst{\ref{inst:ipag}}
}

\institute{Leiden Observatory, Leiden University, PO Box 9513, 2300 RA Leiden, The Netherlands \label{inst:leiden} 
\and European Southern Observatory, Alonso de C\'{o}rdova 3107, Casilla 19001, Vitacura, Santiago, Chile \label{inst:esosantiago}
\and Space Telescope Science Institute, Baltimore 21218, MD, USA \label{inst:stsi}
\and Universit\'e Grenoble Alpes, CNRS, IPAG, 38000 Grenoble, France \label{inst:ipag}
\and Faculty of Aerospace Engineering, Delft University of Technology, Kluyverweg 1, 2629 HS Delft, The Netherlands \label{inst:delftae}
\and Anton Pannekoek Institute for Astronomy, University of Amsterdam, Science Park 904, 1098 XH Amsterdam, The Netherlands \label{inst:pannekoek}
\and ETH Zurich, Institute for Particle Physics and Astrophysics, Wolfgang-Pauli-Strasse 27, 8093 Zurich, Switzerland \label{inst:eth}
\and CRAL, UMR 5574, CNRS, Universit\'e de Lyon, Ecole Normale
Supérieure de Lyon, 46 Allée d’Italie, F-69364 Lyon Cedex 07, France \label{inst:crallyon}
\and Aix Marseille Univ, CNRS, CNES, LAM, Marseille, France \label{inst:lam}
\and Max Planck Institute for Astronomy, K\"onigstuhl 17, D-69117 Heidelberg, Germany \label{inst:mpia}
\and Heidelberg University, Institute of Theoretical Astrophysics, Albert-Ueberle-Str. 2, D-69120 Heidelberg, Germany \label{inst:heidelberg}
\and Universit\'e C\^ote d'Azur, OCA, CNRS, Lagrange, France \label{inst:cotedazur}
\and INAF - Osservatorio Astronomico di Padova, Vicolo della Osservatorio 5, 35122, Padova, Italy \label{inst:inaf}
\and DOTA, ONERA, Universit\'e Paris Saclay, F-91123, Palaiseau France\label{inst:onera}
\and LESIA, Observatoire de Paris, PSL Research University, CNRS, Sorbonne Universit\'es, UPMC, Univ. Paris 06, Univ. Paris Diderot, Sorbonne Paris Cit\'e, 5 place Jules Janssen, 92195 Meudon, France \label{inst:lesia}
\and Geneva Observatory, University of Geneva, Chemin des Mailettes 51, 1290 Versoix, Switzerland \label{inst:geneva}
}

\date{Received TBD / Accepted TBD}

\abstract{
Circumstellar disks and self-luminous giant exoplanets or companion brown dwarfs can be characterized through direct-imaging polarimetry at near-infrared wavelengths. SPHERE/IRDIS at the Very Large Telescope has the capabilities to perform such measurements, but uncalibrated instrumental polarization effects limit the attainable polarimetric accuracy.
}
{
We aim to characterize and correct the instrumental polarization effects of the complete optical system, i.e. the telescope and SPHERE/IRDIS.
}
{
We create a detailed Mueller matrix model in the broadband filters Y-, J-, H- and K$_\mathrm{s}$, and calibrate it using measurements with SPHERE's internal light source and observations of two unpolarized stars. We develop a data-reduction method that uses the model to correct for the instrumental polarization effects, and apply it to observations of the circumstellar disk of T Cha. 
}
{
The instrumental polarization is almost exclusively produced by the telescope and SPHERE's first mirror and varies with telescope altitude angle. The crosstalk primarily originates from the image derotator (K-mirror). At some orientations, the derotator causes severe loss of signal (${>}90\%$ loss in H- and K$_\mathrm{s}$-band) and strongly offsets the angle of linear polarization. With our correction method we reach in all filters a total polarimetric accuracy of ${\lesssim}0.1\%$ in the degree of linear polarization and an accuracy of a few degrees in angle of linear polarization.
}
{
The correction method enables us to accurately measure the polarized intensity and angle of linear polarization of circumstellar disks, and is a vital tool for detecting unresolved (inner) disks and measuring the polarization of substellar companions. We have incorporated the correction method in a highly-automatic end-to-end data-reduction pipeline called IRDAP (IRDIS Data reduction for Accurate Polarimetry) which is publicly available at~\url{https://irdap.readthedocs.io}.
}



\maketitle

%
%

%
%

\section{Introduction}
\label{sec:introduction}



The near-infrared (NIR) polarimetric mode of SPHERE/IRDIS at the Very Large Telescope (VLT), 
which we introduced in de Boer et al. (2019, Paper~I), 
has proven to be very successful for the detection of circumstellar disks in scattered light~\citep{garufi_3yearssphere} and shows much promise for the characterization of exoplanets and companion brown dwarfs~\citep[see][]{vanholstein_exopol}. 
	However, studies of circumstellar disks 
are often limited to analyses of the orientation (position angle and inclination) and morphology (rings, gaps, cavities and spiral arms) of the disks~\citep[e.g.][]{muto_disk, quanz_disk, ginski_diskgaps, deboer_diskcandidates}.
	Quantitative polarimetric measurements of circumstellar disks and substellar companions
	are currently very challenging, 
because existing data-reduction methods do not account for instrumental polarization effects with a sufficiently high accuracy. 
	
Because of instrumental polarization effects, polarized signal arriving at IRDIS' detector is different from that incident on the telescope.
	The two predominant effects are \emph{instrumental polarization} (IP), i.e.~polarization signals produced by the instrument or telescope, and \emph{crosstalk}, i.e.~instrument- or telescope-induced mixing of polarization states.
	IP not only changes the polarization state of an object, but can also make unpolarized sources appear polarized if not accounted for.
	For astronomical targets with a relatively low degree of linear polarization, IP can induce a significant rotation of the angle of linear polarization.
	Crosstalk also causes an offset of the measured angle of linear polarization and can lower the \emph{polarimetric efficiency}, i.e.~the fraction of the incident or true linear polarization that is actually measured.
	We first encountered these instrumental polarization effects when observing the disk around TW Hydrae as described in Paper~I. 

To derive the true polarization state of the light incident on the telescope, we need to calibrate the instrument so that we know the instrumental polarization effects a priori.	
	This will enable us to accurately and quantitatively measure the polarization of circumstellar disks and substellar companions.
	In addition, it will enable accurate mapping of extended objects other than circumstellar disks, such as solar system objects, molecular clouds and galaxies~\citep[e.g.][]{gratadour_irdisgalaxy}, provided the target is sufficiently bright for the adaptive optics correction. 	

	For observations of circumstellar disks (see Paper~I), calibrating the instrument will yield a multitude of improvements.
	Firstly, the calibration will allow for more accurate studies of the orientation and morphology of the disks, especially at the innermost regions (separation ${<0.5''}$).
	In fact, we will be able to deduce the presence of unresolved (inner) disks by measuring the polarization signals of the stars~\citep[see e.g.][]{keppler_pds70}.
	Secondly, the calibration will enable more accurate measurements of the angle of linear polarization. 
	This in turn will allow us to prove the presence of non-azimuthal polarization~\citep{canovas_nonazimuthaldisk} that can be indicative of multiple scattering or the presence of a binary star, and allows for a more in-depth study of dust properties.
	Finally, the calibration will enable more accurate measurements of the polarized intensity, i.e.~the polarized surface brightness of the disk. 
	
	More accurate measurements of the polarized surface brightness will enable us to construct scattering phase functions~\citep[e.g.][]{perrin_gemini, stolker_diskphasefunction, ginski_diskgaps, milli_hr4796}, perform more accurate radiative transfer modeling~\citep[e.g.][]{pinte_radiativetransfer, min_diskradiativetransfer, pohl_tcha, keppler_pds70} and determine dust particle properties~\citep[e.g.][]{min_diskdust, pohl_hd169142, pohl_tcha}. 
	In addition, it will allow accurate measurements of the degree of linear polarization of the disk, enabling us to further constrain dust properties~\citep[e.g.][]{perrin_abaur, perrin_gemini, milli_hr4796a_dolp}. However, before images of the degree of linear polarization can be constructed, an image of the total intensity of the disk needs to be obtained, e.g.~with reference star differential imaging~\citep[RDI;~e.g.][]{canovas_disk} or, for disks viewed edge-on, with angular differential imaging~\citep[ADI;][]{marois_adi}.

To measure polarization signals of young self-luminous giant exoplanets or companion brown dwarfs (see Paper~I), it is of vital importance to calibrate the instrument.
	Based on radiative transfer models, the NIR degree of linear polarization of a companion can be a few tenths of a percent up to several percent~\citep{dekok_exopol, marley_exopol, stolker_exopol}.
	Measurements of these small polarization signals therefore need to be performed with a very high accuracy, which is only possible after careful calibration of the instrumental polarization effects. 

Polarimetric measurements of substellar companions have already been attempted by~\citet{millar_betapic} and \citet{jensen_padi} with the Gemini Planet Imager (GPI), and by~\citet{vanholstein_exopol} with SPHERE/IRDIS (using the calibration results presented in this paper). 
	No polarization signals were detected in these studies.
	Recently, \citet{ginski_cscha} presented the first direct detection of a polarization signal from a substellar companion.
	Using the calibration results presented in this paper, they found the companion to CS Cha to have a NIR degree of linear polarization of $14\%$, which suggests the presence of an unresolved disk and dusty envelope around the companion. \\
\noindent In this paper, we characterize the instrumental polarization effects of the complete optical system of VLT/SPHERE/IRDIS, i.e.~the telescope and the instrument, in the four broadband filters Y, J, H and K$_\mathrm{s}$. 
	Because the complexity of the optical path is comparable to that of solar telescopes and their instruments, we perform a calibration similar to those applied in the field of solar physics~\citep[see e.g.~][]{skumanich_aspmodel, beck_gvttmodel, socasnavarro_dstmodel}.
	For our calibration, we create a detailed Mueller matrix model of the optical path and determine the parameters of the model from measurements with SPHERE's internal light source and observations of two unpolarized stars. 
	 Similar approaches have been adopted for the German Vacuum Tower Telescope~\citep{beck_gvttmodel}, VLT/NACO~\citep{witzel_ipcal} and GPI~\citep{wiktorowicz_gpicalib, millar_gpicalib}.
	We then develop a data-reduction method to correct science measurements for the instrumental polarization effects using the model, and exemplify this correction method and its advantages with polarimetric observations of the circumstellar disk around T Cha from~\citet{pohl_tcha}. 
	This work is Paper~II of a larger study in which Paper~I discusses IRDIS' polarimetric mode, the data reduction and recommendations for observations and instrument upgrades.
	
With our instrument model we aim to achieve in all four broadband filters a \emph{total polarimetric accuracy}, i.e.~the uncertainty in the measured polarization signal, of ${\sim}0.1\%$ in the degree of linear polarization.
	In addition, we aim to attain an accuracy of a few degrees in angle of linear polarization in these filters.
	Reaching these accuracies will enable us to measure the linear polarization of substellar companions (we regard the extremely high degree of linear polarization found by \citet{ginski_cscha} to be an exception). 
	These accuracies also readily suffice for quantitative polarimetry of circumstellar disks, because the degree of linear polarization of disks is typically much higher than that of substellar companions: on the order of percents to several ten percent~\citep[see e.g.][]{perrin_abaur}.
	To attain a total polarimetric accuracy of ${\sim}0.1\%$, an \emph{absolute polarimetric accuracy}, i.e.~the uncertainty in the instrumental polarization (IP), of ${\leq}0.1\%$ and a \emph{relative polarimetric accuracy}, i.e.~the uncertainty that scales with the 
input polarization signal, of ${<}1\%$ is aimed for.

The outline of this paper is as follows.
	In Sect.~\ref{sec:definitions} we present the conventions and definitions used throughout this paper.
	Subsequently, we briefly review the SPHERE/IRDIS optical path and discuss the expected instrumental polarization effects 
in Sect.~\ref{sec:sphere_irdis}.
	We explain the Mueller matrix model describing these effects in Sect.~\ref{sec:model_mathematics}.
	In Sects.~\ref{sec:response_downstream_m4} and \ref{sec:response_telescope_m4} we determine the parameters of the model from measurements with the internal light source and observations of two unpolarized stars, respectively.
	We then discuss the accuracy of the model in Sect.~\ref{sec:model_accuracy}.
	In Sect.~\ref{sec:correction_method} we present our correction method and exemplify it with polarimetric observations of the circumstellar disk of T Cha.
	In the same Section we describe the improvements we attain with respect to conventional data-reduction methods, discuss the limits to and optimization of the polarimetric accuracy, and introduce our data-reduction pipeline that incorporates the correction method.
	Finally, we present conclusions in Sect.~\ref{sec:conclusions}.
	If the reader is only interested in applying our correction method to on-sky data, one could suffice with reading Sects.~\ref{sec:definitions}, \ref{sec:sphere_irdis}, \ref{sec:correction_method} and \ref{sec:conclusions}.

\section{Conventions and definitions}
\label{sec:definitions}

\begin{figure}[!hbtp] 
\centering 
\includegraphics[width=\hsize]{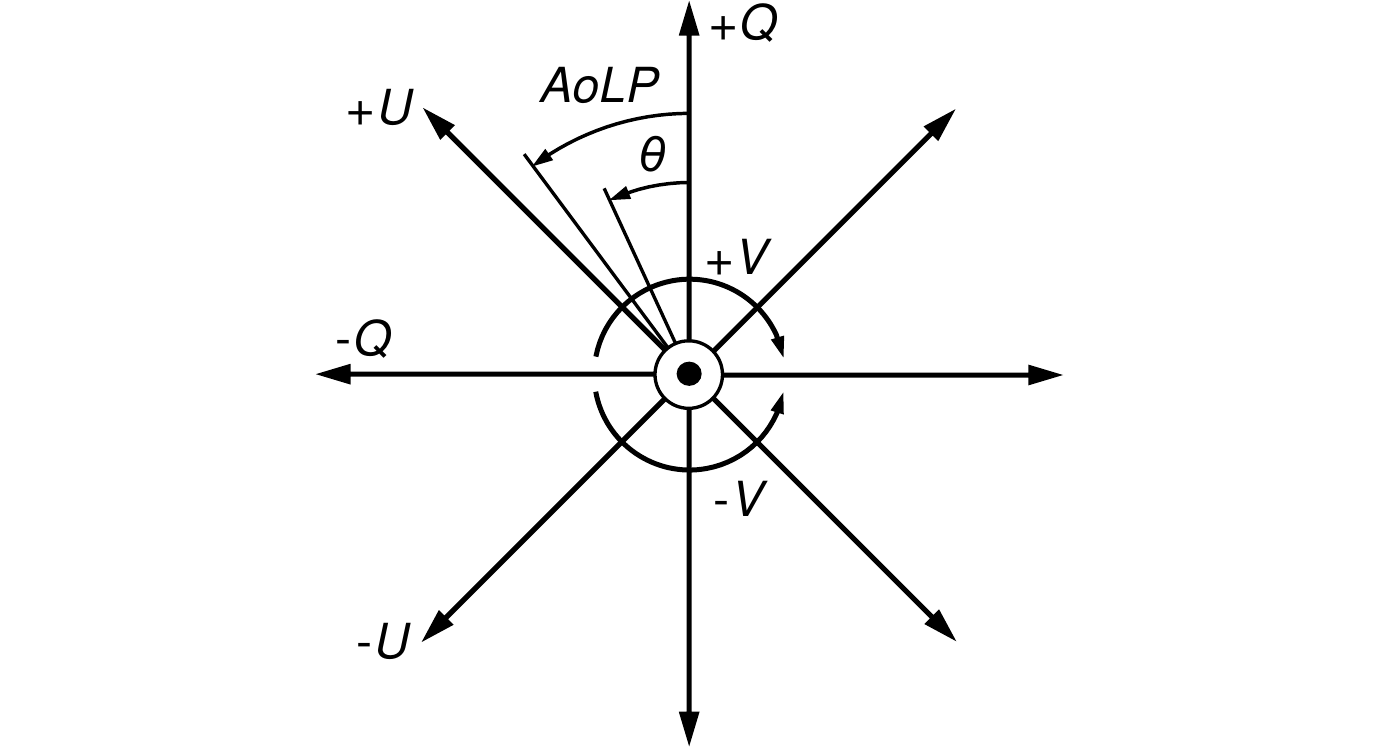}
\caption[Reference frame for the definition of the Stokes parameters.]{Reference frame for the definition of the Stokes parameters describing the oscillation direction of the electric field within a beam of light. 
	The propagation direction of the light beam is out of the paper, towards the reader. Positive and negative Stokes $Q$ are oriented along the vertical ($+Q$) and horizontal ($-Q$) axes, respectively. 	
	Looking into the beam of light, positive Stokes $U$ ($+U$) is oriented \SI{45}{\degree} counterclockwise from positive Stokes $Q$ and positive Stokes $V$ ($+V$) is defined as clockwise rotation. 
	The angle of linear polarization $AoLP$ and the rotation angle $\theta$ of an optical component used in the rotation Mueller matrix 
(see Eqs.~\ref{eq:rotation_mueller_matrix} and~\ref{eq:rotated_mueller_matrix})
	are defined counterclockwise when looking into the beam of light.} 
\label{fig:reference_frame_general} 
\end{figure}

\noindent In this Section we will briefly outline the conventions and definitions used throughout this paper. 
	The total intensity and polarization state of a beam of light can be described by a Stokes vector $\boldsymbol{S}$~\citep[e.g.][]{tinbergen}:
\begin{equation} 
	\boldsymbol{S} = \begin{bmatrix} I \\ Q \\ U \\ V \end{bmatrix},
	\label{eq:stokes_vector}
\end{equation}    
where $I$ is the total intensity (or flux), $Q$ and $U$ describe linear polarization and $V$ represents circular polarization.
	We define these Stokes parameters with respect to the \emph{general reference frame} shown in Fig.~\ref{fig:reference_frame_general}. 
	Positive Stokes $Q$ ($+Q$) and negative Stokes $Q$ ($-Q$) correspond to vertical and horizontal linear polarization, respectively.
	When looking into the beam of light, positive (negative) Stokes $U$ is 
oriented \SI{45}{\degree} counterclockwise (clockwise) from positive Stokes $Q$. 
	Finally, positive (negative) Stokes $V$ is defined as circularly polarized light with clockwise (counterclockwise) rotation when looking into the beam of light. 

We can normalize the Stokes vector of Eq.~\ref{eq:stokes_vector} by dividing each of its Stokes parameters by the total intensity $I$:
\begin{equation} 
	\boldsymbol{S} = \left[1,~q,~u,~v\right]^\mathrm{T},
	\label{eq:normalized_stokes_vector}
\end{equation}    
%
with $q$, $u$ and $v$ the normalized Stokes parameters. 
	From the Stokes parameters we can calculate the linearly polarized intensity ($PI_\mathrm{L}$), degree of linear polarization ($DoLP$) 
	and angle of linear polarization ($AoLP$; 
see Fig.~\ref{fig:reference_frame_general}) as follows:
%
\begin{align} 
	PI_\mathrm{L} &= \sqrt{Q^2 + U^2}, \label{eq:polarized_intensity} \\
	DoLP &= \sqrt{q^2 + u^2},
\label{eq:DoLP} \\
	AoLP &= \dfrac{1}{2}\arctan\left(\dfrac{U}{Q}\right) = \dfrac{1}{2}\arctan\left(\dfrac{u}{q}\right).
\label{eq:AoLP}
\end{align} 
%

%
%

\section{Optical path and instrumental polarization effects of SPHERE/IRDIS}
\label{sec:sphere_irdis}


\subsection{SPHERE/IRDIS optical path}	
\label{sec:sphere_irdis_optical_path}
	
\begin{figure*} 
\centering 
\includegraphics[width=17.0cm]{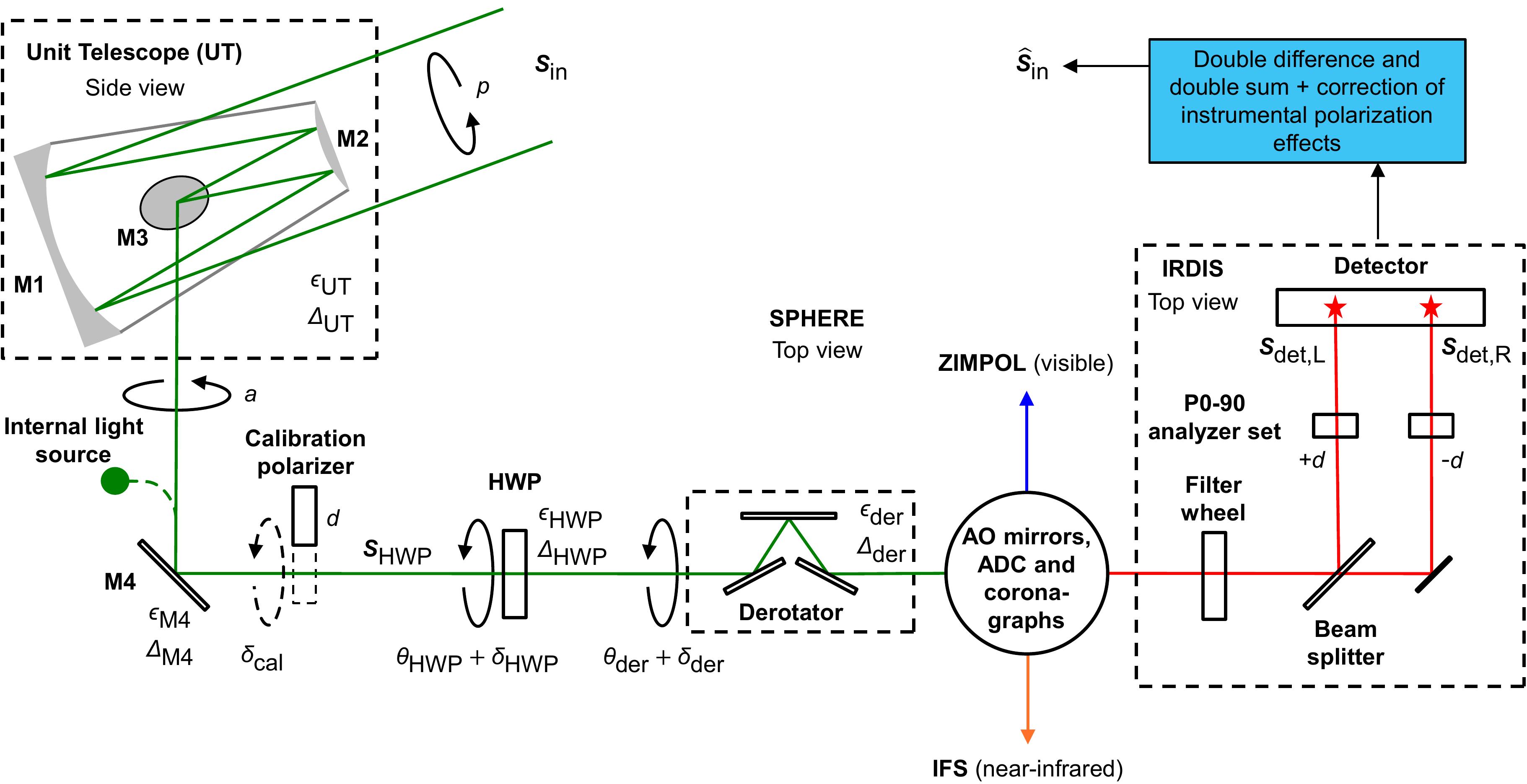} 
\caption[]{Overview of the optical path of the complete optical system, i.e.~the Unit Telescope (UT) and SPHERE/IRDIS, showing only the components relevant for polarimetric measurements (image adapted from Fig.~2 of Paper~I). The names of the (groups of) components are indicated in boldface. The black circular arrows indicate the astronomical target's parallactic angle $p$, the telescope's rotation with the altitude angle $a$, the offset angle of the calibration polarizer $\delta_\mathrm{cal}$, and the rotation of the HWP and image derotator with the angles $\theta_\mathrm{HWP} + \delta_\mathrm{HWP}$ and $\theta_\mathrm{der} + \delta_\mathrm{der}$, respectively. Also shown are the parameters describing the instrumental polarization effects of the (groups of) components: the component diattenuations $\epsilon$, retardances $\varDelta$ and the polarizer diattenuation $d$. The Stokes vectors $\boldsymbol{S}_\mathrm{in}$, $\boldsymbol{S}_\mathrm{HWP}$, $\boldsymbol{S}_\mathrm{det,L}$, $\boldsymbol{S}_\mathrm{det,R}$ and $\skew{2.5}\hat{\boldsymbol{S}}_\mathrm{in}$ used in the instrument model are indicated as well. Finally, the top right of the image shows the data-reduction process that produces the measured (after calibration) Stokes vector incident on the telescope.} 
\label{fig:ut_sphere_irdis_schematic} 
\end{figure*}

Before discussing the instrumental polarization effects expected for SPHERE/IRDIS, in this Section we will first summarize the optical path and the working principle of IRDIS' polarimetric mode.
	As described in detail in Paper~I, SPHERE's optical system is complex and has many rotating components. 	
	A simplified version of the optical path 
is shown in Fig.~\ref{fig:ut_sphere_irdis_schematic}. 
	The model parameters, Stokes vectors and the top right part of the image will be discussed in Sect.~\ref{sec:model_mathematics}.  
%
%

During an observation, light is collected by the altazimuth-mounted Unit Telescope (UT) which consists of three mirrors. 
	The incident light hits the primary mirror (M1) and is subsequently re-focused by the secondary mirror (M2) that is suspended at the top of the telescope tube.
	The flat tertiary mirror (M3) has an angle of incidence of \SI{45}{\degree} and reflects the beam of light to the Nasmyth platform where SPHERE is located.
	When the telescope tracks a target across the sky, the target rotates with the parallactic angle in the pupil of the UT and the UT rotates with the telescope altitude angle with respect to Nasmyth platform.

The light entering SPHERE~\citep{beuzit_sphereoverview} passes a system that can feed the instrument with light from an internal light source to enable internal calibrations~\citep{wildi_sphere_calib, roelfsema_zimpol}. 
	Subsequently, the beam of light hits the flat mirror M4 (the pupil tip-tilt mirror) that like M3 is coated with aluminum and has a \SI{45}{\degree} inclination angle. 
	M4 is the only aluminum mirror in SPHERE; all other mirrors are coated with protected silver. 
	For calibrations, a linear polarizer with its transmission axis aligned vertical, i.e.~perpendicular to the Nasmyth platform, can be inserted after M4~\citep{wildi_sphere_calib}.

The light then reaches the insertable and rotatable half-wave plate (HWP; HWP2 in Paper~I) that can rotate the incident angle of linear polarization.
	The HWP is used to temporally modulate the incident Stokes $Q$ and $U$ 
and to correct for field rotation so that the polarization direction of the source is kept fixed on the detector. 
	 The HWP is followed by the image derotator, which is a rotating assembly of three mirrors (a K-mirror) that rotates both the image and angle of linear polarization for field- or pupil-stabilized observations.
	Before reaching IRDIS, the light passes the mirrors of the adaptive-optics (AO) common path~\citep{fusco_saxo, hugot_spheretorics}, several dichroic mirrors, the rotating atmospheric dispersion corrector (ADC) and the coronagraphs~\citep{carbillet_sphereaplc, guerri_sphereaplc}.

The light beam entering IRDIS~\citep{dohlen_irdis, langlois_irdis} passes a filter wheel containing various color 
filters. 
	In this work, only the four available broadband filters Y, J, H and $\mathrm{K_s}$ are considered (see Table~1 of Paper~I for the central wavelengths and bandwidths).
	After the filter wheel, the light is split into parallel beams by a combination of a non-polarizing beamsplitter plate and a mirror.
	The light beams subsequently pass a pair of insertable linear polarizers (the P0-90 analyzer set) with orthogonal transmission axes at \SI{0}{\degree} (left) and \SI{90}{\degree} (right) with respect to vertical.
	Both beams strike the same detector to form two adjacent images, one on the left and one on the right half of the detector. 

Images of Stokes $Q$ and $U$ and the corresponding total intensities $I$ ($I_Q$ and $I_U$) can then be constructed from the \emph{single difference} and \emph{single sum}, respectively, of the left and right images on the detector (see Paper~I):
\begin{align} 
	X^\pm &= I_\mathrm{det,L} - I_\mathrm{det,R}, \label{eq:single_difference} \\[0.3cm]
	I_{X^\pm} &= I_\mathrm{det,L} + I_\mathrm{det,R}, \label{eq:single_sum} 
\end{align} 
where $X^\pm$ is the single-difference $Q$ or $U$ and $I_{X^\pm}$ is the single-sum intensity $I_Q$ or $I_U$. 
	$I_\mathrm{det,L}$ and $I_\mathrm{det,R}$ are the intensities of the left~(L) and right~(R) images on the detector, respectively. 
	$Q$ and $I_Q$ are measured with the HWP angle switched by \SI{0}{\degree} and $U$ and $I_U$ are measured with the HWP angle switched by \SI{22.5}{\degree}. 
	We call the resulting single differences $Q^+$ and $U^+$ and the corresponding single-sum intensities $I_{Q^+}$ and $I_{U^+}$.
	Additional measurements of $Q$ and $I_Q$, and of $U$ and $I_U$, are taken with the HWP angle switched by \SI{45}{\degree} and \SI{67.5}{\degree}, respectively.
	We call the results $Q^-$, $I_{Q^-}$, $U^-$ and $I_{U^-}$.
	The set of measurements with HWP switch angles equal to \SI{0}{\degree}, \SI{45}{\degree}, \SI{22.5}{\degree} and \SI{67.5}{\degree} are called a HWP or polarimetric cycle.
	The single differences and single sums will be used in Sect.~\ref{sec:polarization_effects_optical_path} to calculate the so-called double difference and double sum.
	Note that Stokes $V$ cannot be measured by IRDIS, as it lacks a quarter-wave plate (however, see the last paragraph of Sect.~\ref{sec:results_discussion_instrument}).

\subsection{Instrumental polarization effects of optical path}	
\label{sec:polarization_effects_optical_path}

In this Section, we will discuss the expected instrumental polarization effects of the optical path of SPHERE/IRDIS.
	Basically all optical components described in Sect.~\ref{sec:sphere_irdis_optical_path} produce instrumental polarization (IP) and crosstalk.
	IP is a result of the optical components' (linear) \emph{diattenuation}, i.e.~it is caused by the different reflectances (e.g.~for the mirrors) or transmittances (e.g.~for the beamsplitter or HWP) of the perpendicular linearly polarized components of an incident beam of light. 
	Crosstalk is created by the optical components' \emph{retardance} (or relative retardation), i.e.~the relative phase shift of the perpendicular linearly polarized 
components. 
	Because IRDIS cannot measure circularly polarized light, crosstalk from linearly polarized to circularly polarized light results in a loss of polarization signal and thus a decrease of the polarimetric efficiency.
	The diattenuation and retardance of an optical component are a function of wavelength and the component's rotation angle.

The diattenuation and retardance are strongest for reflections at large angles of incidence.
	Therefore the largest effects are expected for M3, M4, the derotator, the two reflections at an angle of incidence of \SI{45}{\degree} just upstream of IRDIS and IRDIS' beamsplitter-mirror combination (the non-polarizing beamsplitter is in fact ${\sim}10\%$ polarizing).
	The diattenuation and retardance of M1 and M2 are expected to be small, because these mirrors are rotationally symmetric with respect to the optical axis~\citep[see e.g.~][]{tinbergen}.
	Also the diattenuation and retardance of the ADC and the mirrors of the AO common path are likely small, because these components have small angles of incidence (${<}\SI{10}{\degree}$) and stress birefringence in the ADC is expected to be limited.
	The HWP will create (some) circular polarization because its retardance is not completely achromatic and only approximately half-wave (or \SI{180}{\degree} in phase).

The IP of the non-rotating components downstream of the HWP can be removed by taking advantage of beam switching with the HWP 
and computing the Stokes parameters from the \emph{double difference}~(see Paper~I;~\citealt{bagnulo_polarimetry}):	
\begin{equation} 
	X = \dfrac{1}{2}\left(X^+ - X^-\right), \label{eq:double_difference}
\end{equation} 
where $X$ is the double-difference Stokes $Q$ or $U$, and $X^+$ and $X^-$ are 
computed from Eq.~\ref{eq:single_difference}.
	An additional advantage of the double-difference method is that it suppresses differential effects such as flat-fielding errors and differential aberrations~\citep{tinbergen, canovas_data}.
	The total intensity corresponding to the double-difference $Q$ or $U$ is computed from the \emph{double sum}:
\begin{equation} 
	I_X = \dfrac{1}{2}\left(I_{X^+} + I_{X^-}\right), \label{eq:double_sum}
\end{equation} 
where $I_X$ is the double-sum intensity $I_Q$ or $I_U$, and $I_{X^+}$ and $I_{X^-}$ are 
computed from Eq.~\ref{eq:single_sum}. 
	Finally, we can compute the normalized Stokes parameter $q$ or $u$ (see Eq.~\ref{eq:normalized_stokes_vector}) as:
\begin{equation} 
	x = \dfrac{X}{I_X}.
\label{eq:model_normalized_stokes_parameter} 
\end{equation} 

All reflections downstream of the derotator lie in the horizontal plane, i.e.~parallel to the Nasmyth platform that SPHERE is installed on. 
	These reflections can only produce crosstalk between light linearly polarized at $\SI{\pm45}{\degree}$ with respect to the horizontal plane and circularly polarized light. 
	Light that is linearly polarized in the vertical or horizontal direction is not affected by crosstalk.
	Because the P0-90 analyzer set has vertical/horizontal transmission axes and thus only measures the vertical/horizontal polarization components, crosstalk created downstream of the derotator will not affect the measurements. 
	The P45-135 analyzer set is sensitive to this crosstalk and is therefore not discussed in this work. 
	For polarimetric science observations we strongly advice against using the P45-135 analyzer set.

After computing the double difference, IP from the UT (dominated by M3), M4, the HWP and the derotator remains, because these components are located upstream of the HWP and/or are rotating between the two measurements used in the double difference. 
	In addition, the measurements will be affected by the crosstalk created by these components (IP and crosstalk created by the ADC is found to be negligible). 
	We therefore need to calibrate these instrumental polarization effects.
	To do this, we will start by developing a mathematical model of the complete optical system in the next Section.	
	
%
%

\section{Mathematical description of complete optical system}
\label{sec:model_mathematics}

Before constructing the mathematical model describing the instrumental polarization effects of the optical system, we define two principal reference frames.
	In the \emph{celestial reference frame}, we orient the general reference frame defined in Sect.~\ref{sec:definitions} and Fig.~\ref{fig:reference_frame_general} such that positive Stokes $Q$ is aligned with the local meridian (North up in the sky).
	In the \emph{instrument reference frame}, we orient the general reference frame such that positive Stokes $Q$ corresponds to the vertical direction, i.e.~perpendicular to the Nasmyth platform that SPHERE is installed on.

The goal of our calibration is to obtain a mathematical description of the instrumental polarization effects of the optical system, such that for a given observation we can derive the polarization state of the light incident on the telescope within the required polarimetric accuracy (see Sect.~\ref{sec:introduction} and the top right part of Fig.~\ref{fig:ut_sphere_irdis_schematic}).
	In the general case, we can define the polarimetric accuracy with the following equation~\citep{ichimoto_calib, snik_polreview}:
\begin{equation}
	\skew{2.5}\hat{\boldsymbol{S}}_\mathrm{in} = (\mathbb{I} \pm \Delta Z) \boldsymbol{S}_\mathrm{in},
\label{eq:polarimetric_accuracy_z}
\end{equation}
%
where $\boldsymbol{S}_\mathrm{in}$ is the true Stokes vector incident on the telescope, $\skew{2.5}\hat{\boldsymbol{S}}_\mathrm{in}$ is the measured incident Stokes vector after calibration (after correction for the instrumental polarization effects), $\mathbb{I}$ is the $4 \times 4$ identity matrix and $\Delta Z$ is the $4 \times 4$ matrix describing the polarimetric accuracy.
	Both Stokes vectors in Eq.~\ref{eq:polarimetric_accuracy_z} are defined in the celestial reference frame.
	For a perfect measurement, $\Delta Z$ equals the zero matrix.
	In this work, we write $\Delta Z$ as:	
\begin{equation}
	\Delta Z = \begin{bmatrix} - & - & - & - \\ s_\mathrm{abs} & s_\mathrm{rel} & - & - \\ s_\mathrm{abs} & - & s_\mathrm{rel} & - \\ - & - & - & - \end{bmatrix},	
\label{eq:delta_z}
\end{equation}
with $s_\mathrm{abs}$ and $s_\mathrm{rel}$ the absolute and relative polarimetric accuracies, respectively, as defined in Sect.~\ref{sec:introduction}.
	The values of $s_\mathrm{abs}$ and $s_\mathrm{rel}$ are different for each broadband filter and will be established in Sect.~\ref{sec:model_accuracy} (we will not directly evaluate Eq.~\ref{eq:polarimetric_accuracy_z}, however).
	We do not the determine other elements in Eq.~\ref{eq:delta_z} because for the calibration only a very limited number of different polarization states can be injected into the optical system, and the total intensity is hardly affected by the instrumental polarization effects. \\


\noindent In the following, we will use Mueller calculus~\citep[see e.g.~][]{tinbergen} to construct the model describing the instrumental polarization effects of the complete optical system, i.e. the UT and the instrument. 
	The model parameters and Stokes vectors we will define in the process are displayed in Fig.~\ref{fig:ut_sphere_irdis_schematic}.
	We express the Stokes vector reaching the left (L) or right (R) half of the detector, $\boldsymbol{S}_\mathrm{det,L}$ or $\boldsymbol{S}_\mathrm{det,R}$ (both in the instrument reference frame), in terms of the true Stokes vector incident on the telescope $\boldsymbol{S}_\mathrm{in}$ (in the celestial reference frame) as:
%
\begin{align}
\begin{aligned} 
	\boldsymbol{S}_\mathrm{det,L/R} &= M_\mathrm{sys,L/R} \boldsymbol{S}_\mathrm{in}, \\[0.2cm]
	\begin{bmatrix} I_\mathrm{det,L/R} \\ Q_\mathrm{det,L/R} \\ U_\mathrm{det,L/R} \\ V_\mathrm{det,L/R} \end{bmatrix} &= \begin{bmatrix} I \hspace{-2pt}\rightarrow\hspace{-2pt} I & Q \hspace{-2pt}\rightarrow\hspace{-2pt} I & U \hspace{-2pt}\rightarrow\hspace{-2pt} I & V \hspace{-2pt}\rightarrow\hspace{-2pt} I~~\\~I \hspace{-2pt}\rightarrow\hspace{-2pt} Q & Q \hspace{-2pt}\rightarrow\hspace{-2pt} Q & U \hspace{-2pt}\rightarrow\hspace{-2pt} Q & V \hspace{-2pt}\rightarrow\hspace{-2pt} Q~~\\~I \hspace{-2pt}\rightarrow\hspace{-2pt} U & Q \hspace{-2pt}\rightarrow\hspace{-2pt} U & U \hspace{-2pt}\rightarrow\hspace{-2pt} U & V \hspace{-2pt}\rightarrow\hspace{-2pt} U~~\\~I \hspace{-2pt}\rightarrow\hspace{-2pt} V & Q \hspace{-2pt}\rightarrow\hspace{-2pt} V & U \hspace{-2pt}\rightarrow\hspace{-2pt} V & V \hspace{-2pt}\rightarrow\hspace{-2pt} V~~\end{bmatrix} \begin{bmatrix} I_\mathrm{in} \\ Q_\mathrm{in} \\ U_\mathrm{in} \\ V_\mathrm{in} \end{bmatrix},
\label{eq:mueller_matrix_definition}
\end{aligned}
\end{align} 
	where $M_\mathrm{sys,L/R}$ is the $4\times4$ Mueller matrix describing the instrumental polarization effects of the optical system as seen by the left or right half of the detector.
	The only difference between $M_\mathrm{sys,L}$ and $M_\mathrm{sys,R}$ is the orientation of the transmission axis of the analyzer polarizer.
	In Eq.~\ref{eq:mueller_matrix_definition}, an element $A \hspace{-2pt}\rightarrow\hspace{-2pt} B$ describes the contribution of the incident $A$ into the resulting $B$ Stokes parameter.  
	The optical system is comprised of a sequence of optical components that rotate with respect to each other during an observation.
	To describe the various components and their rotations, we rewrite Eq.~\ref{eq:mueller_matrix_definition} as a multiplication of Mueller matrices~\citep[see e.g.~][]{tinbergen}:
%
\begin{equation} 
	\boldsymbol{S}_\mathrm{det,L/R} = M_n M_{n-1} \cdots M_2 M_1 \boldsymbol{S}_\mathrm{in}.
\label{eq:mueller_matrices_multiplication}
\end{equation}

In Eq.~\ref{eq:mueller_matrices_multiplication}, we do not have to include every separate mirror or component independently. 
	We can combine components which share a fixed reference frame, such as the three mirrors of the derotator. 
	This allows us to create a model with Mueller matrices for only five component groups (see Sect.~\ref{sec:sphere_irdis} and Fig.~\ref{fig:ut_sphere_irdis_schematic}):
\begin{itemize}
\setlength\itemsep{0em}
\item {\makebox[1.2cm]{$M_\mathrm{UT}$,\hfill} the three mirrors of the Unit Telescope (UT),}
\item {\makebox[1.2cm]{$M_\mathrm{M4}$,\hfill} the first mirror of SPHERE (M4),}
\item {\makebox[1.2cm]{$M_\mathrm{HWP}$,\hfill} the half-wave plate (HWP),}
\item {\makebox[1.2cm]{$M_\mathrm{der}$,\hfill} the three mirrors of the derotator,}
\item {\makebox[1.14cm]{$M_\mathrm{CI,L/R}$,\hfill} the optical path downstream of the derotator, in- \makebox[1.3cm]{~}cluding IRDIS and the left or right polarizer of the \makebox[1.3cm]{~}P0-90 analyzer set.}
\end{itemize}
%
$M_\mathrm{M4}$ and $M_\mathrm{CI,L/R}$ are defined in the instrument reference frame, while $M_\mathrm{UT}$, $M_\mathrm{HWP}$ and $M_\mathrm{der}$ have their own (rotating) reference frames.

	The rotations between subsequent reference frames 
can be described by the rotation matrix $T$($\theta$)~\citep[see e.g.~][]{tinbergen}:
\begin{equation} 
	T(\theta) = \begin{bmatrix} 1 & 0 & 0 & 0 \\ 0 & \cos(2\theta) & \sin(2\theta) & 0 \\ 0 & -\sin(2\theta) & \cos(2\theta) & 0 \\ 0 & 0 & 0 & 1 \end{bmatrix},	
	\label{eq:rotation_mueller_matrix} 
\end{equation}
where the component (group) is rotated counterclockwise by an angle $\theta$ when looking into the beam (see Fig.~\ref{fig:reference_frame_general}).
	After applying the Mueller matrix of the optical component $M$ in its own reference frame, the reference frame can be rotated back to the original frame with the rotation matrix $T(-\theta)$:
\begin{equation} 
M_\theta = T(-\theta) M T(\theta),
\label{eq:rotated_mueller_matrix} 	 
\end{equation}  
where $\mathrm{M}_{\theta}$ is the rotated component Mueller matrix.

Taking into account the rotations between the component groups (see Fig.~\ref{fig:ut_sphere_irdis_schematic}), the complete optical system can be described by:
\begin{align}
\boldsymbol{S}_\mathrm{det,L/R} =& ~M_\mathrm{sys,L/R} \boldsymbol{S}_\mathrm{in}, \nonumber \\[0.2cm] 
	\boldsymbol{S}_\mathrm{det,L/R} =& ~M_\mathrm{CI,L/R} T(-\varTheta_\mathrm{der}) M_\mathrm{der} T(\varTheta_\mathrm{der}) T(-\varTheta_\mathrm{HWP}) M_\mathrm{HWP} T(\varTheta_\mathrm{HWP}) \nonumber \\
	&~M_\mathrm{M4}  T(a) M_\mathrm{UT} T(p) \boldsymbol{S}_\mathrm{in},
\label{eq:complete_model} 
\end{align} 
%
where $p$ is the astronomical target's parallactic angle, $a$ is the altitude angle of the telescope, and:
\begin{align}
	\varTheta_\mathrm{HWP} &= \theta_\mathrm{HWP} + \delta_\mathrm{HWP}, \label{eq:theta_hwp}\\[0.3cm]
\varTheta_\mathrm{der} &= \theta_\mathrm{der} + \delta_\mathrm{der}, \label{eq:theta_der}
\end{align} 
with $\theta_\mathrm{HWP}$ the HWP angle, $\theta_\mathrm{der}$ the derotator angle, and $\delta_\mathrm{HWP}$ and $\delta_\mathrm{der}$ the to-be-determined offset angles (due to misalignments) of the HWP and derotator, respectively. 
	$\theta_\mathrm{HWP} = \SI{0}{\degree}$ when the HWP has its fast or slow optic axis vertical
, and $\theta_\mathrm{der} = \SI{0}{\degree}$ when the derotator has its plane of incidence horizontal. 
	The parallactic, altitude, HWP and derotator angles are obtained from the headers of the FITS-files of the measurements (see Appendix~\ref{app:angles_headers}).

Ideally, all 16 elements of the component group Mueller matrices $M_\mathrm{UT}$, $M_\mathrm{M4}$, $M_\mathrm{HWP}$, $M_\mathrm{der}$ and $M_\mathrm{CI,L/R}$ would be determined from calibration measurements that inject a multitude of different polarization states into the system. 
	However, IRDIS' non-rotatable calibration polarizer can only inject light that is nearly 100\% linearly polarized in the positive Stokes $Q$-direction (in the instrument reference frame), and polarized standard stars are limited in number and have a low degree of linear polarization at near-infrared wavelengths. 
	To limit the number of model parameters to determine, we model $M_\mathrm{UT}$, $M_\mathrm{M4}$, $M_\mathrm{HWP}$ and $M_\mathrm{der}$ as a function of their diattenuation ($\epsilon$) and retardance ($\varDelta$)~(see Sect.~\ref{sec:polarization_effects_optical_path}; \citealt{keller_astropolinstrumentation, handbook_optics}):
\begin{equation} 
 	M_\mathrm{com} = \begin{bmatrix} ~1~ & ~\epsilon~ & 0 & 0 \\[1em] ~\epsilon~ & ~1~ & 0 & 0 \\[1em] ~0~ & ~0~ & \sqrt{1 - \epsilon^2}\cos\varDelta & \sqrt{1 - \epsilon^2}\sin\varDelta \\[1em] ~0~ & ~0~ & -\sqrt{1 - \epsilon^2}\sin\varDelta & \sqrt{1 - \epsilon^2}\cos\varDelta \end{bmatrix},
 	\label{eq:component_mueller_matrix} 
\end{equation}    
where we have assumed the transmission of the total intensity, which is a scalar multiplication factor to the matrix, equal to 1.
	The real transmission of the optical system is not important, because we always measure Stokes $Q$ and $U$ relative to the total intensity $I$ and the system transmission cancels out when computing the normalized Stokes parameters and degree and angle of linear polarization (see Eqs.~\ref{eq:normalized_stokes_vector}, \ref{eq:DoLP} and \ref{eq:AoLP}).

For the HWP, $M_\mathrm{com}$ is defined with the positive Stokes $Q$-direction parallel to one of its optic axes. 
	For the other component groups, it is defined with the positive Stokes $Q$-direction perpendicular to the plane of incidence of the mirrors.
	The diattenuation $\epsilon$ has the range $\left[-1, 1 \right]$ and creates IP in the positive Stokes $Q$-direction when $\epsilon > 0$, in the negative $Q$-direction when $\epsilon < 0$ and no IP when $\epsilon = 0$. 
	Ideally, the retardance $\varDelta = \SI{180}{\degree}$, causing no crosstalk and only changing the signs of Stokes $U$ and $V$.
	For other values, an incident Stokes $U$-signal is converted into Stokes $V$ and vice versa. 	
	We use this definition of the retardance for the HWP as well as the other groups containing mirrors, so that we can use the same $M_\mathrm{com}$ for these component groups.
	This is only possible because M4, the UT and the derotator are comprised of an odd number of mirrors; for an even number of mirrors, the signs of Stokes $U$ and $V$ do not change and the ideal $\varDelta$ would be $\SI{0}{\degree}$ with our definition.
	$\epsilon$ and $\varDelta$ depend on the angle of incidence and the wavelength of the light and, for the mirrors, can be computed from the Fresnel equations. 
	
	As outlined in Sect.~\ref{sec:polarization_effects_optical_path}, the effects of the diattenuation and retardance of the optical path downstream of the derotator are negated by the double difference and use of the P0-90 analyzer set, respectively. 
	Therefore, when including the double difference in our mathematical description (see below), $M_\mathrm{CI,L/R}$ only needs to describe the combination of the beamsplitter plate and the left or right linear polarizer of the P0-90 analyzer set.
	To this end, we use Eq.~\ref{eq:component_mueller_matrix}, but set the transmission of the total intensity equal to ${}^1{\mskip -3mu/\mskip -1mu}_2$ and the retardance $\varDelta$ equal to $\SI{0}{\degree}$:
\begin{equation} 
	M_\mathrm{CI,L/R} = \dfrac{1}{2} \begin{bmatrix} 1 & \pm d & 0 & 0 \\ \pm d & 1 & 0 & 0 \\ 0 & 0 & \sqrt{1 - d^2} & 0 \\ 0 & 0 & 0 & \sqrt{1 - d^2} \end{bmatrix},
\label{eq:efficiency_matrix_polarizers}	
\end{equation} 	
where $d$ is the diattenuation of the polarizers that accounts for their imperfect extinction ratios.
	The plus-sign (minus-sign) in Eq.~\ref{eq:efficiency_matrix_polarizers} is used for the left (right) polarizer with the vertical (horizontal) transmission axis.

Because IRDIS uses a non-polarizing beamsplitter with polarizers, rather than a polarizing beamsplitter or Wollaston prism, the transmission of the total intensity of $M_\mathrm{CI,L/R}$ should in reality be set to ${}^1{\mskip -3mu/\mskip -1mu}_4$ rather than ${}^1{\mskip -3mu/\mskip -1mu}_2$.
	However, in practice the reference flux measurements are taken with the polarizers inserted, but are generally not multiplied by a factor 2 to account for the loss of flux. 
	We therefore choose to set the transmission of the total intensity to ${}^1{\mskip -3mu/\mskip -1mu}_2$ to prevent accidental (relative) photometric errors.



As the final step, we will compute the double-difference Stokes $Q$ or $U$ and the corresponding double-sum intensity $I_Q$ or $I_U$ from the Mueller matrix description of the optical path. 
	For this, we first compute $\boldsymbol{S}_\mathrm{det,L}$ and $\boldsymbol{S}_\mathrm{det,R}$ from Eq.~\ref{eq:complete_model} using $+d$ and $-d$, respectively, in Eq.~\ref{eq:efficiency_matrix_polarizers}.
	We then obtain $I_\mathrm{det,L}$ and $I_\mathrm{det,R}$ from the first element of $\boldsymbol{S}_\mathrm{det,L}$ and $\boldsymbol{S}_\mathrm{det,R}$.
	Subsequently,  we use $I_\mathrm{det,L}$ and $I_\mathrm{det,R}$ to compute the single differences $X^\pm$ and corresponding single sums $I_{X^\pm}$ from Eqs.~\ref{eq:single_difference} and \ref{eq:single_sum}, respectively.	
	After computing the single difference and single sum for two measurements, 
we compute the double-difference $X$ and corresponding double-sum $I_X$ (see Eqs.~\ref{eq:double_difference} and \ref{eq:double_sum}, respectively) as:
\begin{align} 
	X &= \dfrac{1}{2} \left[ X^+(p^+, a^+, \theta_\mathrm{HWP}^+, \theta_\mathrm{der}^+) - X^-(p^-, a^-, \theta_\mathrm{HWP}^-, \theta_\mathrm{der}^-) \right], \label{eq:model_double_difference}\\[0.3cm]
	I_X &= \dfrac{1}{2} \left[ I_{X^+}(p^+, a^+, \theta_\mathrm{HWP}^+, \theta_\mathrm{der}^+) + I_{X^-}(p^-, a^-, \theta_\mathrm{HWP}^-, \theta_\mathrm{der}^-) \right], \label{eq:model_double_sum} 
\end{align} 
where we explicitly show that $X^\pm$ and $I_{X^\pm}$ are functions of the parallactic, altitude, HWP and derotator angles of the first (superscript $+$) and second (superscript $-$) measurement. 
	Finally, we compute the normalized Stokes parameter $x$ from Eq.~\ref{eq:model_normalized_stokes_parameter}.

The rotation laws of the derotator and HWP in field- and pupil-tracking mode 
are such that for an ideal optical system, $X$ (or $x$) in the instrument reference frame would correspond to $Q_\mathrm{in}$ ($q_\mathrm{in}$) and $U_\mathrm{in}$ ($u_\mathrm{in}$) in the celestial reference frame for HWP switch angle combinations $\left[\SI{0}{\degree}, \SI{45}{\degree}\right]$ and $\left[\SI{22.5}{\degree}, \SI{67.5}{\degree}\right]$, respectively\footnote{For pupil-tracking observations this is true since January 22, 2019, when the new HWP rotation law was implemented~\citep[see also][]{vanholstein_exopol}.}.
	However, the optical system is not ideal.
	We therefore need to determine the model parameters of the five component group Mueller matrices ($\epsilon$'s, $\varDelta$'s and $d$) and the HWP and derotator offset angles $\delta_\mathrm{HWP}$ and $\delta_\mathrm{der}$ (see Fig.~\ref{fig:ut_sphere_irdis_schematic}).
	When we have the values of these model parameters, we can mathematically describe any measurement 
and invert the equations to derive $\skew{2.5}\hat{\boldsymbol{S}}_\mathrm{in}$, the estimate of the true incident Stokes vector $\boldsymbol{S}_\mathrm{in}$.

%
%

\section{Instrumental polarization effects of instrument downstream of M4}
\label{sec:response_downstream_m4}

\subsection{Calibration measurements and determination of model parameters}
\label{sec:measurements_downstream_m4}


With the Mueller matrix model of the telescope and instrument defined, we can now determine the model parameters describing the optical path downstream of M4. 
	To this end, we have taken measurements with the internal light source (see Fig.~\ref{fig:ut_sphere_irdis_schematic}) using the Y-, J-, H- and K$_\mathrm{s}$-band filters.
	The following data sets were obtained:	
\begin{itemize}
\setlength\itemsep{0em}
\item On August 15, 2015, a total of 528 exposures were taken \emph{with} the calibration polarizer inserted, injecting light that is nearly 100\% linearly polarized in the vertical direction (in the positive $Q$-direction in the instrument reference frame). 
	The derotator and HWP were rotated between the exposures with $\theta_\mathrm{der}$ ranging from \SI{0}{\degree} to \SI{90}{\degree} and $\theta_\mathrm{HWP}$ ranging from \SI{0}{\degree} to \SI{101.25}{\degree} (varying step sizes).
	This data, hereafter called the \emph{polarized source measurements}, is used to determine for each broadband filter the retardances of the derotator and HWP ($\varDelta_\mathrm{der}$ and $\varDelta_\mathrm{HWP}$), the offset angles of the derotator and HWP ($\delta_\mathrm{der}$ and $\delta_\mathrm{HWP}$) and the diattenuation of the polarizers ($d$).
\item On June 12 and 13, 2016, a total of 400 exposures were taken \emph{without} the calibration polarizer inserted, so that almost completely unpolarized light was injected.
	The derotator and HWP were rotated between the exposures with $\theta_\mathrm{der}$ and $\theta_\mathrm{HWP}$ ranging from \SI{0}{\degree} to \SI{101.25}{\degree} with a step size of \SI{11.25}{\degree}. 
	This data, hereafter called the \emph{unpolarized source measurements}, 
is used to fit for each broadband filter the diattenuations of the derotator and HWP ($\epsilon_\mathrm{der}$ and $\epsilon_\mathrm{HWP}$).
	The light injected is actually weakly polarized, because it is reflected off M4 before reaching the HWP.
	We therefore also fit the injected normalized Stokes parameters $q_\mathrm{in,unpol}$ and $u_\mathrm{in,unpol}$.
\end{itemize}

We pre-process the data by applying dark subtraction, flat fielding and bad-pixel correction according to Paper~I.
	Subsequently, we construct double-difference and double-sum images from Eqs.~\ref{eq:double_difference} and \ref{eq:double_sum}, respectively, using pairs of exposures with the same $\theta_\mathrm{der}$ and with $\theta_\mathrm{HWP}^+$ (first measurement) and $\theta_\mathrm{HWP}^-$ (second measurement) differing \SI{45}{\degree}. 
	In this case the images do not always correspond to $Q$-, $U$-, $I_Q$- and $I_U$-images in the instrument reference frame, because HWP angles different from \SI{0}{\degree}, \SI{45}{\degree}, \SI{22.5}{\degree} and \SI{67.5}{\degree} have been used as well.
	The only model parameter that cannot be determined from these double-difference and double-sum images is the derotator diattenuation $\epsilon_\mathrm{der}$, because with the constant derotator angle the derotator's induced polarization is removed in the double difference.
	Therefore, the unpolarized source measurements are used to create additional double-difference and double-sum images by pairing exposures with the same $\theta_\mathrm{HWP}$ (rather than $\theta_\mathrm{der}$) and with $\theta_\mathrm{der}^+$ (first measurement) and $\theta_\mathrm{der}^-$ (second measurement) differing \SI{45}{\degree}.

The flux in most of the produced images is not uniform, but displays a gradient (for a detailed description see Appendix~\ref{app:structure_flux}).
	To take into account the resulting uncertainty in the normalized Stokes parameters, we compute the median of the double-difference and double-sum images in nine apertures (100 pixel radii, arranged $3 \times 3$) located throughout almost the complete frame. 
	Subsequently, we calculate the normalized Stokes parameters according to Eq.~\ref{eq:model_normalized_stokes_parameter}. 
	This yields a total of 6696 data points with nine data points for every derotator and HWP angle combination. 
	We will determine the model parameters based on all these data points together such that our model is valid over the complete field of view.
	
To describe the measurements, we use Eq.~\ref{eq:model_normalized_stokes_parameter} and insert the model equations of Sect.~\ref{sec:model_mathematics}. 
	This set of equations comprises the model function.
	We apply only the part of Eq.~\ref{eq:complete_model} without the UT and M4:
%
\begin{align}
	\boldsymbol{S}_\mathrm{det,L/R} =& ~M_\mathrm{CI,L/R} T(-\varTheta_\mathrm{der}) M_\mathrm{der} T(\varTheta_\mathrm{der}) \nonumber \\
	&~T(-\varTheta_\mathrm{HWP}) M_\mathrm{HWP} T(\varTheta_\mathrm{HWP}) \boldsymbol{S}_\mathrm{HWP}, \label{eq:model_internal_calibration}
\end{align} 
where $\boldsymbol{S}_\mathrm{HWP}$ is the Stokes vector injected upstream of the HWP (in the instrument reference frame; see Fig.~\ref{fig:ut_sphere_irdis_schematic}).
	For the polarized source measurements, it is difficult to discern the diattenuation (due to the imperfect extinction ratio) of the calibration polarizer from that of the analyzer polarizers.
	Therefore, we assume the diattenuations of the calibration and analyzer polarizers to be identical and write $\boldsymbol{S}_\mathrm{HWP} = T(-\delta_\mathrm{cal}) \left[1, d, 0, 0\right]^\mathrm{T}$, with $\delta_\mathrm{cal}$ the offset angle of the calibration polarizer that we will also fit from the measurements (see Fig.~\ref{fig:ut_sphere_irdis_schematic}). 
	For the unpolarized source measurements, the incident light will be weakly polarized due to the reflection off M4.
	We therefore write $\boldsymbol{S}_\mathrm{HWP} = [1, q_\mathrm{in, unpol}, u_\mathrm{in, unpol}, 0]^\mathrm{T}$, with $q_\mathrm{in, unpol}$ and $u_\mathrm{in, unpol}$ the to-be-determined injected normalized Stokes parameters, assuming that no circularly polarized light will be produced. 
	Note that there are no degeneracies among the model parameters with the above definitions of $\boldsymbol{S}_\mathrm{HWP}$, because the derotator, HWP, calibration polarizer and M4 each have their own independent (local) references frames.

With the description of the measurements complete, we determine the model parameters by fitting the model function to the data points using non-linear least squares (with sequential least squares programming as implemented in the Python function \emph{scipy.optimize.minimize}).
	The HWP and derotator angles required for this are obtained from the headers of the FITS-files of the measurements (see Appendix~\ref{app:angles_headers}).
	To prevent the values of $\epsilon_\mathrm{HWP}$ and $\epsilon_\mathrm{der}$ from being dominated by the polarized source measurements (which have larger residuals), we fit the data of the polarized and unpolarized source measurements sequentially and repeat the two fits until convergence.
	The graphs of the model fits including the residuals can be found in Appendix~\ref{app:graphs_internal}.
			

\subsection{Results and discussion for internal source calibrations}
\label{sec:results_discussion_instrument}

The resulting values for the model parameters are shown in Table~\ref{tab:parameters_instrument}. 
	The $1\sigma$-uncertainties of the parameters are also tabulated and are computed from the residuals of fit using a linear approximation (see Appendix~\ref{app:uncertainty_parameters}).
	For this calculation it was necessarily assumed that the determined model parameters are uncorrelated and that they do not contain systematic errors.
	The systematic errors are likely very small, because the residuals of fit are close to normally distributed (see Figs.~\ref{fig:STOKES_POL_BB_H}, \ref{fig:STOKES_UNPOL_BB_H} and \ref{fig:STOKES_UNPOL_BB_H_DERROT}).  
%
\begin{table*}
\caption{Determined parameters and their errors of the part of the model describing the instrument downstream of M4 in Y-, J-, H- and K$_\mathrm{s}$-band. 
	The retardances of the derotator and HWP ($\varDelta_\mathrm{der}$ and $\varDelta_\mathrm{HWP}$, respectively)
 cause the strongest instrumental polarization effects (i.e.~crosstalk) and are indicated in red. 
The polarizer diattenuations $d$ correspond to extinction ratios (computed as ${(1 + d)} \hspace{1pt} / \hspace{1pt} {(1 - d)}$) of 100:1, 189:1, 447:1 and 126:1 in Y-, J-, H- and K$_\mathrm{s}$-band, respectively.}
\centering
\setlength{\tabcolsep}{1pt}
\begin{tabular}{l c c r c l c r c l c r c l c r c l}
\hline\hline
\multicolumn{2}{c}{Parameter} & \hspace{16pt} & \multicolumn{3}{c}{BB\_Y} & \hspace{16pt} &  \multicolumn{3}{c}{BB\_J} & \hspace{16pt} &  \multicolumn{3}{c}{BB\_H} & \hspace{16pt} &  \multicolumn{3}{c}{BB\_$\mathrm{K_s}$} \T\B \\ 
\hline
$\epsilon_\mathrm{HWP}$ & & & -0.00021 & $\pm$ & $2\cdot10^{-5}$ & & -0.000433 & $\pm$ & $4\cdot10^{-6}$ & & -0.000297 & $\pm$ & $7\cdot10^{-6}$ & & -0.000415 & $\pm$ & $8\cdot10^{-6}$ \T \\
\color{red}$\varDelta_\mathrm{HWP}$ & \color{red}$(^\circ)$ & & \color{red}184.2 & \color{red}$\pm$ & \color{red}0.2 & & \color{red}177.5 & \color{red}$\pm$ & \color{red}0.2 & & \color{red}170.7 & \color{red}$\pm$ & \color{red}0.1 & & \color{red}177.6 & \color{red}$\pm$ & \color{red}0.1 \\
$\delta_\mathrm{HWP}$ & (\si{\degree}) & & -0.6132 & $\pm$ & 0.0007 & & -0.6132 & $\pm$ & 0.0007 & & -0.6132 & $\pm$ & 0.0007 & & -0.6132 & $\pm$ & 0.0007 \\
$\epsilon_\mathrm{der}$ & & & -0.00094 & $\pm$ & $2\cdot10^{-5}$ & & -0.008304 & $\pm$ & $6\cdot10^{-6}$ & & -0.002260 & $\pm$ & $7\cdot10^{-6}$ & & 0.003552 & $\pm$ & $7\cdot10^{-6}$ \\
\color{red}$\varDelta_\mathrm{der}$ & \color{red}$(^\circ)$ & & \color{red}126.1 & \color{red}$\pm$ & \color{red}0.1 & & \color{red}156.1 & \color{red}$\pm$ & \color{red}0.1 & & \color{red}99.32 & \color{red}$\pm$ & \color{red}0.06 & & \color{red}84.13 & \color{red}$\pm$ & \color{red}0.05 \\
$\delta_\mathrm{der}$ & (\si{\degree}) & & 0.50007 & $\pm$ & $6\cdot10^{-5}$ & & 0.50007 & $\pm$ & $6\cdot10^{-5}$ & & 0.50007 & $\pm$ & $6\cdot10^{-5}$ & & 0.50007 & $\pm$ & $6\cdot10^{-5}$ \\
$d$ & & & 0.9802 & $\pm$ & 0.0004 & & 0.9895 & $\pm$ & 0.0002 & & 0.9955 & $\pm$ & 0.0002 & & 0.9842 & $\pm$ & 0.0003 \B \\
\hline
$q_\mathrm{in, unpol}$ & (\%) & & 1.789 & $\pm$ & 0.001 & & 1.2150 & $\pm$ & 0.0003 & & 0.9480 & $\pm$ & 0.0005 & & 0.8352 & $\pm$ & 0.0006 \T \\
$u_\mathrm{in, unpol}$ & (\%) & & 0.061 & $\pm$ & 0.002 & & 0.0585 & $\pm$ & 0.0004 & & 0.0406 & $\pm$ & 0.0007 & & 0.0589 & $\pm$ & 0.0008 \\
\B $\delta_\mathrm{cal}$ & (\si{\degree}) & & -1.542 & $\pm$ & 0.001 & & -1.542 & $\pm$ & 0.001 & & -1.542 & $\pm$ & 0.001 & & -1.542 & $\pm$ & 0.001 \\
\hline
\end{tabular}
\label{tab:parameters_instrument}
\end{table*}
%
	
\begin{figure}
\centering 
\includegraphics[width=\hsize]{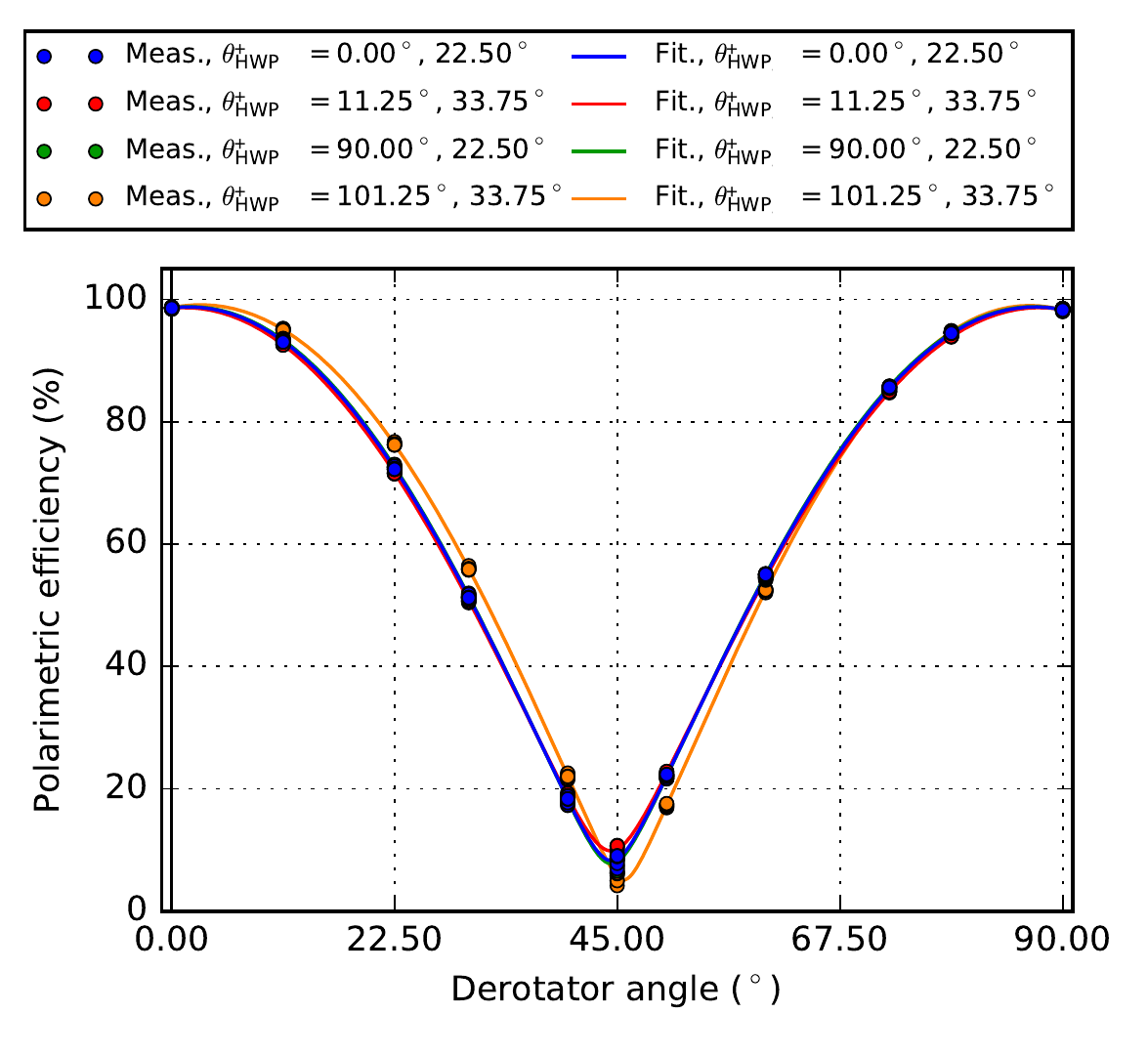} 
\caption{Measured and fitted polarimetric efficiency of the instrument downstream of M4 as a function of HWP and derotator angle in H-band. The legend only shows the $\theta_\mathrm{HWP}^+$-values of each data point or curve; it is implicit that the corresponding values for $\theta_\mathrm{HWP}^-$ differ \SI{45}{\degree} from those of $\theta_\mathrm{HWP}^+$. Note that the measurement points and fitted curves for $\theta_\mathrm{HWP}^+ = \SI{0.00}{\degree}, \SI{22.50}{\degree}$ (blue) and $\theta_\mathrm{HWP}^+ = \SI{90.00}{\degree}, \SI{22.50}{\degree}$ (green) are overlapping.} 
\label{fig:DOLP_POL_BB_H} 
\end{figure}

	To visualize the effect of the parameters determined from the polarized source measurements, we plot the measured and fitted degree of linear polarization of the H-band polarized source measurements as a function of HWP and derotator angle in Fig.~\ref{fig:DOLP_POL_BB_H}.	
	Recall that the data points created in Sect.~\ref{sec:measurements_downstream_m4} are normalized Stokes parameters computed from the double difference and double sum using \emph{pairs of exposures} with $\theta_\mathrm{HWP}^+$ (first exposure) and $\theta_\mathrm{HWP}^-$ (second exposure) differing \SI{45}{\degree}.
	The degree of linear polarization (see Eq.~\ref{eq:DoLP}) is computed from \emph{pairs of data points} with values for $\theta_\mathrm{HWP}^+$ (and therefore also values for $\theta_\mathrm{HWP}^-$) that differ \SI{22.5}{\degree} or \SI{67.5}{\degree} from each other.
	The effect of the gradient in the measured flux (see Appendix~\ref{app:structure_flux}) appears to be limited, because the nine data points of each HWP and derotator angle combination in Fig.~\ref{fig:DOLP_POL_BB_H} are relatively close together, within a few percent.
%
	For these polarized source measurements, which have nearly 100\% polarized light incident, we interpret the degree of linear polarization as the polarimetric efficiency, i.e.~the fraction of the incident or true linear polarization that is actually measured.
%

For an ideal instrument, the polarimetric efficiency is 100\%. 
	However, in Fig.~\ref{fig:DOLP_POL_BB_H} a dramatic decrease in polarimetric efficiency is seen around $\theta_\mathrm{der} = \SI{45}{\degree}$, reaching values as low as $5\%$. 
	This low efficiency indicates severe loss of polarization signal and is due to the derotator retardance strongly deviating from the ideal value of \SI{180}{\degree}. 
	With $\varDelta_\mathrm{der} = \SI{99.32}{\degree}$, the derotator acts almost as a quarter-wave plate for which $\varDelta = \SI{90}{\degree}$.
	Around $\theta_\mathrm{der} = \SI{45}{\degree}$, the derotator therefore produces strong crosstalk and almost all incident linearly polarized light is converted into circularly polarized light for which the P0-90 analyzer set is not sensitive. 
	We already encountered the strongly varying polarimetric efficiency in Fig.~3 of Paper~I.

The retardance of the HWP has a much smaller effect on the polarimetric efficiency than the retardance of the derotator, as $\varDelta_\mathrm{HWP} = \SI{170.5}{\degree}$ in H-band, relatively close to the ideal value of \SI{180}{\degree}.
	In Fig.~\ref{fig:DOLP_POL_BB_H} the effect of the HWP retardance is visible as the changing skewness of the fitted curves for different HWP angles.
	The offset angles $\delta_\mathrm{HWP}$, $\delta_\mathrm{der}$ and $\delta_\mathrm{cal}$ also contribute a small shift of the curves. 
	Finally, the diattenuation of the polarizers $d$ determines the maximum values of the curves around $\theta_\mathrm{der} = \SI{0}{\degree}$ and $\theta_\mathrm{der} = \SI{90}{\degree}$.
	
\begin{figure}
\centering 
\includegraphics[width=\hsize]{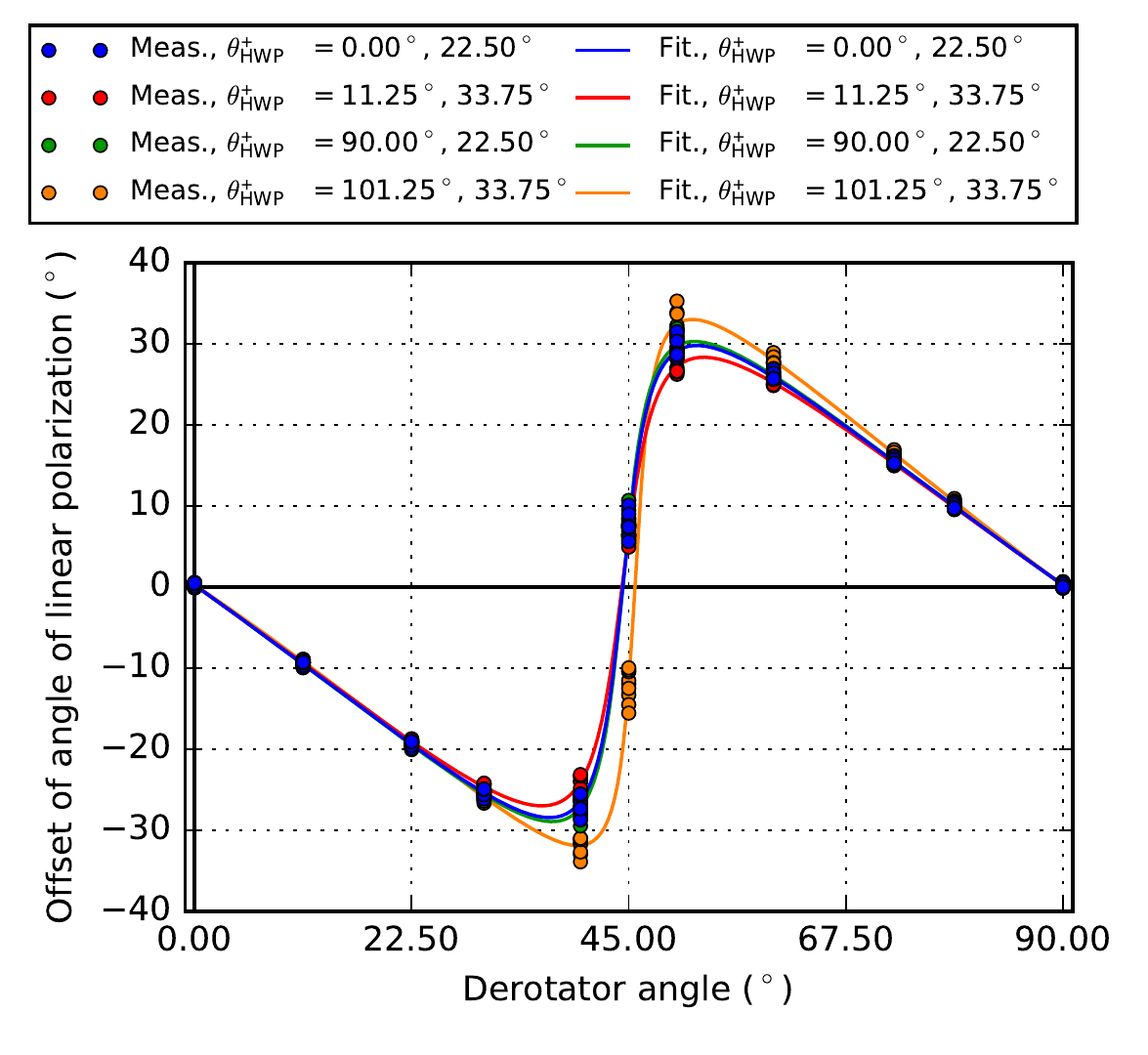} 
\caption{Measured and fitted offset of the angle of linear polarization induced by the instrument downstream of M4 as a function of HWP and derotator angle in H-band. The legend only shows the $\theta_\mathrm{HWP}^+$-values of each data point or curve; it is implicit that the corresponding values for $\theta_\mathrm{HWP}^-$ differ \SI{45}{\degree} from those of $\theta_\mathrm{HWP}^+$. Note that the measurement points and fitted curves for $\theta_\mathrm{HWP}^+ = \SI{0.00}{\degree}, \SI{22.50}{\degree}$ (blue) and $\theta_\mathrm{HWP}^+ = \SI{90.00}{\degree}, \SI{22.50}{\degree}$ (green) are overlapping.} 
\label{fig:AOLP_POL_BB_H} 
\end{figure}

The crosstalk produced by the derotator and HWP not only deteriorates the polarimetric efficiency, but also induces an offset in the measurement of the angle of linear polarization, as is illustrated by the varying Stokes $Q$- and $U$-images in Fig.~3 of Paper~I. 
	Figure~\ref{fig:AOLP_POL_BB_H} (of this paper) shows the measured and fitted offsets of the angle of linear polarization corresponding to the curves of Fig.~\ref{fig:DOLP_POL_BB_H}.
	The offsets are computed as the actually measured angle of linear polarization (see Eq.~\ref{eq:AoLP}) minus the angle that would be measured in case the optical system were ideal.
	Figure~\ref{fig:AOLP_POL_BB_H} shows that the measured angle of linear polarization varies around the ideal angle, with a maximum deviation of $\SI{34}{\degree}$ and the strongest rotation rate around $\theta_\mathrm{der} = \SI{45}{\degree}$. \\
%
%
%
\begin{figure*}
\centering 
\includegraphics[width=14.3cm]{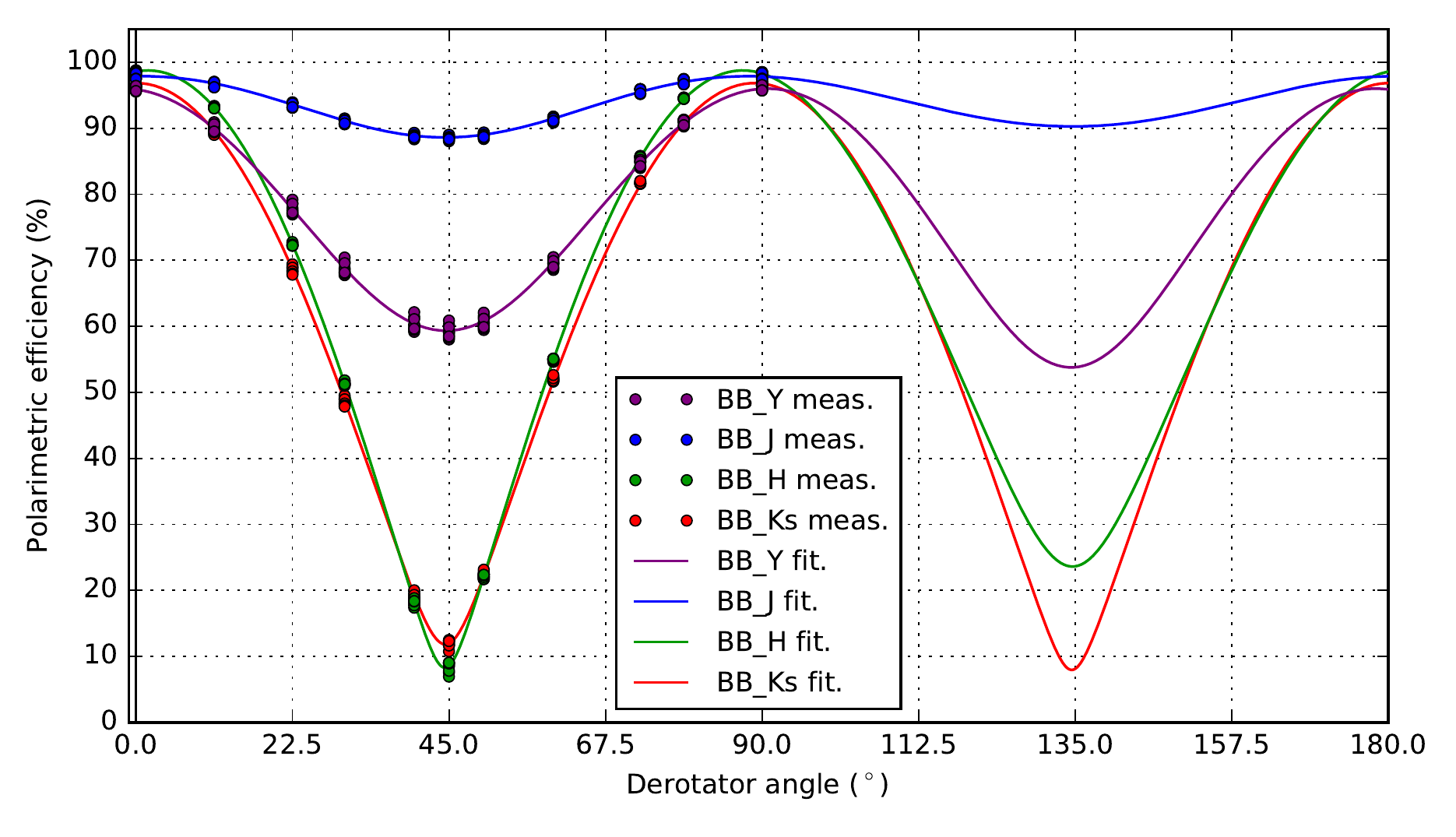} 
\caption{
Measured and fitted polarimetric efficiency of the instrument downstream of M4 with $\theta_\mathrm{HWP}^+ = \SI{0}{\degree}, \SI{22.5}{\degree}$ (and therefore $\theta_\mathrm{HWP}^- = \SI{45}{\degree}, \SI{67.5}{\degree}$) as a function of derotator angle in Y-, J-, H- and K$_\mathrm{s}$-band.}
\label{fig:DOLP_POL_ALL_BBF} 
\end{figure*} 
%
%
\begin{figure*}
\centering 
\includegraphics[width=14.3cm]{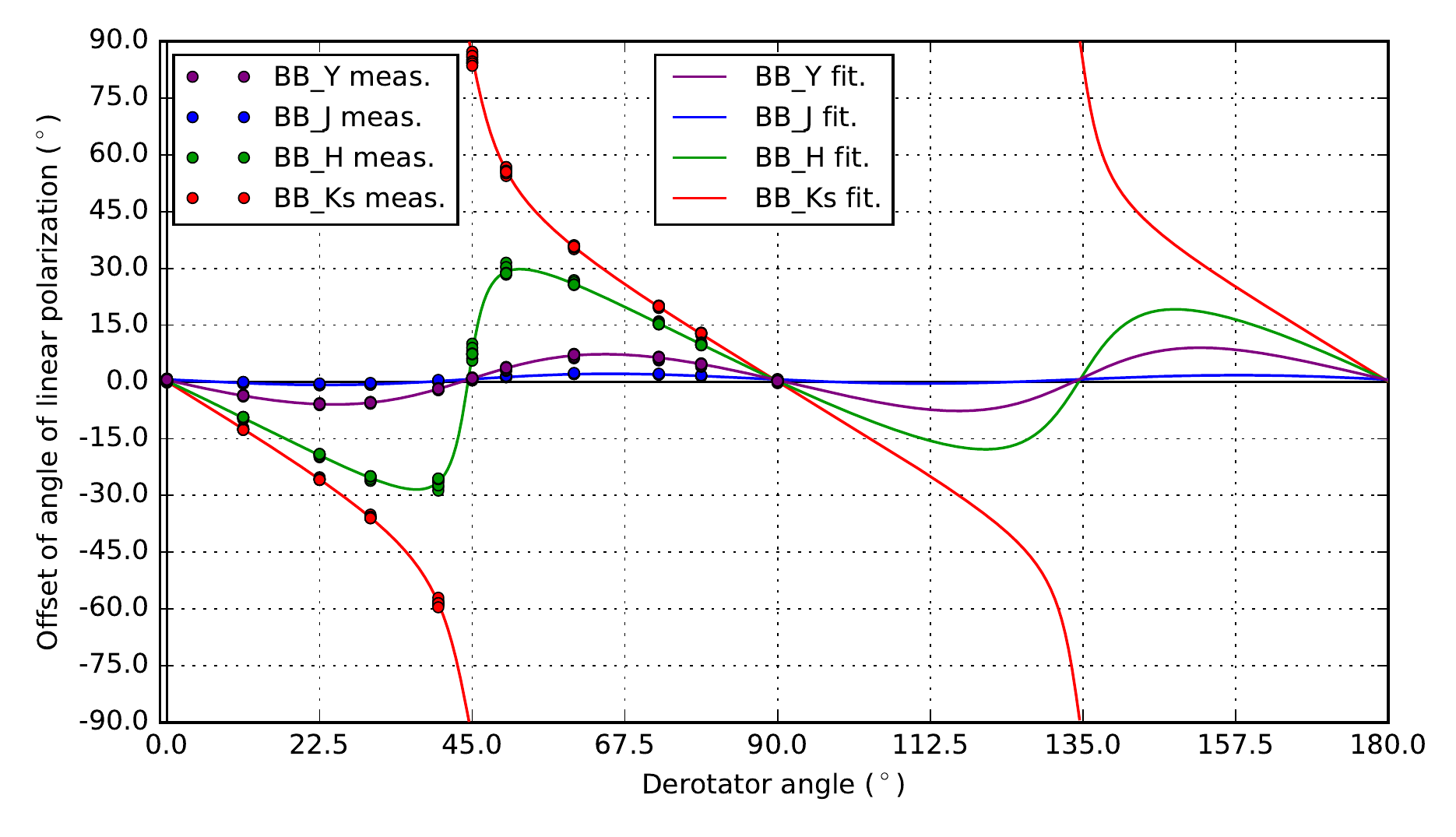} 
\caption{Measured and fitted offset of angle of linear polarization induced by the instrument downstream of M4 with $\theta_\mathrm{HWP}^+ = \SI{0}{\degree}, \SI{22.5}{\degree}$ (and therefore $\theta_\mathrm{HWP}^- = \SI{45}{\degree}, \SI{67.5}{\degree}$) as a function of derotator angle in Y-, J-, H- and K$_\mathrm{s}$-band.}
\label{fig:AOLP_POL_ALL_BBF} 
\end{figure*} 

\noindent Fig.~\ref{fig:DOLP_POL_ALL_BBF} shows the polarimetric efficiency in the four broadband filters Y, J, H and K$_\mathrm{s}$. 
	The curves displayed are for $\theta_\mathrm{HWP}^+ = \SI{0}{\degree}$ and \SI{22.5}{\degree} and the derotator angle ranges from \SI{0}{\degree} to \SI{180}{\degree} (the curves repeat for $\theta_\mathrm{der} > \SI{180}{\degree}$). 
	We have also taken measurements in the range $\SI{0}{\degree} \leq \theta_\mathrm{der} \leq \SI{180}{\degree}$ (not shown) that confirm the curves for $\theta_\mathrm{der} > \SI{90}{\degree}$. 
	However, we do not use these measurements to determine the model parameters, because neutral density filters were inserted which appear to depolarize the light by a few percent.
	Because the nine data points of each HWP and derotator angle combination are relatively close together, we conclude that the effect of the gradient in the measured flux is small for all filters.

From Fig.~\ref{fig:DOLP_POL_ALL_BBF} it follows that for all filters, the efficiency is minimum around $\theta_\mathrm{der} = \SI{45}{\degree}$ and $\theta_\mathrm{der} = \SI{135}{\degree}$.
	The minimum values of the curves differ substantially among the filters, because the derotator retardance varies strongly with wavelength (see Table~\ref{tab:parameters_instrument}).
	The exact shape and minimum values of the curves depend on the HWP angles used (see Fig.~\ref{fig:DOLP_POL_BB_H}), because the HWP retardance deviates slightly from the ideal value of \SI{180}{\degree} in all filters (strongest in H-band; see Table~\ref{tab:parameters_instrument}).
	The asymmetry with respect to $\theta_\mathrm{der} = \SI{90}{\degree}$ visible in Fig.~\ref{fig:DOLP_POL_ALL_BBF} is also due to the non-ideal HWP retardance.

The absolute minimum polarimetric efficiency is the lowest in H-band for which it is $5\%$.
	Also K$_\mathrm{s}$-band (efficiency $\geq 7\%$) shows a strongly varying performance, while in Y-band (${\geq}54\%$) and especially J-band (${\geq}89\%$) the polarimetric efficiency is much less affected by the derotator angle.
	The polarimetric efficiency during science observations, and an observation strategy in which the derotator angle is optimized to prevent observing at a low polarimetric efficiency are discussed in Paper~I. 	

Figure~\ref{fig:AOLP_POL_ALL_BBF} shows the offsets of the angle of linear polarization corresponding to the polarimetric efficiency curves of Fig.~\ref{fig:DOLP_POL_ALL_BBF}. 
	Also in this case the non-ideal HWP retardance causes an asymmetry with respect to $\theta_\mathrm{der} = \SI{90}{\degree}$ and variations of the exact shape and maximum values of the curves with HWP angle (see Fig.~\ref{fig:AOLP_POL_BB_H}).
	While the variation around the ideal value is marginal in J-band, with a maximum deviation of $\SI{4}{\degree}$, the offset of the angle of linear polarization is $\leq \SI{11}{\degree}$ in Y-band and $\leq \SI{34}{\degree}$ in H-band. 
	For K$_\mathrm{s}$-band, the angle of linear polarization does not even return to the ideal value around $\theta_\mathrm{der} = \SI{45}{\degree}$ and $\theta_\mathrm{der} = \SI{135}{\degree}$, but continues rotating beyond \SI{\pm90}{\degree} (where a rotation of $+\SI{90}{\degree}$ is indistinguishable from \SI{-90}{\degree}).

\begin{figure} 
\centering 
\includegraphics[width=0.83\hsize]{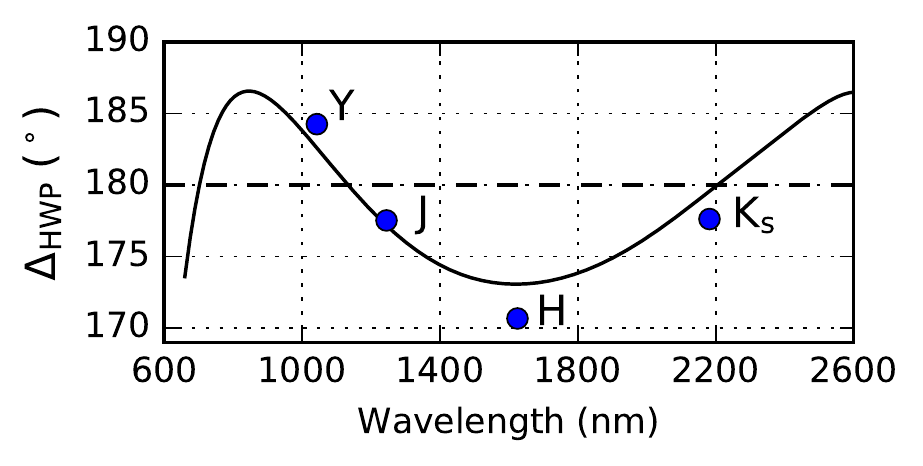} 
\caption{HWP retardance as a function of wavelength as specified by the manufacturer$^{\ref{fn:hwp}}$ compared to the determined HWP retardance ($\varDelta_\mathrm{HWP}$) in Y-, J-, H- and K$_\mathrm{s}$-band.}
\label{fig:hwp_halle} 
\end{figure}

To validate the determined HWP retardances in the four filters, the values are compared to the retardance as specified by the manufacturer in Fig.~\ref{fig:hwp_halle}.
	The error bars on the determined HWP retardances are smaller than the size of the symbols used.
	It follows that the determined HWP retardances are accurate, since they follow the general shape of the curve and are well within the 4\% manufacturing tolerance 
as specified by the manufacturer\footnote{\label{fn:hwp}B. Halle Nachfl. GmbH, \url{http://www.b-halle.de/products/Retarders/Achromatic_Retarders.html}, consulted November 21, 2017.}. \\
	
\noindent For the unpolarized source measurements, the light incident on the HWP is primarily linearly polarized in the positive $Q$-direction as follows from the determined values of $q_\mathrm{in,unpol}$ and $u_\mathrm{in,unpol}$. 
	The degree of linear polarization decreases with increasing wavelength (from Y- to K$_\mathrm{s}$-band). 
	This polarization signal must be IP from M4 that is in between the internal light source and the HWP (see Fig.~\ref{fig:ut_sphere_irdis_schematic}). 
	The determined values of $q_\mathrm{in,unpol}$ are also in good agreement with the determined diattenuations of M4 (see Fig.~\ref{fig:M3M4_IP} and the discussion in Sect.~\ref{sec:response_telescope_results_discussion}), and shows that the light from the internal light source is almost completely unpolarized until it reaches M4.
	
The polarization signals induced by the HWP and the derotator are very small, since $\epsilon_\mathrm{HWP}$ and $\epsilon_\mathrm{der}$ are very close to the ideal value of 0 in all filters (with the largest deviation for the derotator in J-band; see Table~\ref{tab:parameters_instrument}). 
	The low diattenuation of the derotator is as expected, because its main surface coating is protected silver that is highly reflective. 
	However, considering that the derotator has its plane of incidence horizontal when $\theta_\mathrm{der} = \SI{0}{\degree}$, one would naively expect $\epsilon_\mathrm{der}$ to be positive in all filters (producing polarization in the positive $Q$-direction) while it turns out to be negative (producing polarization in the negative $Q$-direction) in three of the four filters.
	This behavior of the diattenuation with wavelength is likely due to the complex combination of coatings on the derotator mirrors. \\
	
\noindent The strong crosstalk produced by the derotator in H- and K$_\mathrm{s}$-band can also be used to our advantage. 
	In these filters, the retardance of the derotator is close to that of a quarter-wave plate (close to \SI{90}{\degree}; see Table~\ref{tab:parameters_instrument}).
	At $\theta_\mathrm{der} = \SI{45}{\degree}$ and \SI{135}{\degree}, the derotator will not only convert almost all incident linearly polarized into circularly polarized light (problematic for the polarimetric efficiency), but will also convert almost all incident circularly polarized light into linearly polarized light that can then be measured by the P0-90 analyzer set. 
	Hence by using the derotator as a quarter-wave plate to modulate Stokes $V$, we can measure circularly polarized light, for example from molecular clouds.
	The development of a technique to measure circularly polarized light with IRDIS is beyond the scope of this paper and will be left for future work.

%
%

\section{Instrumental polarization effects of telescope and M4}
\label{sec:response_telescope_m4}

\subsection{Calibration measurements and determination of model parameters}
\label{sec:measurements_telescope_m4}

Now that we have a validated description of the optical path downstream of M4, we can complete our instrument model by determining the model parameters describing the UT and M4 (see Fig.~\ref{fig:ut_sphere_irdis_schematic}).
	On June 15, 2016, we therefore observed the unpolarized standard star HD~176425~(\citealt{turnshek_standardstars}; $0.020\pm0.009\%$ polarized in B-band) at different telescope altitude angles using the four broadband filters Y, J, H and K$_\mathrm{s}$ under program ID 60.A-9800(S).	
	Because M1 and M3 were re-aluminized between April 3 and April 16, 2017, we repeated the calibration measurements on August 21, 2018 with the unpolarized star HD~217343 under program ID 60.A-9801(S).
	Although HD~217343 is not an unpolarized standard star, it is located at only $31.8$ pc from Earth~\citep{gaia_dr2} and therefore the probability of it being polarized by interstellar dust is very low. 

The two data sets are used to determine the diattenuations of the UT and M4 ($\epsilon_\mathrm{UT}$ and $\epsilon_\mathrm{M4}$) before and after the re-aluminization of M1 and M3. 
	The retardances of the UT and M4 ($\varDelta_\mathrm{UT}$ and $\varDelta_\mathrm{M4}$) are assumed to be equal for both data sets and are computed analytically because 
their limited effect does not justify dedicated calibration measurements (see Sect.~\ref{sec:response_telescope_results_discussion}).
	In addition the 
degree of linear polarization of polarized standard stars at near-infrared wavelengths is too low 
to accurately determine the retardances, and observations of the polarized daytime sky~\citep[see e.g~][]{harrington_skycal, boer_naco, harrington_skycallimits} are very time consuming.
		
	
During the observations of HD~176425 (2016), the derotator was fixed with its plane of incidence horizontal ($\theta_\mathrm{der} = \SI{0}{\degree}$) to ensure a polarimetric efficiency close to 100\%. 
	The adaptive optics were turned off (open-loop) to 
	reach a large total photon count per detector integration time, minimizing read-out noise.
	The calibration polarizer was out of the beam.
	For every filter, 10 HWP cycles (measurements with $\theta_\mathrm{HWP} = \SI{0}{\degree}$ and \SI{45}{\degree} for Stokes $Q$, and with $\theta_\mathrm{HWP} = \SI{22.5}{\degree}$ and \SI{67.5}{\degree} for Stokes $U$; see Sect.~\ref{sec:sphere_irdis_optical_path}) were taken at different altitude and parallactic angle combinations. 
	In this way, the effect of the diattenuations of the UT and M4 and a possible (but unlikely) stellar polarization signal can be distinguished when fitting the data to the model.
	The HWP cycles were kept short (\SI{{\sim}140}{\second}) to limit the parallactic and altitude angle variations of the data points themselves.

For the observations of HD~217343 (2018) we took 12 HWP cycles per filter with a similar instrument setup as used for HD~176425. 
	The most important difference between the two setups is that this time we (accidentally) observed in field-tracking mode.
	In this mode the derotator is rotating continuously and therefore the polarimetric efficiency varies during the measurements.
	Because we did not optimize the derotator angle as recommended (see Paper I), the polarimetric efficiency reached a value as low as $31\%$ for the last measurement in K$_\mathrm{s}$-band.

Both data sets are processed by applying dark subtraction, flat fielding, bad-pixel correction and centering with a Moffat function as described in Paper~I. 
	Subsequently, we construct the double-difference $Q$- and $U$-images from Eq.~\ref{eq:double_difference} and the double-sum $I_Q$- and $I_U$-images from Eq.~\ref{eq:double_sum}.
	Finally, we calculate the normalized Stokes parameters $q$ and $u$ by dividing the sum in an aperture in the $Q$- and $U$-images by the sum in the same aperture in the corresponding $I_Q$- and $I_U$-images (see Eq.~\ref{eq:model_normalized_stokes_parameter}). 
	For an elaboration on the extraction of the normalized Stokes parameters and the selected aperture sizes see Appendix~\ref{app:graphs_unpolstar}.

To describe the measurements, we use Eq.~\ref{eq:model_normalized_stokes_parameter} with the model equations of Sect.~\ref{sec:model_mathematics} inserted (together the model function). 
	We use the complete Eq.~\ref{eq:complete_model} and fill in the values of the determined parameters $\epsilon_\mathrm{HWP}$ to $d$ from Table~\ref{tab:parameters_instrument}.
	We compute the retardances of the UT (actually M3 since M1 and M2 are rotationally symmetric) and M4 using the Fresnel equations with the complex refractive index of aluminum obtained from~\citet{rakic_aluminum}. 
	This computation needs to be performed before determining the diattenuations, because the retardance of M4 affects the measurement of the IP produced by the UT.
	Because we observed unpolarized (standard) stars, we write $\boldsymbol{S}_\mathrm{in} = \left[1, 0, 0, 0\right]^\mathrm{T}$.

We determine the diattenuations of the UT and M4 independently for both data sets by fitting the model function to the data points using non-linear least squares.
	The parallactic, altitude, HWP and derotator angles required for this are obtained from the headers of the FITS-files of the measurements (see Appendix~\ref{app:angles_headers}).
	We have tested fitting the incident Stokes vectors in addition to the diattenuations (writing $\boldsymbol{S}_\mathrm{in} = [1, q_\mathrm{in}, u_\mathrm{in}, 0]^\mathrm{T}$), and found that the degree of linear polarization of the stars is indeed insignificant ($<0.1\%$) in all filters.
	We therefore choose not to fit the incident Stokes vectors and assume the stars to be completely unpolarized. 
	Graphs of the model fits and the residuals can be found in Appendix~\ref{app:graphs_unpolstar}.

\subsection{Results and discussion for unpolarized star calibrations}
\label{sec:response_telescope_results_discussion}

The determined diattenuations and calculated retardances of the UT and M4 for both data sets are shown in Table~\ref{tab:parameters_telescope}. 
	The listed $1\sigma$-uncertainties of the diattenuations are computed from the residuals of fit (see Appendix~\ref{app:uncertainty_parameters}) under the same assumptions as described in Sect.~\ref{sec:results_discussion_instrument}.
\begin{table*}
\caption{Determined diattenuations with their errors and computed retardances of the part of the model describing the telescope and M4 in Y-, J-, H- and K$_\mathrm{s}$-band. The second column shows when the parameters are valid, i.e. before and/or after the re-aluminization of M1 and M3 that took place between April 3 and April 16, 2017. The diattenuations of the UT and M4 that are valid before April 16, 2017 are determined from the observations of HD~176425 in 2016, and those valid after April 16, 2017 are determined from the observations of HD~217343 in 2018.}
\centering
\setlength{\tabcolsep}{1pt}
\begin{tabular}{l l c c c r c l c r c l c r c l c r c l}
\hline\hline
\multicolumn{2}{c}{Parameter} & \hspace{16pt} & \begin{tabular}{@{}c@{}} Valid before or after \T \\ April 16, 2017 \B \end{tabular} & \hspace{16pt} & \multicolumn{3}{c}{BB\_Y} & \hspace{16pt} &  \multicolumn{3}{c}{BB\_J} & \hspace{16pt} &  \multicolumn{3}{c}{BB\_H} & \hspace{16pt} &  \multicolumn{3}{c}{BB\_$\mathrm{K_s}$} \T\B \\ 
\hline
$\epsilon_\mathrm{UT}$ & & & \hspace{7pt}before & & 0.0236 & $\pm$ & 0.0002 & & 0.0167 & $\pm$ & 0.0001 & & 0.01293 & $\pm$ & $8\cdot10^{-5}$ & & 0.0106 & $\pm$ & 0.0003 \T \\
 & & & after & & 0.0175 & $\pm$ & 0.0003 & & 0.0121 & $\pm$ & 0.0002 & & 0.0090 & $\pm$ & 0.0001 & & 0.0075 & $\pm$ & 0.0005 \\
$\epsilon_\mathrm{M4}$ & & & \hspace{7pt}before & & 0.0182 & $\pm$ & 0.0002 & & 0.0128 & $\pm$ & 0.0001 & & 0.00985 & $\pm$ & $8\cdot10^{-5}$  & & 0.0078 & $\pm$ & 0.0003 \\
 & & & after & & 0.0182 & $\pm$ & 0.0003 & & 0.0130 & $\pm$ & 0.0002 & & 0.0092 & $\pm$ & 0.0001 & & 0.0081 & $\pm$ & 0.0005 \\
$\varDelta_\mathrm{UT}$ & (\si{\degree}) & & before and after & & 171.9 & & & & 173.4 & & & & 175.0 & & & & 176.3 & & \\
$\varDelta_\mathrm{M4}$ & (\si{\degree}) & & before and after & & 171.9 & & & & 173.4 & & & & 175.0 & & & & 176.3 & & \B \\ 
\hline
\end{tabular}
\label{tab:parameters_telescope}
\end{table*}

The calculated values of $\varDelta_\mathrm{UT}$ and $\varDelta_\mathrm{M4}$ are close to the ideal value of \SI{180}{\degree} and therefore the crosstalk produced by the UT and M4 is very limited.
	In all filters, the combined polarimetric efficiency of the UT and M4 is $>98\%$ and the corresponding offset of the angle of linear polarization is at most a few tenths of a degree (largest effect in Y-band).
	Because of the limited crosstalk, any realistic deviation of the real retardances from the computed ones will result in very small errors only.
	This also implies that the systematic error on $\epsilon_\mathrm{UT}$ due to using an analytical rather than a measured value of $\varDelta_\mathrm{M4}$ is very small.

\begin{figure*} 
\centering 
\includegraphics[width=14.53cm]{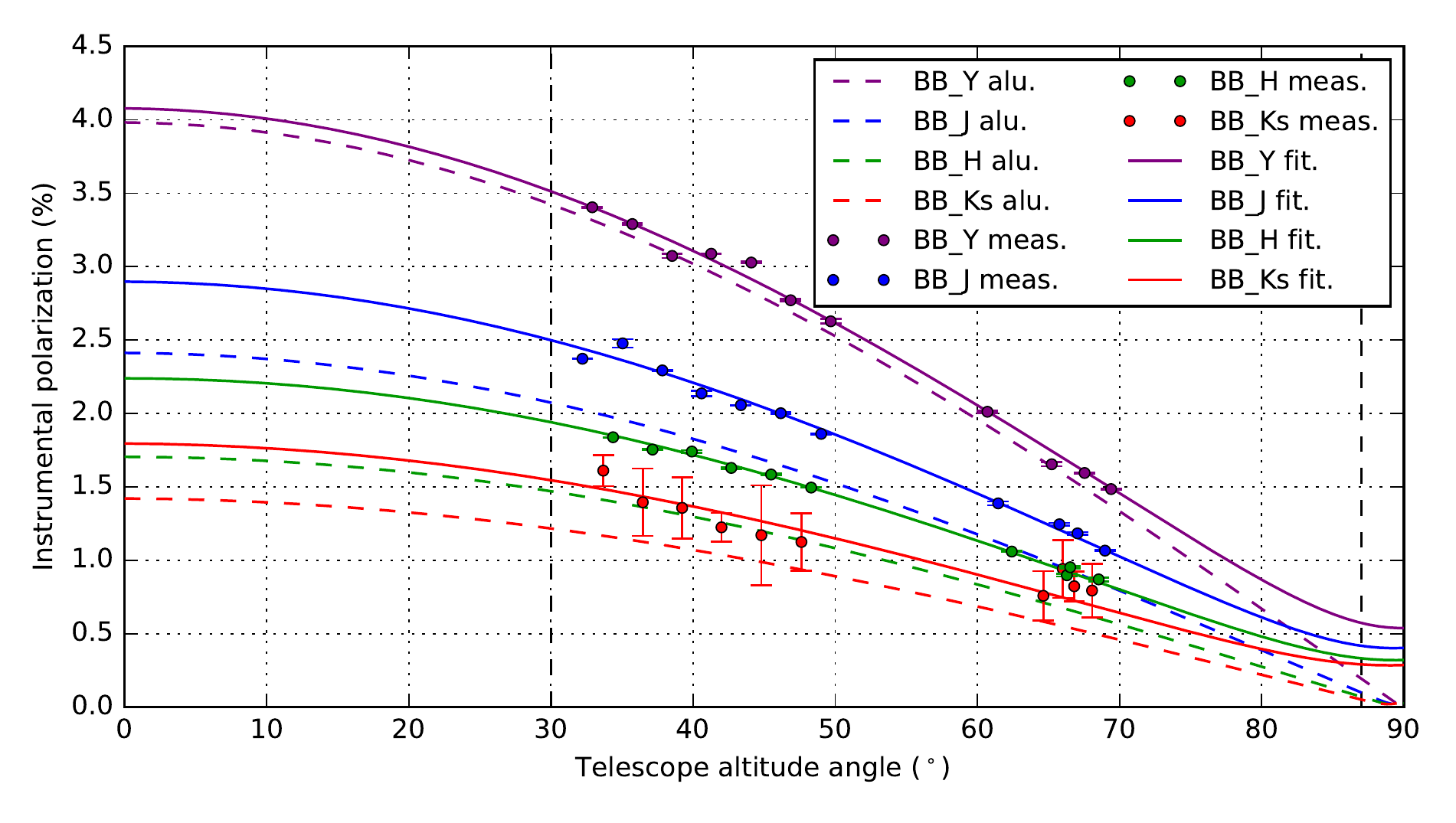} 
\caption{Analytical (aluminum), measured (including error bars) and fitted instrumental polarization (IP) of the telescope and M4 as a function of telescope altitude angle in Y-, J-, H- and K$_\mathrm{s}$-band from the measurements of HD~176425 taken in 2016 before the re-aluminization of M1 and M3. Note that for science observations the telescope altitude angle is restricted to $\SI{30}{\degree} \leq a \leq \SI{87}{\degree}$.} 
\label{fig:M3_M4_DoLP} 
\end{figure*}

\begin{figure*} 
\centering 
\includegraphics[width=14.53cm]{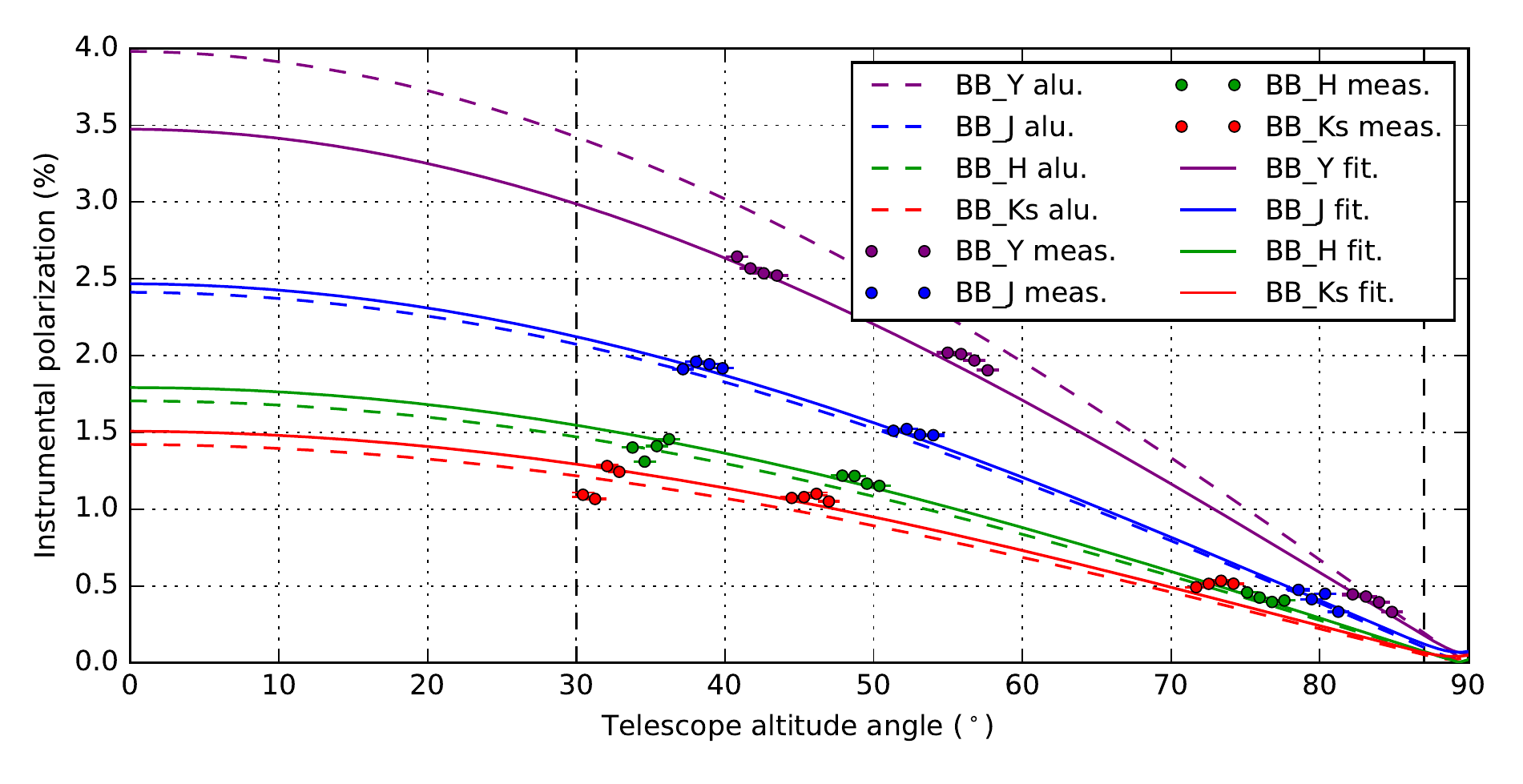} 
\caption{Analytical (aluminum), measured (including error bars) and fitted instrumental polarization (IP) of the telescope and M4 as a function of telescope altitude angle in Y-, J-, H- and K$_\mathrm{s}$-band from the measurements of HD~217343 taken in 2018 after the re-aluminization of M1 and M3. Note that for science observations the telescope altitude angle is restricted to $\SI{30}{\degree} \leq a \leq \SI{87}{\degree}$.}
\label{fig:M3_M4_DoLP_2018} 
\end{figure*}

To understand the effect of the determined diattenuations, we plot the measured and fitted degree of linear polarization (see Eq.~\ref{eq:DoLP}) as a function of telescope altitude angle for the observations of HD~176425 (2016) and HD~217343 (2018) in Fig.~\ref{fig:M3_M4_DoLP} and \ref{fig:M3_M4_DoLP_2018}, respectively.
	The degree of linear polarization can in this case be interpreted as the IP of the UT and M4. 
	The Figures also show analytical curves that are constructed by computing the diattenuations from the Fresnel equations and assuming that the aluminum coatings of the UT (M3) and M4 have the same properties.
	The error bars on the measurements are calculated as half the difference between the degree of linear polarization determined from apertures with radii 50 pixels larger and smaller than that used for the data points themselves (see Appendix~\ref{app:graphs_unpolstar}).
	The error bars show the uncertainty in the degree of linear polarization due to the dependency of the measured values on the chosen aperture radius.
	The uncertainty is 
small for all measurements except for those of HD~176425 (2016) taken in K$_\mathrm{s}$-band. 
	The latter measurements are less certain because of difficulties in removing the thermal background signal (see Appendix~\ref{app:graphs_unpolstar}).
	Note that for science observations the telescope altitude angle is restricted to $\SI{30}{\degree} \leq a \leq \SI{87}{\degree}$.
%
%
%

Figure~\ref{fig:M3_M4_DoLP} shows that the IP increases with decreasing altitude angle and that before the re-aluminization of M1 and M3 the maximum IP (at $a = \SI{30}{\degree}$) 
is equal to approximately 3.5\%, 2.5\%, 1.9\% and 1.5\% in Y-, J-, H- and K$_\mathrm{s}$-band, respectively. 
	The corresponding minimum values (at $a = \SI{87}{\degree}$) 
are 0.58\%, 0.42\%, 0.33\% and 0.29\%, respectively.
	Ideally, we would expect the IP of M3 to completely cancel that of M4 when the reflection planes of the mirrors are crossed at $a = \SI{90}{\degree}$ (analytical curves).
	However, because the determined $\epsilon_\mathrm{UT}$ and $\epsilon_\mathrm{M4}$ are not identical, this is not the case.
	This discrepancy is probably caused by differences in the coating or aluminum oxide layers of the mirrors~\citep[see][]{harten_mirror}.

Figure~\ref{fig:M3_M4_DoLP_2018} shows that the IP after the re-aluminization of M1 and M3 is significantly smaller than before.
	The maximum values (at $a = \SI{30}{\degree}$) are now equal to approximately 3.0\%, 2.1\%, 1.5\% and 1.3\% in Y-, J-, H- and K$_\mathrm{s}$-band, respectively, and the corresponding minimum values (at $a = \SI{87}{\degree}$) are 0.18\%, 0.12\%, 0.07\%, 0.06\%, respectively.
	This decrease of IP is due to the lower diattenuation of the UT (see Table~\ref{tab:parameters_telescope}).
	In fact, after re-aluminization the diattenuation of the UT is comparable to that of M4, leading to almost complete cancellation of the IP at \SI{90}{\degree} altitude angle\footnote{\label{fn:ZIMPOL}ZIMPOL~\citep{schmid_zimpoloverview}, the visible imaging polarimeter of SPHERE, has an additional HWP in between M3 and M4 that is used to rotate the IP produced by M3 such that it is ideally completely canceled by M4 at any altitude angle~\citep{roelfsema_zimpol}. However, also at visible wavelengths the diattenuations of M3 and M4 were probably not equal before the re-aluminization of M1 and M3, so that some IP originating from the UT and M4 must have remained for ZIMPOL. After the re-aluminization, the IP of ZIMPOL is most likely close to zero because the diattenuations are much more comparable.}.
	Because the measurements were taken in field-tracking mode, the data points shown have been corrected for the polarimetric efficiency (the residuals for the two data points in K$_\mathrm{s}$-band close to $a = \SI{30}{\degree}$ are considerably enhanced because of this correction). 
	Finally, note that during the observations of HD~217343 we did not switch filter after every HWP cycle as we did for HD~176425 (compare Figs.~\ref{fig:M3_M4_DoLP} and \ref{fig:M3_M4_DoLP_2018}).
	Therefore the measurement points are less spread out over the range of altitude angles, making them constrain the model function somewhat less.

\begin{figure} 
\centering 
\includegraphics[width=\hsize]{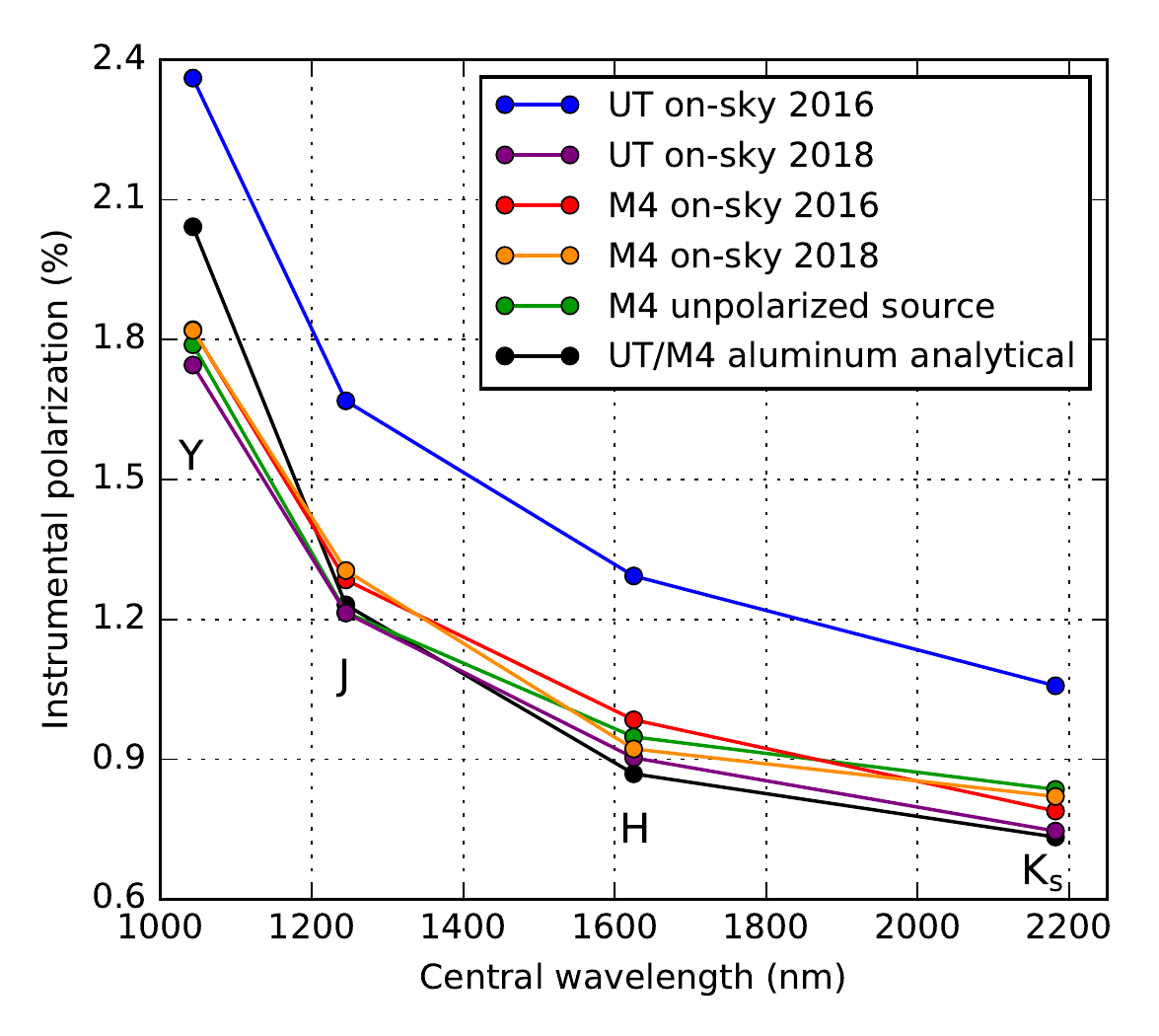} 
\caption{Instrumental polarization (IP) of the UT and M4 separately, as determined from the various measurements, versus central wavelength of the Y-, J-, H- and K$_\mathrm{s}$-band. The curves show the IP of the UT and M4 from the observations of the unpolarized stars HD~176425 (on-sky 2016) and HD~217343 (on-sky 2018), the IP of M4 from the unpolarized source measurements and the IP of the UT and M4 computed from the Fresnel equations (aluminum analytical).} 
\label{fig:M3M4_IP} 
\end{figure}

The IP created by the UT or M4 separately, as determined from the various measurements, is shown as a function of central wavelength of the Y-, J-, H- and K$_\mathrm{s}$-band in Fig.~\ref{fig:M3M4_IP}.
 	The IP created is equal to the diattenuation of the mirror(s) when assuming that the incident light is completely unpolarized (see Eq.~\ref{eq:component_mueller_matrix}).
	Figure~\ref{fig:M3M4_IP} shows that before the re-aluminization of M1 and M3, the IP of the UT is significantly larger than that of M4 (on-sky 2016).
	After the re-aluminization, the IP of the UT has decreased and differs less than 0.1\% from that of M4 in all filters (on-sky 2018).
	This indicates that the coatings of M3 and M4 are much more similar after the re-aluminization.
	Between the observations of the unpolarized stars in 2016 and 2018, the IP of M4 (which has not been re-aluminized) differs less than 0.07\% in all filters, showing that the diattenuation does not significantly change in time. 

	Figure~\ref{fig:M3M4_IP} also shows the IP of M4 as determined from the unpolarized source measurements, i.e.~$q_\mathrm{in, unpol}$ from Table~\ref{tab:parameters_instrument} (ignoring $u_\mathrm{in, unpol}$, which is close to zero in all filters).
 	 Clearly, the observations of the unpolarized stars are in good agreement with the measurements with the internal light source. 
 	 The small differences among the values determined from the measurements of the unpolarized stars and the internal light source could be due to the different spectra of the stars and the internal light source, the calibration unit producing some polarization or the finite precision of the measurements.
	Finally, Fig.~\ref{fig:M3M4_IP} shows the IP produced by the UT or M4 as computed from the Fresnel equations (aluminum analytical).
	We conclude that the determined IP agrees well with the theoretical expectation.																														
%


%
%

\section{Polarimetric accuracy of instrument model} 
\label{sec:model_accuracy}

In this Section we determine for each broadband filter the total polarimetric accuracy 
of 
our completed instrument model and compare it to the aims we set in Sect.~\ref{sec:introduction}.
	As first step to calculate the accuracy of the model, we compute the accuracies of fitting the model parameters to the calibration data.
	These accuracies of fit are calculated as the corrected sample standard deviation of the residuals in Appendix~\ref{app:uncertainty_parameters} and show the random errors of the measurements.
	The systematic errors of the model fits are likely small, because the residuals of fit are close to normally distributed (see Figs.~\ref{fig:STOKES_POL_BB_H}, \ref{fig:STOKES_UNPOL_BB_H}, \ref{fig:STOKES_UNPOL_BB_H_DERROT}, \ref{fig:M3_M4_STOKES_BB_H}, \ref{fig:M3_M4_STOKES_BB_Ks} and \ref{fig:M3_M4_STOKES_BB_H_2018}).

To compute the total polarimetric accuracy from the residuals of fit, we need to compute the absolute and relative polarimetric accuracies $s_\mathrm{abs}$ and $s_\mathrm{rel}$ (see Eqs.~\ref{eq:polarimetric_accuracy_z} and \ref{eq:delta_z}).
	For the absolute polarimetric accuracy we compute separate values before and after the re-aluminization of M1 and M3. 
	The absolute polarimetric accuracy is calculated 
as $s_\mathrm{abs} = \surd(s_\mathrm{unpol}^2 + s_\mathrm{star}^2)$, with $s_\mathrm{unpol}$ the accuracy of fit of the unpolarized source measurements and $s_\mathrm{star}$ the accuracy of fit of the observations of the unpolarized star under consideration (see Appendix~\ref{app:uncertainty_parameters}).
	We take the relative polarimetric accuracy $s_\mathrm{rel}$ (valid before and after the re-aluminization) 
equal to the accuracy of fit of the polarized source measurements.
	The resulting absolute and relative polarimetric accuracies in Y-, J-, H- and K$_\mathrm{s}$-band are shown in Table~\ref{tab:accuracy_complete_system}.

From Table~\ref{tab:accuracy_complete_system} we conclude that the absolute polarimetric accuracies before and after the re-aluminization of M1 and M3 are comparable and that the requirements on the absolute and relative polarimetric accuracies (${\leq}0.1\%$ and ${<}1\%$, respectively) are 
met for all filters.
	The values of $s_\mathrm{abs}$ are consistent with the ${\sim}0.05\%$ absolute difference among the independent estimates of the IP of M4 from the observations of the unpolarized stars and the unpolarized source measurements (see Fig.~\ref{fig:M3M4_IP}).
	Because the residuals of fit are close to normally distributed, the absolute and relative polarimetric accuracies can probably be improved by obtaining calibration measurements with a higher signal-to-noise ratio.
	However, 
the accuracy we attain when correcting science observations appears to be limited by systematic errors (see Sect.~\ref{sec:accuracy_after_correction}).
%
\begin{table}
\caption{Absolute and relative polarimetric accuracies in Y-, J-, H- and K$_\mathrm{s}$-band.  For the absolute polarimetric accuracy separate values have been calculated before and after the re-aluminization of M1 and M3 that ended on April 16, 2017.} 
\centering 
\begin{tabular}{l c c c} 
\hline\hline 
\T\B Filter & \begin{tabular}{@{}c@{}} $s_\mathrm{abs}$ (\%) (before \T \\ April 16, 2017) \B \end{tabular} & \begin{tabular}{@{}c@{}} $s_\mathrm{abs}$ (\%) (after \T \\ April 16, 2017) \B \end{tabular} & $s_\mathrm{rel}$ (\%) \\
\hline 
BB\_Y & 0.062 & 0.068 & 0.73 \T \\
BB\_J & 0.047 & 0.072 & 0.41 \\
BB\_H & 0.026 & 0.030 & 0.58 \\
BB\_K$_\mathrm{s}$ & 0.10~~ & 0.093 & ~0.54 \B \\
\hline 
\end{tabular} 
\label{tab:accuracy_complete_system} 
\end{table} 

With the absolute and relative polarimetric accuracies calculated, we can now compute the total polarimetric accuracies in Stokes $Q$ and $U$, $s_\mathrm{Q}$ and $s_\mathrm{U}$, respectively, as:    
\begin{align} 
	s_\mathrm{Q} &= s_\mathrm{abs} \skew{2.5}\hat{I}_{Q,\mathrm{in}} + s_\mathrm{rel} \left| \skew{2.5}\hat{Q}_\mathrm{in} \right|, \label{eq:accuracy_sq} \\[0.3cm]
	s_\mathrm{U} &= s_\mathrm{abs} \skew{2.5}\hat{I}_{U,\mathrm{in}} + s_\mathrm{rel} \left| \skew{2.5}\hat{U}_\mathrm{in} \right|, \label{eq:accuracy_su}
\end{align} 
where $\skew{2.5}\hat{I}_{Q,\mathrm{in}}$, $\skew{2.5}\hat{I}_{U,\mathrm{in}}$, $\skew{2.5}\hat{Q}_\mathrm{in}$ and $\skew{2.5}\hat{U}_\mathrm{in}$ are the measured Stokes $I_Q$, $I_U$, $Q$ and $U$ incident on the telescope after correcting the instrumental polarization effects with the model (see Sect.~\ref{sec:model_correction}).
	Eqs.~\ref{eq:accuracy_sq} and \ref{eq:accuracy_su} are derived from Eqs.~\ref{eq:polarimetric_accuracy_z} and \ref{eq:delta_z} by substituting $\skew{2.5}\hat{Q}_\mathrm{in}$ and $\skew{2.5}\hat{U}_\mathrm{in}$ for the true incident $Q_\mathrm{in}$ and $U_\mathrm{in}$.
	We can determine the total polarimetric accuracy in the degree and angle of linear polarization ($s_\mathrm{DoLP}$ and $s_\mathrm{AoLP}$) as:
%
\begin{align} 	
	s_\mathrm{DoLP} &= \sqrt{\frac{\skew{2.5}\hat{q}_\mathrm{in}^2 {s_\mathrm{q}}^2 + \skew{2.5}\hat{u}_\mathrm{in}^2 {s_\mathrm{u}}^2} {\skew{2.5}\hat{q}_\mathrm{in}^2 + \skew{2.5}\hat{u}_\mathrm{in}^2}}, \label{eq:s_DoLP} \\[0.3cm]
	s_\mathrm{AoLP} &= \frac{\sqrt{\skew{2.5}\hat{u}_\mathrm{in}^2 {s_\mathrm{q}}^2 + \skew{2.5}\hat{q}_\mathrm{in}^2 {s_\mathrm{u}}^2}}{2\left(\skew{2.5}\hat{q}_\mathrm{in}^2 + \skew{2.5}\hat{u}_\mathrm{in}^2\right)}, \label{eq:s_AoLP}
\end{align} 
%
where $\skew{2.5}\hat{q}_\mathrm{in} = \skew{2.5}\hat{Q}_\mathrm{in} \hspace{1pt} / \hspace{1pt} \skew{2.5}\hat{I}_{Q,\mathrm{in}}$, $s_\mathrm{q} = s_\mathrm{Q} \hspace{1pt} / \hspace{1pt} \skew{2.5}\hat{I}_{Q,\mathrm{in}}$, $\skew{2.5}\hat{u}_\mathrm{in} = \skew{2.5}\hat{U}_\mathrm{in} \hspace{1pt} / \hspace{1pt} \skew{2.5}\hat{I}_{U,\mathrm{in}}$ and $s_\mathrm{u} = s_\mathrm{U} \hspace{1pt} / \hspace{1pt} \skew{2.5}\hat{I}_{U,\mathrm{in}}$.
	We have derived Eqs.~\ref{eq:s_DoLP} and \ref{eq:s_AoLP} from Eqs.~\ref{eq:DoLP}, \ref{eq:AoLP}, \ref{eq:accuracy_sq} and \ref{eq:accuracy_su} by applying standard error propagation and assuming Gaussian statistics, zero uncertainty in $\skew{2.5}\hat{I}_{Q,\mathrm{in}}$ and $\skew{2.5}\hat{I}_{U,\mathrm{in}}$, and no correlation between $s_\mathrm{Q}$ and $s_\mathrm{U}$.
	In case $\skew{2.5}\hat{I}_{Q,\mathrm{in}}$ and $\skew{2.5}\hat{I}_{U,\mathrm{in}}$ contain substantial flux from the central star, $\skew{2.5}\hat{Q}_\mathrm{in}$, $\skew{2.5}\hat{U}_\mathrm{in}$, $s_\mathrm{Q}$ and $s_\mathrm{U}$ should be divided by the intensity from the source we are interested in (e.g.~a circumstellar disk or substellar companion) when computing $s_\mathrm{DoLP}$ and $s_\mathrm{AoLP}$.
	Note that corrections need to be applied to Eqs.~\ref{eq:s_DoLP} and \ref{eq:s_AoLP} in case the signal-to-noise ratio in the degree of linear polarization is very low, i.e.~lower than ${\sim}3$~\citep[see][]{sparks_polarimetry, patat_polarimetryerror}.

Table~\ref{tab:accuracy_companion_disk} shows the polarimetric accuracies of measuring the degree and angle of linear polarization of a 1\% polarized substellar companion and a 30\% polarized circumstellar disk in \mbox{Y-,} J-, H- and K$_\mathrm{s}$-band before the re-aluminization of M1 and M3 (the results after the aluminization are comparable). 
	The accuracies are computed from Eqs.~\ref{eq:s_DoLP} and \ref{eq:s_AoLP} under the assumption that $\skew{2.5}\hat{I}_{Q,\mathrm{in}}$ and $\skew{2.5}\hat{I}_{U,\mathrm{in}}$ contain no starlight.
	The accuracies weakly depend on the angle of linear polarization of the incident light (the specific values of $\skew{2.5}\hat{q}_\mathrm{in}$ and $\skew{2.5}\hat{u}_\mathrm{in}$) and so the worst case 
is shown.
	From Table~\ref{tab:accuracy_companion_disk} it follows that for increasing degrees of linear polarization of the source, the error on the degree of linear polarization increases.	
	For sources with a low degree of linear polarization (up to a few percent) the error is 
	nearly equal to the absolute polarimetric accuracy $s_\mathrm{abs}$, while for sources with a high degree of linear polarization (several tens of percent) the contribution of the relative polarimetric accuracy $s_\mathrm{rel}$ dominates.
	Table~\ref{tab:accuracy_companion_disk} also shows that the error on the angle of linear polarization decreases with increasing degree of linear polarization of the source, because the polarization components $Q$ and $U$ are measured with a higher relative accuracy.
	This also means that for sources with a very low degree of linear polarization (${\sim}0.1\%$) the error on the angle of linear polarization can be as large as \SI{10}{\degree} or more.
\begin{table}
\caption{Polarimetric accuracy of measuring the degree and angle of linear polarization of a 1\% polarized substellar companion and a 30\% polarized circumstellar disk in Y-, J-, H- and K$_\mathrm{s}$-band before the re-aluminization of M1 and M3. The results after the re-aluminization are comparable.} 
\centering 
\begin{tabular}{l c c c c} 
\hline\hline 
Filter & \begin{tabular}{@{}c@{}} $s_\mathrm{DoLP}$ (\%) \T \\ 1\% pol. \\ companion \B \end{tabular} & \begin{tabular}{@{}c@{}} $s_\mathrm{AoLP}$ (\si{\degree}) \T \\ 1\% pol. \\ companion \end{tabular} & \begin{tabular}{@{}c@{}} $s_\mathrm{DoLP}$ (\%) \T \\ 30\% pol. \\ disk \end{tabular} & \begin{tabular}{@{}c@{}} $s_\mathrm{AoLP}$ (\si{\degree}) \T \\ 30\% pol. \\ disk \end{tabular} \\
\hline 
BB\_Y & 0.069 & 1.9~~ & 0.28 & ~0.21 \T \\
BB\_J & 0.051 & 1.4~~ & 0.17 & 0.13 \\
BB\_H & 0.032 & 0.86 & 0.20 & 0.14 \\
BB\_K$_\mathrm{s}$ & 0.11~~ & 3.0~~ & 0.26 & ~0.20 \B \\
\hline 
\end{tabular} 
\label{tab:accuracy_companion_disk} 
\end{table} 

Assuming that Gaussian statistics apply and that systematic errors are small, Table~\ref{tab:accuracy_companion_disk} shows that the polarization signal of a 1\% polarized substellar companion can be measured in all filters with the required total polarimetric accuracy of ${\sim}0.1\%$ in the degree of linear polarization and an accuracy of a few degrees in angle of linear polarization.
	For the 30\% polarized circumstellar disk, the attainable accuracies in degree of linear polarization are below $0.3\%$ in all filters, which is amply sufficient for quantitative polarimetry.
	Note that for real measurements the attained accuracies are generally somewhat worse because of for example measurement noise and varying atmospheric conditions (see Sect.~\ref{sec:accuracy_after_correction}).
	In addition, note that the accuracy of measuring a circumstellar disk's degree of linear polarization itself is limited by the accuracy with which the total intensity of the disk can be obtained.
	
%
%



\section{Correction of science observations}
\label{sec:correction_method}

\subsection{Correction method}
\label{sec:model_correction}

\begin{figure}[!hbtp] 
\centering 
\includegraphics[width=\hsize]{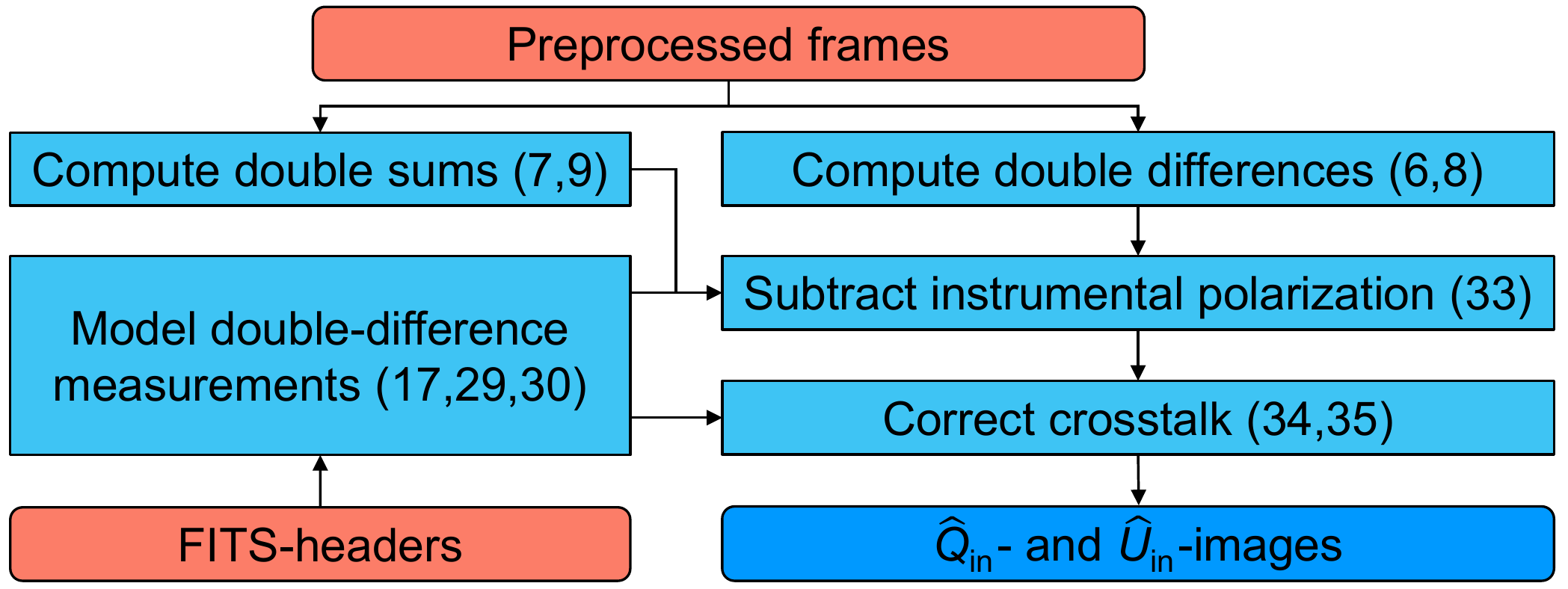} 
\caption{Flow diagram showing the steps to construct the incident $\skew{2.5}\hat{Q}_\mathrm{in}$- and $\skew{2.5}\hat{U}_\mathrm{in}$-images from field-tracking observations using the instrument model. The numbers of the Equations used for the various steps are indicated in parentheses.} 
\label{fig:flow_chart} 
\end{figure}

In this Section, we explain the data-reduction method we have developed to correct science measurements for the instrumental polarization effects of the complete optical system using our instrument model.
	The goal of the correction method is to obtain from the measurements the $\skew{2.5}\hat{Q}_\mathrm{in}$- and $\skew{2.5}\hat{U}_\mathrm{in}$-images, i.e. the estimates of the true $Q_\mathrm{in}$- and $U_\mathrm{in}$-images incident on the telescope (see top right part of Fig.~\ref{fig:ut_sphere_irdis_schematic}). 
	A flow diagram of our correction method for field-tracking observations 
is shown in Fig.~\ref{fig:flow_chart}.
		
%
%

Before applying our correction method, we pre-process the raw data by performing dark subtraction, flat fielding, bad-pixel correction and centering (see Sect.~\ref{sec:model_correction_tcha} and Paper~I).
	Subsequently, we construct for each HWP cycle the $Q$- and $U$-images from the double difference~(Eq.~\ref{eq:double_difference}) and the corresponding $I_Q$- and $I_U$-images from the double sum~(Eq.~\ref{eq:double_sum}).
	We denote the $n$ double-difference images ($Q$ or $U$) by $X_i$ and the corresponding double-sum images ($I_Q$ or $I_U$) by $I_{X,i}$, with $i = 1, 2,\dots,n$.
We construct the $\skew{2.5}\hat{I}_{Q,\mathrm{in}}$-and $\skew{2.5}\hat{I}_{U,\mathrm{in}}$-images, i.e.~the $I_Q$- and $I_U$-images incident on the telescope, simply by computing the mean (or median) of the double-sum $I_{Q,i}$- and $I_{U,i}$-images, respectively.


To construct the $\skew{2.5}\hat{Q}_\mathrm{in}$- and $\skew{2.5}\hat{U}_\mathrm{in}$-images 
 we use our instrument model.
	The instrumental polarization effects are different for each measurement, because the parallactic, altitude, HWP and derotator angles change continuously as the telescope tracks the target.
	To describe these changing instrumental polarization effects, we compute the vector equivalents of the single and double difference (Eqs.~\ref{eq:single_difference} and \ref{eq:model_double_difference}) using our instrument model.
	To this end, we obtain the date, filter and the parallactic, altitude, HWP and derotator angles of each measurement from the headers of the FITS-files of the data (see Appendix~\ref{app:angles_headers}).	
	We then take the model parameters corresponding to the filter from Tables~\ref{tab:parameters_instrument} (parameters $\epsilon_\mathrm{HWP}$ to $d$) and \ref{tab:parameters_telescope}, taking into account the date of the observations for the latter.
	For each measurement, we compute $M_\mathrm{sys,L}$ and $M_\mathrm{sys,R}$ from Eq.~\ref{eq:complete_model} using $+d$ and $-d$ in $M_\mathrm{CI,L/R}$ (Eq.~\ref{eq:efficiency_matrix_polarizers}), respectively.
	Similar to Sect.~\ref{sec:model_mathematics}, where we computed the single difference from the top elements of $\boldsymbol{S}_\mathrm{det,L}$ and $\boldsymbol{S}_\mathrm{det,R}$ (i.e. $I_\mathrm{det,L}$ and $I_\mathrm{det,R}$), we now compute the single difference from the top rows of $M_\mathrm{sys,L}$ and $M_\mathrm{sys,R}$ (which we call $\boldsymbol{I}_\mathrm{sys,L}$ and $\boldsymbol{I}_\mathrm{sys,R}$):
%
\begin{equation} 
	\boldsymbol{D}^\pm = \boldsymbol{I}_\mathrm{sys,L} - \boldsymbol{I}_\mathrm{sys,R},
\end{equation} 
%
where $\boldsymbol{D}^\pm$ is the single-difference row vector.
	Subsequently, we compute for every double-difference image $X_i$ the double-difference row vector $\boldsymbol{D}_i$ as:
\begin{align}
	\boldsymbol{D}_i &= \dfrac{1}{2} \left[\boldsymbol{D}^+\left(p_i^+, a_i^+, \theta_{\mathrm{HWP},i}^+, \theta_{\mathrm{der},i}^+ \right) - \boldsymbol{D}^-\left(p_i^-, a_i^-, \theta_{\mathrm{HWP},i}^-, \theta_{\mathrm{der},i}^-\right)\right], \nonumber \\[0.2cm]
	&=	\begin{bmatrix} (I \hspace{-2pt}\rightarrow\hspace{-2pt} X)_i & (Q \hspace{-2pt}\rightarrow\hspace{-2pt} X)_i & (U \hspace{-2pt}\rightarrow\hspace{-2pt} X)_i & (V \hspace{-2pt}\rightarrow\hspace{-2pt} X)_i 
\end{bmatrix},	\label{eq:demodulation_matrix_dd}	
\end{align}
where $\boldsymbol{D}^+$ and $\boldsymbol{D}^-$ are a function of the parallactic, altitude, HWP and derotator angles of the first (superscript $+$) and second (superscript $-$) measurements used to compute the double difference, respectively.

To describe the $i$-th double-difference measurement, we can write:
\begin{align} 
	X_i  &= \boldsymbol{D}_i \cdot \boldsymbol{S}_\mathrm{in}, \\[0.2cm]
		 &= \left(I \hspace{-2pt}\rightarrow\hspace{-2pt} X\right)_i I_\mathrm{in} + \left(Q \hspace{-2pt}\rightarrow\hspace{-2pt} X\right)_i Q_\mathrm{in} + \left(U \hspace{-2pt}\rightarrow\hspace{-2pt} X\right)_i U_\mathrm{in} + \left(V \hspace{-2pt}\rightarrow\hspace{-2pt} X\right)_i V_\mathrm{in}. \nonumber
	\label{eq:correction_double_difference_full}
\end{align} 
	We can ignore the element $\left(V \hspace{-2pt}\rightarrow\hspace{-2pt} X\right)_i$, i.e.~assume $V_\mathrm{in} = 0$, because we do not expect circularly polarized signals from the targets we are interested in. 
	In addition, we can assume that 
the measured double-sum intensities $I_{X,i}$ are equal to the incident intensity $I_\mathrm{in}$	(the resulting maximum relative error is ${\sim}10^{-4}$). 
	Therefore, we can describe the $i$-th double-difference measurement as: 
\begin{equation} 
	X_i = \left(I \hspace{-2pt}\rightarrow\hspace{-2pt} X\right)_i I_{X,i} + \left(Q \hspace{-2pt}\rightarrow\hspace{-2pt} X\right)_i Q_\mathrm{in} + \left(U \hspace{-2pt}\rightarrow\hspace{-2pt} X\right)_i U_\mathrm{in}.
\label{eq:correction_double_difference}
\end{equation} 

The elements $\left(I \hspace{-2pt}\rightarrow\hspace{-2pt} X\right)_i$ describe the instrumental polarization (IP) of the complete optical system for each measurement.
	We remove the IP from each double-difference image $X_i$ by scaling the corresponding double-sum intensity image $I_{X,i}$ with this element and subtracting the result from the double-difference image:
\begin{equation} 
	X_{\mathrm{IPS},i} = X_i - \left(I \hspace{-2pt}\rightarrow\hspace{-2pt} X\right)_i I_{X,i},
\label{eq:ip_subtraction}
\end{equation} 
where $X_{\mathrm{IPS},i}$ is the $i$-th IP-subtracted double-difference image.

The elements $\left(Q \hspace{-2pt}\rightarrow\hspace{-2pt} X\right)_i$ and $\left(U \hspace{-2pt}\rightarrow\hspace{-2pt} X\right)_i$ in Eq.~\ref{eq:correction_double_difference} account for the crosstalk (and thus for the polarimetric efficiency and offset of the angle of linear polarization) of the complete optical system for each measurement.
	To correct for the crosstalk, we set up a system of equations as follows:
\begin{align} 
	\boldsymbol{Y} &= A \left[Q_\mathrm{in}, U_\mathrm{in}\right]^T, \nonumber \\[0.2cm]
	\begin{bmatrix} X_{\mathrm{IPS},1} \\ X_{\mathrm{IPS},2} \\[-3pt] \vdots \\ X_{\mathrm{IPS},n} \end{bmatrix} &= 
	\begin{bmatrix} (Q \hspace{-2pt}\rightarrow\hspace{-2pt} X)_1 & (U \hspace{-2pt}\rightarrow\hspace{-2pt} X)_1 \\ 
	(Q \hspace{-2pt}\rightarrow\hspace{-2pt} X)_2 & (U \hspace{-2pt}\rightarrow\hspace{-2pt} X)_2 \\[-3pt] 
	\vdots & \vdots \\
	(Q \hspace{-2pt}\rightarrow\hspace{-2pt} X)_n & (U \hspace{-2pt}\rightarrow\hspace{-2pt} X)_n \end{bmatrix}
	\begin{bmatrix} Q_\mathrm{in} \\ U_\mathrm{in} \end{bmatrix},
\end{align} 
with $\boldsymbol{Y}$ a column vector containing the $i = 1, 2, \dots, n$ IP-subtracted double-difference images, $Q_\mathrm{in}$ and $U_\mathrm{in}$ the true $Q$- and $U$-images incident on the telescope and $A$ the $n\times2$ system matrix containing the elements $\left(Q \hspace{-2pt}\rightarrow\hspace{-2pt} X\right)_i$ and $\left(U \hspace{-2pt}\rightarrow\hspace{-2pt} X\right)_i$ of each double difference.
	We obtain the $\skew{2.5}\hat{Q}_\mathrm{in}$- and $\skew{2.5}\hat{U}_\mathrm{in}$-images, i.e.~the estimates of the true incident $Q_\mathrm{in}$- and $U_\mathrm{in}$-images, by solving for every pixel the system of equations using linear least squares:
\begin{equation} 
	\left[\skew{2.5}\hat{Q}_\mathrm{in}, \skew{2.5}\hat{U}_\mathrm{in}\right]^T = (A^\mathrm{T} A)^{-1} A^\mathrm{T} 
\boldsymbol{Y}.
\label{eq:lsq}
\end{equation} 
%
	Alternatively, we can obtain the incident $\skew{2.5}\hat{Q}_\mathrm{in}$- and $\skew{2.5}\hat{U}_\mathrm{in}$-images by solving the system of equations for each pair of IP-subtracted double-difference $Q$- and $U$-images (each HWP cycle) separately, and then computing the median or trimmed mean over all resulting $\skew{2.5}\hat{Q}_\mathrm{in}$- and $\skew{2.5}\hat{U}_\mathrm{in}$-images. 
	Computing the median or trimmed mean has the advantage that any bad pixels still visible in the images are removed, 
but using Eq.~\ref{eq:lsq} is expected to generally yield more accurate results.
	In place of Eq.~\ref{eq:lsq} we can also use weighted linear least squares, wherein the weight matrix takes into account the signal-to-noise ratio of the images or the polarimetric efficiency as predicted by the instrument model.
	Note that the correction method (using Eq.~\ref{eq:lsq}) can be applied to data sets having an unequal number of double-difference Q and U measurements.

The instrument model is valid for any combination of parallactic, altitude, HWP and derotator angles and does not require the use of a particular rotation control law for the HWP and derotator.
	However, for observations not taken in field-tracking mode (e.g.~pupil-tracking mode)
, the derotator does not keep the image orientation constant.
	We therefore need to derotate with our pipeline the images after subtracting the IP and before correcting the crosstalk. 
	The adapted correction method for pupil-tracking observations, which in addition combines polarimetry with angular differential imaging (ADI), is presented in \citet{vanholstein_exopol}.
  					
	
\subsection{Correction of images of circumstellar disk of T Cha}
\label{sec:model_correction_tcha}

%

The correction method presented in Sect.~\ref{sec:model_correction} has already been successfully applied to over 70 polarimetric data sets, including HR 8799 and PZ Tel~\citep{vanholstein_exopol}, TW Hydrae (Paper~I), T Cha~\citep{pohl_tcha}, DZ Cha~\citep{canovas_dzcha}, TWA7~\citep{olofsson_twa7}, PDS 70~\citep{keppler_pds70} and CS Cha~\citep{ginski_cscha}. 
	In this Section, we will demonstrate our correction method with the H-band polarimetric observations of the circumstellar disk of T Chamaeleontis (T Cha) as published in~\citet{pohl_tcha}. 
	The transition disk around T Cha consists of a coplanar inner and outer disk separated by a large gap, and is viewed close to edge-on with an inclination of ${\sim}\SI{69}{\degree}$~\citep{olofsson_tcha2, pohl_tcha, hendler_tcha}.
	While the outer disk can easily be spatially resolved with SPHERE, the very narrow and close-in inner disk cannot (its extent is only ${<}0.2$ pixel on the IRDIS detector).
\begin{figure*}
\centering 
\includegraphics[width=\hsize]{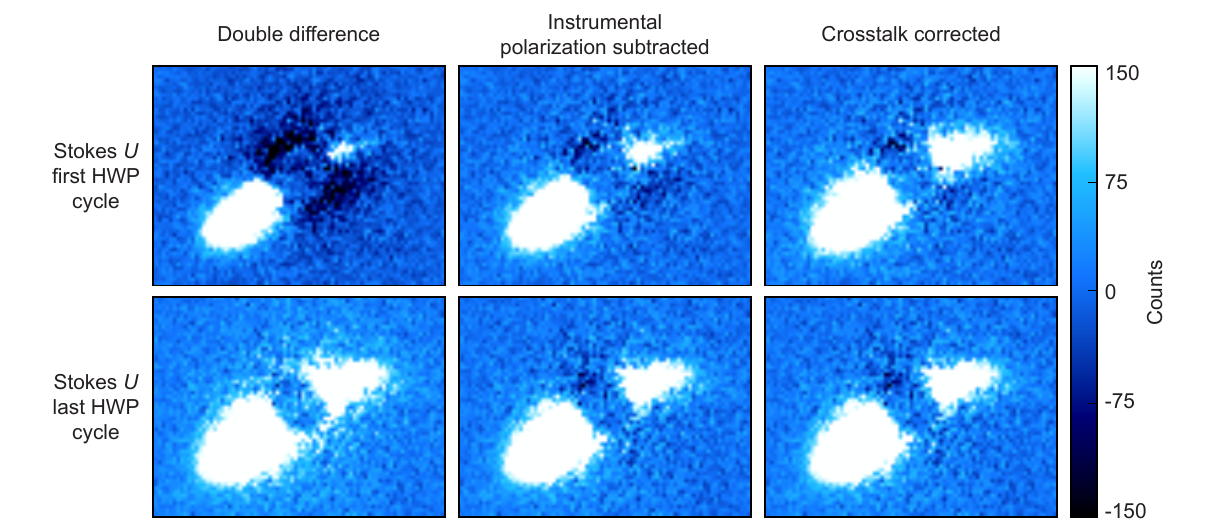} 
\caption{The effect of the data-reduction steps of our correction method on the Stokes $U$-images of the first and last (30th) HWP cycle of the observations of the circumstellar disk of T Cha.}
\label{fig:model_correction} 
\end{figure*} 

The data of T Cha was taken on February 20, 2016 under program ID 096.C-0248(C).
	It consists of a total of 30 HWP cycles with HWP switch angles \SI{0}{\degree}, \SI{45}{\degree}, \SI{22.5}{\degree} and \SI{67.5}{\degree} to measure Stokes $Q$ and $U$ (see Sect.~\ref{sec:sphere_irdis_optical_path}).
	During the observations, the parallactic and altitude angles varied from \SI{5.8}{\degree} to \SI{27.3}{\degree}, and from \SI{35.1}{\degree} to \SI{34.1}{\degree}, respectively.
	We pre-process the raw data by performing dark subtraction, flat fielding, bad-pixel correction and centering with the star center frames as described in Paper~I and \citet{pohl_tcha}.
	We then construct the $Q$- and $U$-images from the double difference (Eq.~\ref{eq:double_difference}) and the $I_Q$- and $I_U$-images from the double sum (Eq.~\ref{eq:double_sum}).
	The $Q$- and $U$-images show a weak detector artifact comprised of continuous vertical bands.
	We remove this artifact by subtracting, for every pixel column, the median value of the 60 pixels at the top and bottom of that column (see Paper~I).	
	The resulting double-difference $U$-images of the first and last (30th) HWP cycle are shown in the left column of Fig.~\ref{fig:model_correction}.
	The pronounced differences between the two images are predominantly caused by IP that evolves from negative to positive $U$ during the 78 min total observing time.

We now apply our correction method (using the diattenuations of the UT and M4 valid before April 16, 2017) and subtract the IP from the double-difference $Q$- and $U$-images (see Eq.~\ref{eq:ip_subtraction}).
	The resulting IP-subtracted $U$-images of the first and last HWP cycle are shown in the center column of Fig.~\ref{fig:model_correction}. 
	The resulting images are much more similar compared to the original double-difference images (left column).
	However, the optical system's crosstalk makes the disk brighter in Stokes $U$ and fainter in Stokes $Q$ during the course of the observations.
	This is because the crosstalk transfers part of the flux in Stokes $Q$ to Stokes $U$ or vice versa, i.e.~it introduces an offset in the angle of linear polarization (see Fig.~\ref{fig:AOLP_POL_BB_H}). 
	In addition the crosstalk converts part of the linearly polarized light into circularly polarized light that the P0-90 analyzer set is not sensitive to, entailing a loss of signal as quantified by the polarimetric efficiency (see Fig.~\ref{fig:DOLP_POL_BB_H}).
	These two effects are also seen in Fig.~3 of Paper~I as variations in the Stokes $Q$- and $U$-images.
	Although the polarimetric efficiency during the observations of T Cha is not very low (minimum of $88\%$), the offset of the angle of linear polarization reaches values as large as $\SI{13}{\degree}$.
	This shows that even for observations taken at a reasonably high polarimetric efficiency, there is still significant transfer of signal between the Stokes $Q$ and $U$-images (recall that the orientations of $Q$ and $U$ differ by \SI{45}{\degree}).

\begin{figure} 
\centering 
\includegraphics[width=\hsize]{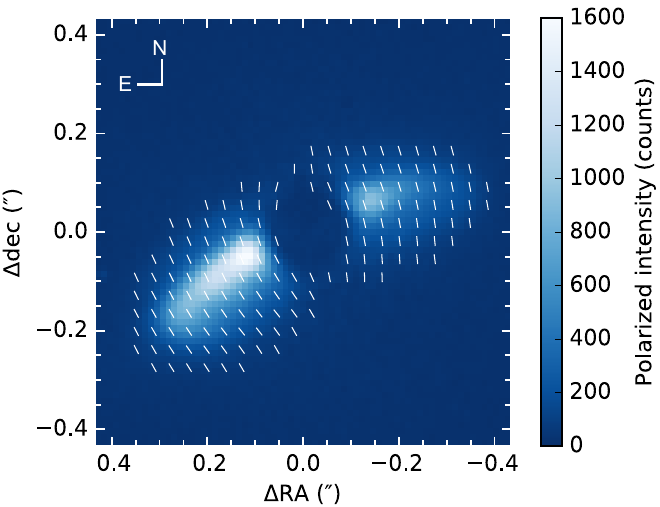} 
\caption{Polarized intensity and angle of linear polarization of the circumstellar disk of T Cha after applying the correction method. The white lines indicating the angle of linear polarization have arbitrary length and are only shown where the polarized intensity is higher than 50 counts.}
\label{fig:Ipol_AoLP} 
\end{figure}

We correct for the crosstalk using linear least squares (see Eq.~\ref{eq:lsq}), directly yielding the $\skew{2.5}\hat{Q}_\mathrm{in}$- and $\skew{2.5}\hat{U}_\mathrm{in}$-images.
	The right column of Fig.~\ref{fig:model_correction} shows the resulting $\skew{2.5}\hat{U}_\mathrm{in}$-images of the first and last HWP cycle after solving the system of equations for each HWP cycle separately.
	It follows that after crosstalk correction the disk has a very similar surface brightness distribution in all images. 
	The integrated signal of the disk only varies by a few percent among the images, which is due to varying atmospheric conditions 
during the observations (e.g.~seeing and sky transparency).
	Although by correcting the crosstalk we compensate for the polarimetric efficiency, we note that this does not increase the signal-to-noise ratio (as clearly visible in Fig.~8 of Paper~I). 
	Next, we subtract the constant polarized background in the $\skew{2.5}\hat{Q}_\mathrm{in}$- and $\skew{2.5}\hat{U}_\mathrm{in}$-images after determining it from a large star-centered annulus with inner and outer radii of 360 and 420 pixels, respectively.
	Finally, we use the resulting images and Eqs.~\ref{eq:polarized_intensity} and \ref{eq:AoLP} to compute the polarized intensity and angle of linear polarization of the disk as shown in Fig.~\ref{fig:Ipol_AoLP}.  
%
%


\subsection{Improvements attained with correction method}
\label{sec:model_correction_tcha_improvements}

In this Section we will show the improvements attained with our correction method by comparing the model-corrected $\skew{2.5}\hat{Q}_\mathrm{in}$- and $\skew{2.5}\hat{U}_\mathrm{in}$-images of T Cha with $Q$- and $U$-images generated with the conventional IP-subtraction method as presented by~\citet{canovas_data}.
	In Paper~I we made a similar comparison using data of the (nearly) face-on viewed disk of TW Hydrae.
	While that data set could in principle be reduced using conventional data-reduction methods, in this Section we show that the correction method is essential to accurately reduce data of an inclined disk and that it enables us to detect non-azimuthal polarization and the polarization of the starlight. 
	
\begin{figure} 
\centering 
\includegraphics[width=\hsize]{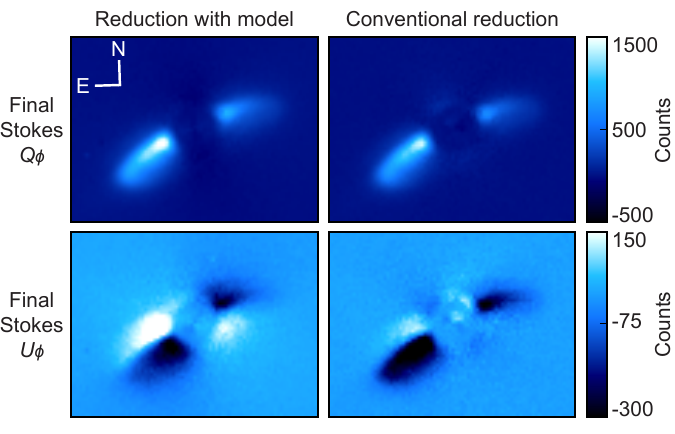} 
\caption{Final azimuthal Stokes $Q_\phi$- and $U_\phi$-images of the circumstellar disk of T Cha after applying our correction method compared to the images generated with the conventional IP-subtraction method from~\protect\citet{canovas_data}. Positive $Q_\phi$ indicates linear polarization in the azimuthal direction and $U_\phi$ shows the linear polarization at $\pm\SI{45}{\degree}$ from this direction. Note that the color scales of the top and bottom row are different, i.e. the signals in $Q_\phi$ are almost 10 times larger than the signals in $U_\phi$.}
\label{fig:reductions_comparison} 
\end{figure}

To construct the $Q$- and $U$-images with the conventional IP-subtraction method, we compute the mean of the double-difference $Q$- and $U$- and double-sum $I_Q$- and $I_U$-images, and subtract the IP following the steps described in Sect.~4.1 of Paper~I.
	We convert these and the model-corrected images into images of the azimuthal Stokes parameters $Q_\phi$ and $U_\phi$ (see Sect.~4.2 and Eqs.~15 to 17 of Paper~I) to ease the comparison and interpretation of the images.
	The resulting images are shown in Fig.~\ref{fig:reductions_comparison}.

The model-corrected images are more accurate than the images generated with the conventional IP-subtraction method.
	With our correction method the instrumental polarization effects are known a priori and are corrected with an absolute polarimetric accuracy of ${\sim}0.1\%$ or better (see Table~\ref{tab:accuracy_complete_system} and Sect.~\ref{sec:accuracy_after_correction}).
	The conventional IP-subtraction method on the other hand does not correct the crosstalk and estimates the IP from the science data under the assumption that the starlight is unpolarized, resulting in errors in the polarized intensity and angle of linear polarization.

Comparing the left and right columns of Fig.~\ref{fig:reductions_comparison}, it follows that the disk in the model-corrected $Q_\phi$-image is ${\sim}20\%$ brighter.
	This increase in brightness is largely due to the crosstalk correction, i.e. the correction of the polarimetric efficiency and transfer of signal between the $Q_\phi$- and $U_\phi$-images (or $Q$- and $U$-images).
	As a result of the correction, the polarized surface brightness distribution, orientation and morphology of the disk are more accurately retrieved in the model-corrected images.

Fig.~\ref{fig:reductions_comparison} also shows that both reduction methods yield non-zero $U_\phi$-signals, but with significant differences.
	Our correction method corrects for the IP and crosstalk without an assumption on the polarization of the star (as in the conventional IP-subtraction method) or the angle of linear polarization over the disk (as in the $U_\phi$-minimization method, see Paper~I).
	Therefore our correction method is truly sensitive to non-azimuthal polarization and yields the accurate $U_\phi$-image.
	From Fig.~\ref{fig:Ipol_AoLP} and the model-corrected $U_\phi$-image of Fig.~\ref{fig:reductions_comparison}, we can conclude that away from the brightness region of the disk the angle of linear polarization deviates from the azimuthal direction.
	\citet{pohl_tcha} primarily attribute this non-azimuthal polarization to multiple scattering starting in the inner disk. \\

\begin{figure} 
\centering 
\includegraphics[width=\hsize]{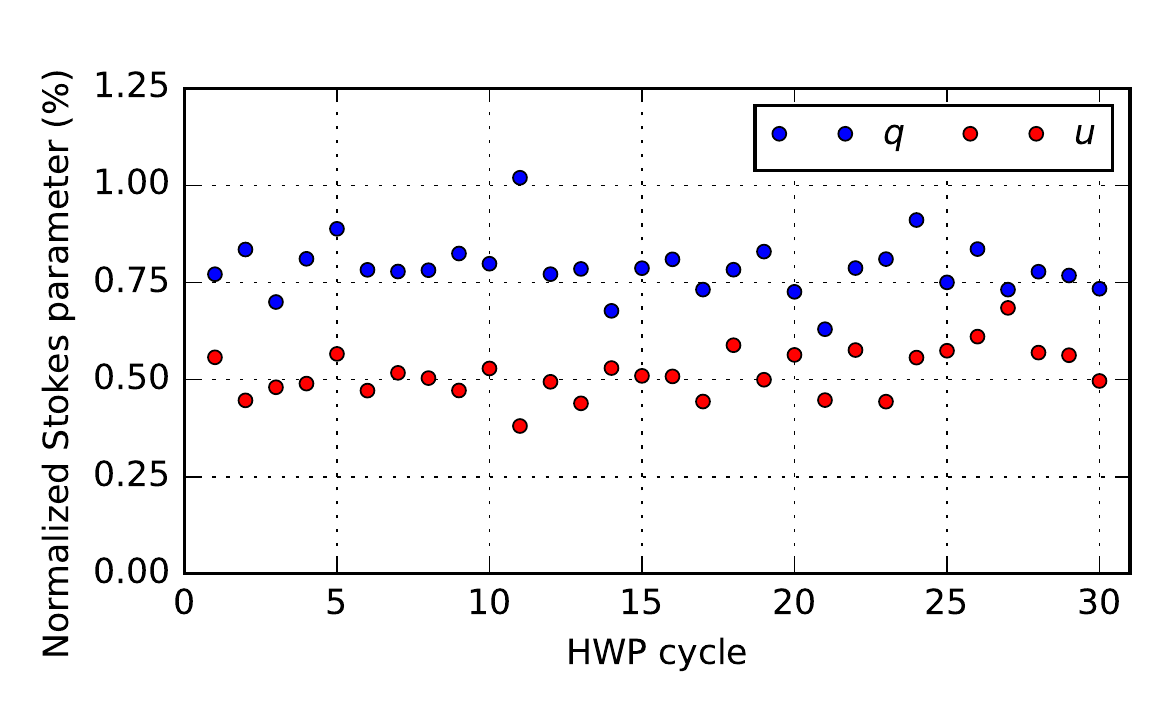} 
\caption{Normalized Stokes parameters of the measured stellar polarization of T Cha as function of HWP cycle after applying our correction method.} 
\label{fig:star_polarization_HWP_cycle} 
\end{figure}

\noindent A clear disadvantage of the conventional IP-subtraction method is that it substantially over-subtracts the IP when the star is polarized, because it cannot discern IP from polarized starlight. 
	Figure~\ref{fig:star_polarization_HWP_cycle} shows for each individual HWP cycle the polarization signal as measured from the AO residuals in the model-corrected $\skew{2.5}\hat{Q}_\mathrm{in}$- and $\skew{2.5}\hat{U}_\mathrm{in}$-images. 
	The Figure shows that the measured polarization signal, and therefore the angle of linear polarization, is constant in time.
	This indicates that the starlight is polarized, because 
any uncorrected IP would have changed with the variation in parallactic and altitude angle during the observations.
%

From Fig.~\ref{fig:star_polarization_HWP_cycle}, and using the variation of the data points for the uncertainties, we find that the star has a degree and angle of linear polarization of $0.94 \pm 0.07 \%$ and $17 \pm 2\si{\degree}$, respectively.
	This stellar polarization signal is most likely not caused by interstellar dust, because T Cha is located in front of, and not in, the Cha I dark cloud~\citep{murphy_epsiloncha, gaia_dr2} and the angle of linear polarization differs by ${\sim}\SI{80}{\degree}$ with respect to the average in the cloud~\citep{covino_chaidarkcloud}.
	Because the measured angle of linear polarization is approximately perpendicular to the position angle of the outer circumstellar disk (see the top left image of Fig.~\ref{fig:Ipol_AoLP}), the stellar polarization signal most likely originates from the coplanar, unresolved inner disk and/or part of the outer disk viewed close to the star.
	Indeed, the model-corrected images of Fig.~\ref{fig:reductions_comparison}, which still contain the stellar polarization signal, correspond much better to radiative transfer models than the images generated with the conventional IP-subtraction method~(see~\citealt{pohl_tcha}; also \citealt{keppler_pds70}).

It appears to be quite common for stars that host a circumstellar disk to be polarized, because in about half of the more than 70 data sets we have applied our correction method to we measure significant stellar polarization. 
	If interstellar dust can be excluded as the origin, the stellar polarization can indicate the presence of an unresolved (inner) disk, in particular for a circumstellar disk with a low to moderate inclination~\citep[see e.g.][]{keppler_pds70}.
	The position angle of an inner disk can then be determined from the measured angle of linear polarization.
	For a detailed example on how to determine whether the stellar polarization is caused by interstellar dust, see~\citet{ginski_cscha}.	
	Note that to measure the small polarization signals of substellar companions, measuring the polarization of the star is imperative to prove that the companion's polarization is intrinsic and is not caused by over-subtraction of disk-induced stellar polarization or interstellar dust. 

\subsection{Limits to and optimization of polarimetric accuracy}
\label{sec:accuracy_after_correction}

The polarimetric accuracy of measuring Stokes $Q$ and $U$ and the degree and angle of linear polarization after applying our correction method can be computed from Eqs.~\ref{eq:accuracy_sq} to \ref{eq:s_AoLP}. 
	However, with real measurements the uncertainty on these physical quantities is generally somewhat worse than the computed accuracies.
	The accuracies of Table~\ref{tab:accuracy_complete_system} should therefore be considered lower limits.
	In general, for stars that are not polarized because of their circumstellar disk or interstellar dust, a polarization signal of ${\sim}0.1\%$ remains after applying our correction method.
	The higher uncertainty on the measured polarization is likely due to limitations of the instrument model, measurement noise and varying atmospheric conditions.
	In this Section we will elaborate on these limiting factors and discuss means to optimize the polarimetric accuracy.
	

	A first limitation of the instrument model is that we assume the instrumental polarization effects to be fixed for a given broadband filter.
	However, because the instrumental polarization effects vary with wavelength (see e.g.~Figs.~\ref{fig:DOLP_POL_ALL_BBF} and \ref{fig:M3_M4_DoLP}), any spectral differences between the science object and the calibration sources used to determine the model parameters will introduce small errors in the correction of the IP and crosstalk.
	We can limit these errors by comparing the spectra of the science object and calibration sources and interpolating the values of the model parameters over the wavelength domain.
	Such an interpolation will be quite accurate for the diattenuations of the UT and M4 and the retardance of the HWP, because their spectral dependency is smooth and is 
known from theory and the manufacturer, respectively (see e.g.~Fig.~\ref{fig:hwp_halle}).
	The largest interpolation errors are expected for the retardance of the derotator, because we need to guess the shape of the function from the four measured data points.
	By interpolating the model parameters we will also be able to correct measurements taken with the narrowband filters.

A second limitation of the instrument model is that the instrumental polarization effects are taken constant over the field of view.
	We know the instrumental polarization effects have spatial dependence, because the images of the internal calibration measurements display a gradient (see Appendix~\ref{app:structure_flux}).
	However, contrary to the polarimetric imaging mode of FORS~\citep{patat_polarimetryerror}, this spatial dependence is very small as demonstrated by the relative proximity of the nine data points taken throughout the image for each HWP and derotator angle combination in Figs.~\ref{fig:DOLP_POL_BB_H} to \ref{fig:AOLP_POL_ALL_BBF} and \ref{fig:STOKES_POL_BB_H} to \ref{fig:STOKES_UNPOL_BB_H_DERROT}.
	The main reason for the limited spatial dependence is that the light beams within SPHERE have much larger f-numbers than those within FORS, i.e.~the beams converge and diverge much more slowly within SPHERE.
	Because we have determined the model parameters from all these data points together (see Sect.~\ref{sec:measurements_downstream_m4}), the spatial dependence 
downstream of M4 is accounted for in the polarimetric accuracy of the model.
	Nevertheless, we can increase the accuracy of the model by determining a separate set of model parameters from each of the nine apertures used, because the nine data points do not vary randomly around their average value but show a relation with position on the detector.
	We do not expect the diattenuations and retardances of the UT and M4 to be strongly spatially dependent, because spatial variations generally originate from transmissive optics near a focal plane.

A third limitation of the model is that the instrumental polarization effects are assumed to be constant in time.
	At least some temporal variation is expected for the diattenuation and retardance of the UT, because the UT is open to the atmosphere and therefore the amount of contamination (e.g.~dust) on the mirrors varies~\citep[see][]{snik_polreview}.
	However, as M1 and M3 are cleaned with $\mathrm{CO_2}$ on a monthly basis, this variation is most likely small.
	For the other optical components we do not expect temporal variations due to contamination because they are located within SPHERE.
	Aging of these components is most likely also limited, because the model parameters describing the optical path downstream of M4 seem not to have changed since the internal calibration measurements of 2016, and the determined diattenuation of M4 has not significantly changed between the observations of the unpolarized stars in 2016 and 2018 (see Sect.~\ref{sec:response_telescope_results_discussion}).
	To optimize the accuracy of our correction, we can recalibrate the diattenuation of the UT and M4 during the same night as the science observations, preferably with an unpolarized star that has a spectrum as similar as possible to that of the science object(s).

To keep the instrument model accurate over time, new calibration measurements need to be taken 
when a modification is made to the optical path that affects the polarimetry.
	Examples of such modifications are the insertion of a new optical component, the replacement or removal of an existing component or the re-coating of a mirror (e.g. the re-aluminization of M1 and M3 as performed between April 3 and April 16, 2017).
	Because the mathematical description of our model includes the double difference, changes to the optical path downstream of the derotator generally do not require new calibration measurements. \\

\noindent The polarimetric accuracy we can really attain is also affected by measurement noise. 
	In Eq.~\ref{eq:polarimetric_accuracy_z}, the polarimetric accuracy is defined for infinite sensitivity, i.e.~without any noise or spurious signals present in the data.
	However, in general the combined photon, speckle, (sky) background and read-out noise of a measurement is much larger than the polarimetric accuracy of the instrument model.
	Therefore, when stating uncertainties of measured polarization signals, we recommend to always compare the polarimetric accuracy as computed from Eqs.~\ref{eq:accuracy_sq} to \ref{eq:s_AoLP} with the measurement noise.
	The criteria to reach a polarimetric sensitivity, in addition to a polarimetric accuracy, of ${\leq}0.1\%$ with IRDIS for the measurement of polarization signals of substellar companions are discussed in~\citet{vanholstein_exopol}.
	
With the double-difference method, spurious polarization signals created when the atmospheric seeing or sky transparency changes between measurements is removed to first order.
	Some spurious signals remain, because these atmospheric variations prevent the effect of the diattenuation of the components downstream from the derotator to be completely removed.
	When the variations in seeing and sky transparency are large, the spurious signals can be suppressed by computing Stokes $Q$ and $U$ from the `normalized' double difference (compare to Eq.~\ref{eq:double_difference}):
\begin{equation} 
	X = \dfrac{1}{2}\left(\dfrac{X^+}{I_{X^+}} - \dfrac{X^-}{I_{X^-}}\right) \cdot I_X,
\label{eq:normalized_double_difference} 
\end{equation} 
with $I_X$ computed from Eq.~\ref{eq:double_sum}.
	
The accurate polarized intensity images that we obtain with our correction method enable the construction of images of the degree of linear polarization of circumstellar disks.
	To construct such an image, an image of the total intensity of the disk is required.
	In principle such an image can be obtained by subtracting the point spread function of a reference star~\citep[e.g.][]{canovas_disk} or by using angular differential imaging for disks seen close to edge-on~\citep[e.g.][]{perrin_gemini}. 
	However, these techniques have proven to be challenging and residual speckles from the star will remain in the total intensity image of the disk.
	Therefore the accuracy of measuring the degree of linear polarization of circumstellar disks will be limited by the accuracy of 
the total intensity image rather than the accuracy of the instrument model. \\


	
\subsection{Data-reduction pipeline including correction method}
\label{sec:irdap}

We have incorporated our correction method in a highly-automated end-to-end data-reduction pipeline called IRDAP (IRDIS Data reduction for Accurate Polarimetry). 
	IRDAP is publicly available and handles data taken both in field- and pupil-tracking mode and using the broadband filters Y, J, H and K$_\mathrm{s}$.
	Data taken with the narrowband filters can be reduced as well, although with a lower accuracy, by using the correction method of the broadband filters.
	For pupil-tracking observations IRDAP can additionally apply angular differential imaging.

Reducing data with IRDAP is very straightforward and does not require the user to do any coding. 
	IRDAP is simply run from a terminal with only a few commands and uses a configuration file with a limited number of input parameters. 
	For an average-sized data set and using a modern computer, IRDAP performs a complete data reduction from raw data to final data products within a few minutes.

The documentation of IRDAP, including the installation and user instructions, can be found at~\url{https://irdap.readthedocs.io}.
	We plan to regularly add functionalities and make improvements to IRDAP.
	Among others, we plan to calibrate the instrument in the narrowband filters to also enable the accurate reduction of data taken in these filters.
		
%
%

\section{Summary and conclusions}
\label{sec:conclusions}

We have created a detailed Mueller matrix model describing the instrumental polarization effects of the Unit Telescope (UT) and SPHERE/IRDIS in the broadband filters Y, J, H and K$_\mathrm{s}$. 
	To determine the parameters of the model, we have taken measurements with SPHERE's internal light source and have observed two unpolarized stars. 
	We have developed a data-reduction method that uses the model to correct for the instrumental polarization and crosstalk. 
	We have exemplified this correction method with observations of the circumstellar disk of T Cha and have shown the improvements compared to conventional data-reduction and analysis methods. \\
	
\noindent The instrumental polarization (IP) of the optical system 
primarily originates from the UT and SPHERE's first mirror (M4) and increases with decreasing telescope altitude angle.
	The IP is different for observations taken before and after the re-aluminization of the primary and tertiary mirrors of the UT (M1 and M3). 
	Before the re-aluminization (i.e. before April 16, 2017), the maximum IP (at an altitude angle of \SI{30}{\degree}) is approximately equal to 3.5\%, 2.5\%, 1.9\% and 1.5\% in Y-, J-, H- and K$_\mathrm{s}$, respectively.
	After the re-aluminization (i.e. after April 16, 2017), the maximum IP in the same filters is approximately 3.0\%, 2.1\%, 1.5\% and 1.3\%, respectively.
	
The crosstalk of the optical system 
is strongly wavelength dependent and is primarily produced by the derotator (K-mirror).
	The crosstalk 
decreases the polarimetric efficiency, 
because it converts linearly polarized light into circularly polarized light that IRDIS cannot measure.
	The polarimetric efficiency is the lowest when the reflection plane of the derotator is at approximately $\pm\SI{45}{\degree}$ from the vertical direction and has minimum values equal to $54\%$, $89\%$, $5\%$ and $7\%$ in Y-, J-, H- and K$_\mathrm{s}$, respectively.
	The crosstalk also causes an offset of the angle of linear polarization in these filters, with maximum deviations equal to $\SI{11}{\degree}$, $\SI{4}{\degree}$, $\SI{34}{\degree}$ and \SI{90}{\degree}, respectively. 
	In Paper~I, we 
present a strategy to prevent observing at a low polarimetric efficiency by optimizing the derotator angle.

		
	In all broadband filters, the instrument model has an absolute and relative polarimetric accuracy 
of ${\leq}0.1\%$ 
and ${<}1\%$, respectively. 
	With these accuracies we can measure the polarization signals of substellar companions with a total polarimetric accuracy of ${\sim}0.1\%$ in the degree of linear polarization and an accuracy of a few degrees in angle of linear polarization.
	These accuracies are amply sufficient for quantitative polarimetry of circumstellar disks, because these objects are typically polarized a few tens of percent.
	The uncertainty on the measured polarization after applying our correction method to science observations is generally somewhat worse 
than the accuracies of the model itself due to limitations of the model, varying atmospheric conditions and measurement noise. \\

\noindent With our correction method the IP and crosstalk are known a priori and for weakly polarized sources are corrected with an absolute polarimetric accuracy of ${\sim}0.1\%$ or better.
	This is contrary to conventional data-reduction methods that do not correct the crosstalk and estimate the IP from the (noisy) science data. 
	Using our correction method we can therefore more accurately measure the polarized intensity and angle of linear polarization.
	With the correction method we can also measure the polarization of the star, which enables us to detect unresolved (inner) disks and prove that the measured polarization signal of a substellar companion is intrinsic to the companion.
	The method can be applied to measurements taken both in field- and pupil-tracking mode. 

We have incorporated our correction method in a highly-automated end-to-end data-reduction pipeline called IRDAP (IRDIS Data reduction for Accurate Polarimetry).
	IRDAP is publicly available and the documentation, including the installation and user instructions, can be found at~\url{https://irdap.readthedocs.io}.
	To achieve the highest polarimetric accuracy, it is recommended to always use IRDAP for the reduction of IRDIS polarimetric data.
	Even for observations of nearly face-on circumstellar disks or measurements taken at a high polarimetric efficiency (e.g.~when the derotator is kept at a favorable angle or observations are performed in J-band), our correction method makes a significant correction to the angle of linear polarization and increases the signal-to-noise ratio in the final images.

\begin{acknowledgements}
A significant part of this work was performed when RGvH, JHG and JdB were affiliated to ESO.
RGvH and JdB thank ESO for the studentship at ESO Santiago during which this project was started.
Many thanks go out to the SPHERE team and the instrument scientists and operators of the ESO Paranal observatory for their support during the calibration measurements. 
The research of JdB and FS leading to these results has received funding from the European Research Council under ERC Starting Grant agreement 678194 (FALCONER). 
This research has made use of the SIMBAD database, operated at the CDS, Strasbourg, France. 
This research has made use of NASA’s Astrophysics Data
System Bibliographic Services.
SPHERE is an instrument designed and built by a consortium consisting of IPAG (Grenoble, France), MPIA (Heidelberg, Germany), LAM (Marseille, France), LESIA (Paris, France), Laboratoire Lagrange (Nice, France), INAF - Osservatorio di Padova (Italy), Observatoire de Gen\`{e}ve (Switzerland), ETH Zurich (Switzerland), NOVA (Netherlands), ONERA (France), and ASTRON (Netherlands) in collaboration with ESO. SPHERE was funded by ESO, with additional contributions from the CNRS (France), MPIA (Germany), INAF (Italy), FINES (Switzerland) and NOVA (Netherlands). 
SPHERE also received funding from the European Commission Sixth and Seventh Framework Programs as part of the Optical Infrared Coordination Network for Astronomy (OPTICON) under grant number RII3-Ct-2004-001566 for FP6 (2004-2008), grant number 226604 for FP7 (2009-2012), and grant number 312430 for FP7 (2013-2016).
\end{acknowledgements}

%
%

\bibliographystyle{aa}   
\bibliography{C:/Users/Rob/Documents/PhD/CentralFiles/biblio}   

%
%

\begin{appendix}
\section{Computation of parallactic, altitude, HWP and derotator angles from FITS-headers}
\label{app:angles_headers}

The parallactic, altitude, HWP and derotator angles needed for the instrument model can be retrieved from the headers of the FITS-files of the measurements.
	However, even during a measurement these angles are continuously changing as the telescope tracks the target.
	For each measurement, we therefore compute the mean value of these angles from the start and end values specified in the FITS-headers. 
	We note that for angles we cannot simply use the arithmetic mean, and instead use the mean of circular quantities:
\begin{equation} 
	\mathrm{mean}\left(\theta_\mathrm{s},~ \theta_\mathrm{e} \right) = \mathrm{atan2}\left(\sin\theta_\mathrm{s} + \sin\theta_\mathrm{e}, \cos\theta_\mathrm{s} + \cos\theta_\mathrm{e} \right),
\end{equation} 
where $\theta_\mathrm{s}$ and $\theta_\mathrm{e}$ are the angles at the start and end of the measurement, respectively.

The parallactic angle $p$ and HWP angle $\theta_\mathrm{HWP}$ are obtained from the FITS-headers as:
\begin{align} 	
	p =&~\mathrm{mean}\left(\mathrm{TEL~PARANG~START},~
	\mathrm{TEL~PARANG~END}\right), \\
	\theta_\mathrm{HWP} =&~\mathrm{mean}\left(\mathrm{INS4~DROT3~BEGIN},~ \mathrm{INS4~DROT3~END}\right) \nonumber \\ 
	&~- \SI{152.15}{\degree}.
\end{align}
	For observations in field-tracking mode
, the derotator angle $\theta_\mathrm{der}$ is computed as:
\begin{equation} 
	\theta_\mathrm{der} = \mathrm{mean}\left(\mathrm{INS4~DROT2~BEGIN},~ \mathrm{INS4~DROT2~END}\right).
\end{equation} 
	For pupil-tracking observations~\citep[see][]{vanholstein_exopol}, the derotator angle is calculated as:
\begin{align} 
	\theta_\mathrm{der} =&~\mathrm{mean}\left(\mathrm{INS4~DROT2~BEGIN},~ \mathrm{INS4~DROT2~END}\right) \\ \nonumber
	&~+ \dfrac{1}{2}\eta_\mathrm{pupil},
\end{align} 
where $\eta_\mathrm{pupil} = 135.99 \pm 0.11\si{\degree}$ is the fixed position angle offset of the image~\citep[see][
]{maire_sphere_astrometric_calib}.
	This offset is used to align a mask added to the Lyot stop (the `spider mask') with the diffraction pattern of the support structure of the UT's secondary mirror. 
	For the altitude angle $a$, only the start value is available from the header TEL~ALT.
	Therefore we use spline interpolation to compute the mean altitude angle during a measurement.
	
\section{Gradient in flux of internal calibration measurements}
\label{app:structure_flux}

\begin{figure} 
\centering 
\includegraphics[width=\hsize]{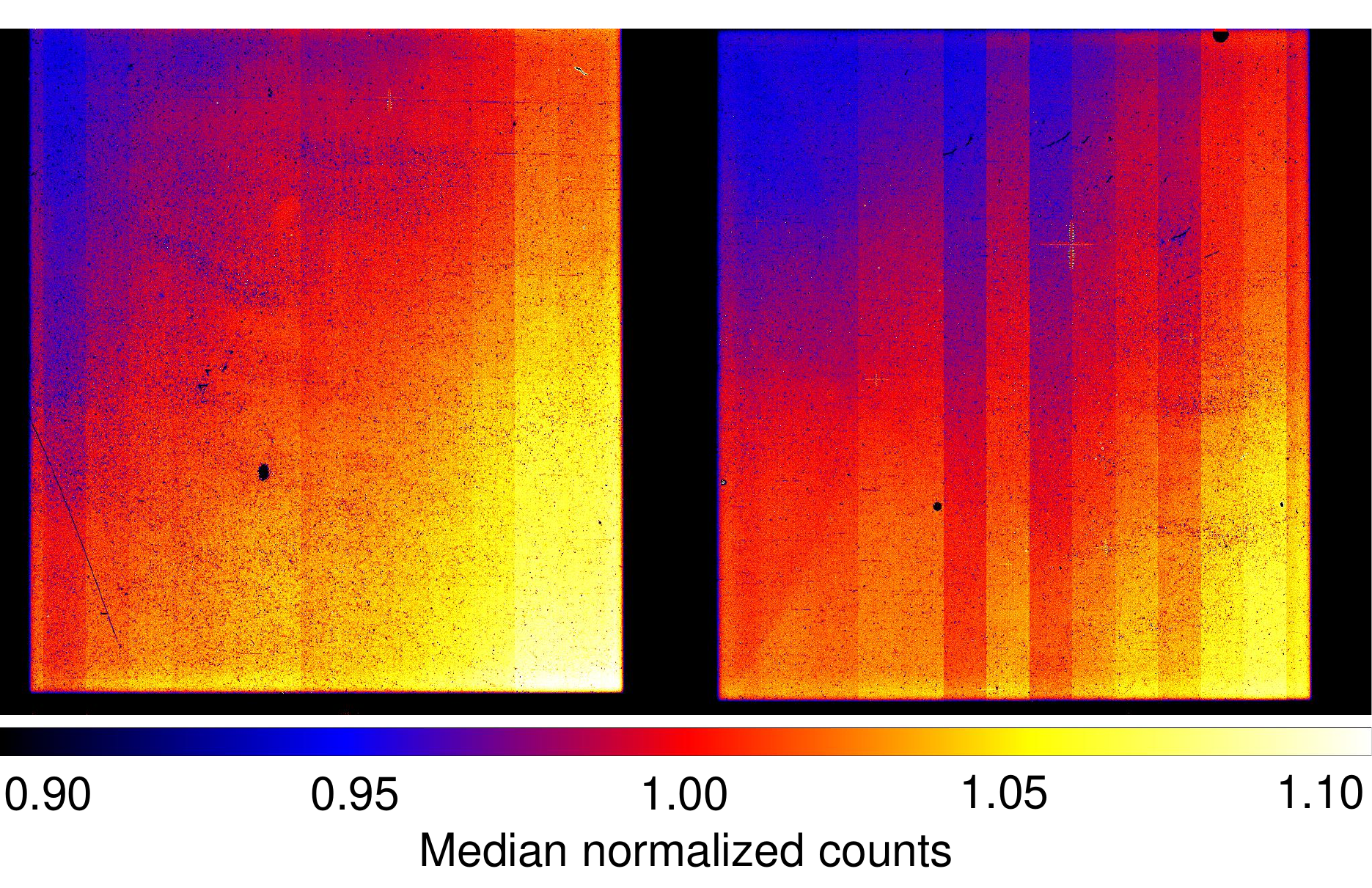} 
\caption{Dark-subtracted and bad-pixel-filtered flat-field frame in K$_\mathrm{s}$-band showing the total-intensity-dependent gradient in the left and right images on the detector.} 
\label{fig:master_flat_Ks} 
\end{figure}

\begin{figure}
\centering 
\includegraphics[width=\hsize]{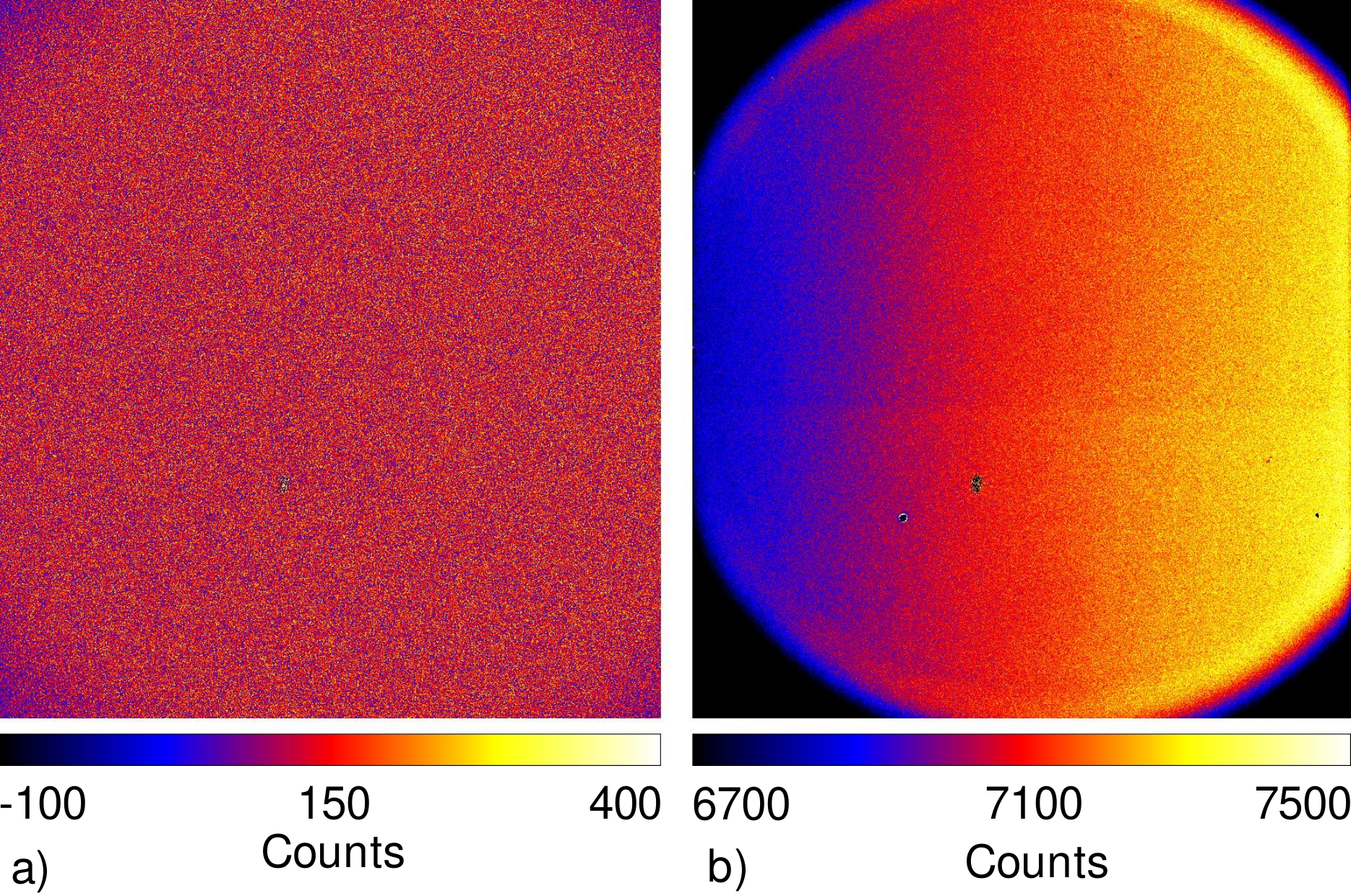} 
\caption{Double-difference images of the unpolarized source (a) and polarized source measurements (b) in K$_\mathrm{s}$-band showing that the double difference completely removes the total-intensity-dependent gradient, but does not remove the polarization-dependent-gradient.} 
\label{fig:gradient_DD} 
\end{figure}

The flux in most of the images taken with the internal light source is not uniform, but shows a gradient. 
	This structure appears to consist of two components: a gradient that depends on the total intensity of the incident light and a gradient that depends on the polarization state of the incident light. 
	The total-intensity-dependent gradient (see Fig.~\ref{fig:master_flat_Ks}) has a different strength and orientation for every broadband filter, and is most prominent in K$_\mathrm{s}$-band.
	It must originate downstream of the derotator, since it does not depend on the derotator or HWP angle.
	The gradient may be due to imperfect alignment of optical components or differences in transmission or reflectivity over the surface of the components. 
	As the gradient is also present in the lamp flat frames, the flat-field correction applied to the exposures suppresses the gradient.
	In the double-difference images (actually already in the single-difference images), the total-intensity-dependent gradient is completely removed (see Fig.~\ref{fig:gradient_DD}a).
	However, it is still visible in the double-sum images. 	
	Therefore, the normalized Stokes parameters determined from these images depend on the position of the apertures from which they are computed. 

In the polarized source measurements, the double difference removes the total-intensity-dependent gradient, but a polarization-dependent-gradient remains (see Fig.~\ref{fig:gradient_DD}b). 
	This gradient is different in strength and orientation for each exposure and therefore seems to depend on the orientation of the HWP and/or derotator. 
	Because the HWP is close to a focal plane, a likely cause of the polarization-dependent-gradient is that the retardance of the HWP varies over the surface of the HWP. 
	The gradient is not visible in the unpolarized source measurements, because the incident light is only very weakly polarized in that case.
%

\section{Graphs of model fits of internal calibration measurements}
\label{app:graphs_internal}

\begin{figure*}[!b]
\centering 
\includegraphics[width=14.3cm]{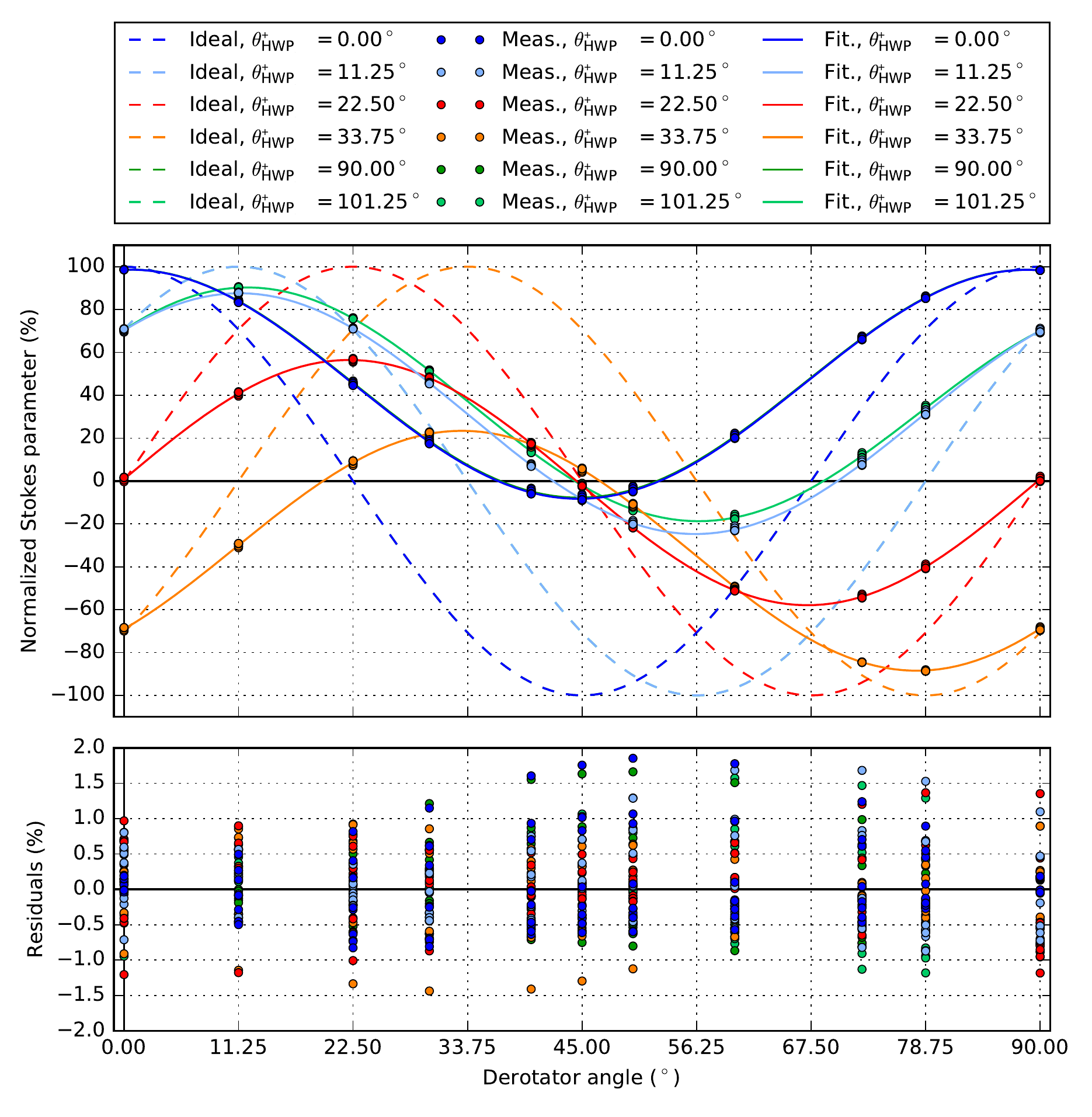} 
\caption{Ideal, measured and fitted normalized Stokes parameters of the polarized source measurements in H-band as a function of HWP and derotator angle. The legend only shows the $\theta_\mathrm{HWP}^+$-value of each data point or curve; it is implicit that the corresponding value for $\theta_\mathrm{HWP}^-$ differs \SI{45}{\degree} from that of $\theta_\mathrm{HWP}^+$.}
\label{fig:STOKES_POL_BB_H} 
\end{figure*}

\begin{figure*}
\centering 
\includegraphics[width=14.3cm, trim={0 0 0 1.15cm}, clip]{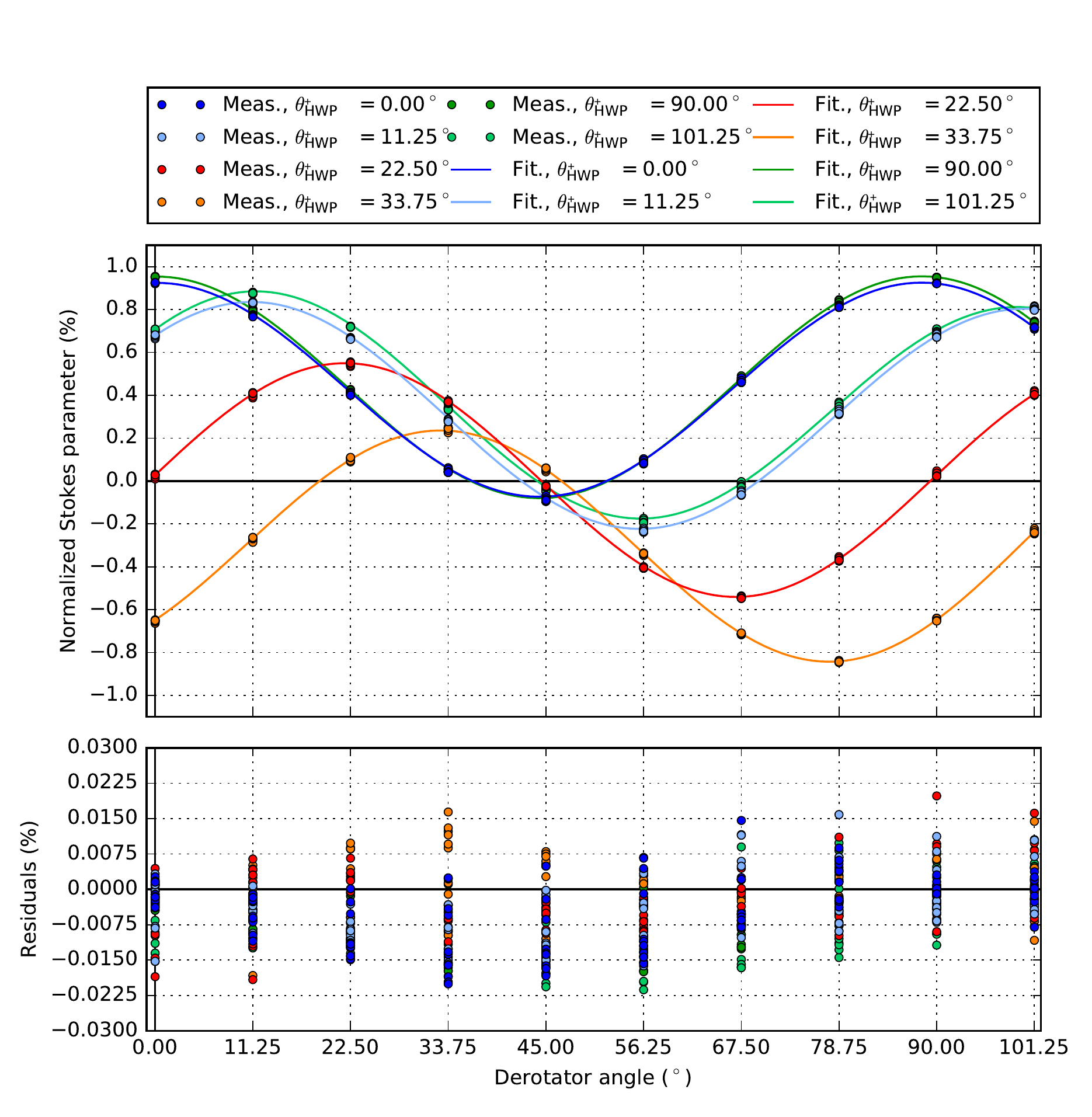} 
\caption{Measured and fitted normalized Stokes parameters of the unpolarized source measurements in H-band (normal double difference with the two HWP angles differing \SI{45}{\degree}) as a function of HWP and derotator angle. The legend only shows the $\theta_\mathrm{HWP}^+$-value of each data point or curve; it is implicit that the corresponding value for $\theta_\mathrm{HWP}^-$ differs \SI{45}{\degree} from that of $\theta_\mathrm{HWP}^+$.}
\label{fig:STOKES_UNPOL_BB_H} 
\end{figure*}

\begin{figure*}
\centering 
\includegraphics[width=14.3cm, trim={0 0 0 1.15cm}, clip]{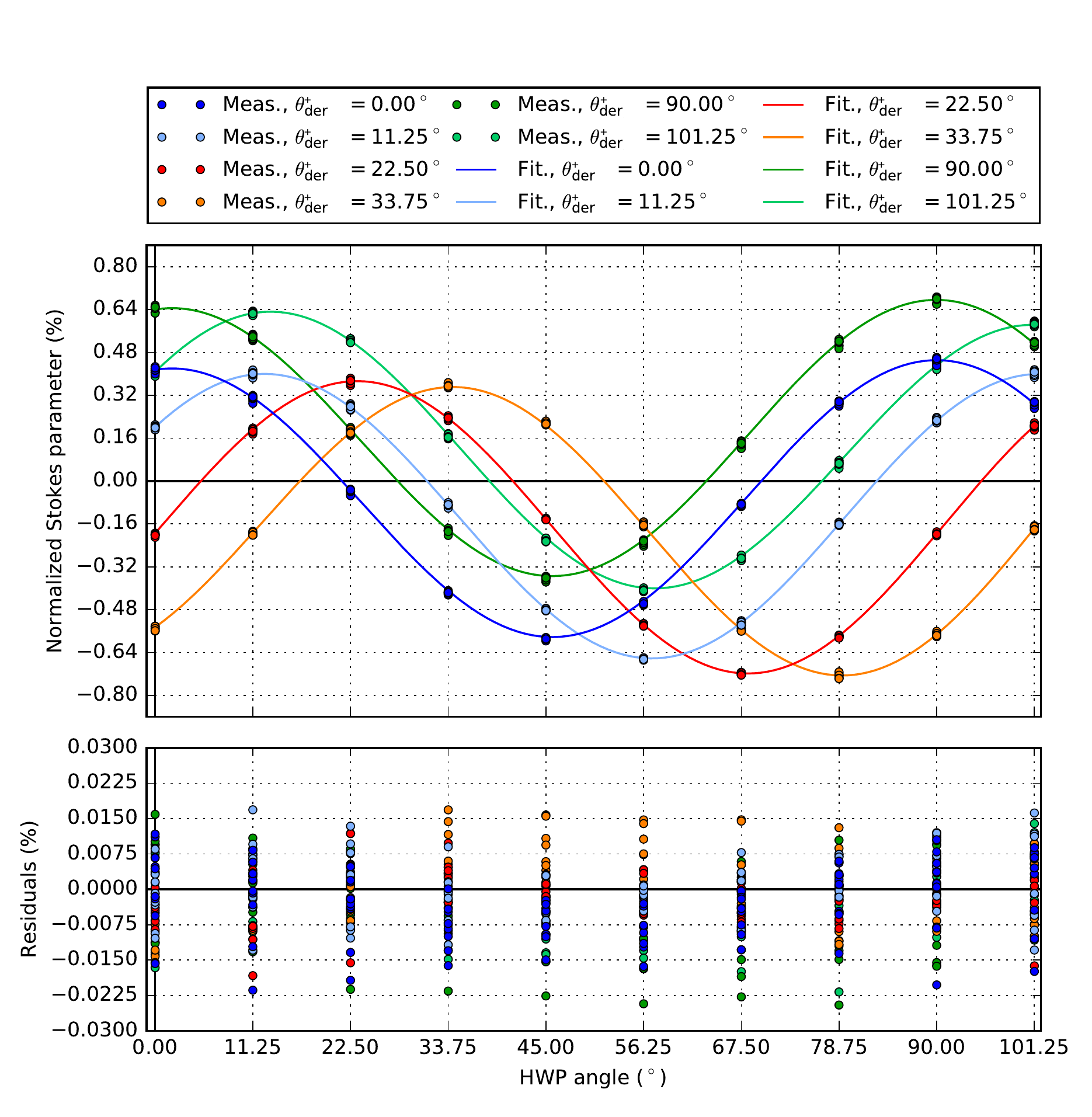} 
\caption{Measured and fitted normalized Stokes parameters of the unpolarized source measurements in H-band (modified double difference with the two derotator angles, rather than the HWP angles, differing \SI{45}{\degree}) as a function of derotator and HWP angle. The legend only shows the $\theta_\mathrm{der}^+$-value of each data point or curve; it is implicit that the corresponding value for $\theta_\mathrm{der}^-$ differs \SI{45}{\degree} from that of $\theta_\mathrm{der}^+$. Note that the x-axis displays the HWP angle and not the derotator angle as in Figs~\ref{fig:STOKES_POL_BB_H} and \ref{fig:STOKES_UNPOL_BB_H}.}
\label{fig:STOKES_UNPOL_BB_H_DERROT} 
\end{figure*}

Figure~\ref{fig:STOKES_POL_BB_H} shows the ideal, measured and fitted normalized Stokes parameters of the polarized source measurements in H-band as a function of HWP and derotator angle, including the residuals of fit. 
	The ideal curves are computed with the HWP and derotator retardances equal to \SI{180}{\degree}, no angle offsets and the diattenuation of the polarizers equal to 1.
	The measured and fitted normalized Stokes parameters of the unpolarized source measurements in H-band are displayed in Figs.~\ref{fig:STOKES_UNPOL_BB_H} (normal double difference) and \ref{fig:STOKES_UNPOL_BB_H_DERROT} (modified double difference with the derotator angles, rather than the HWP angles, differing \SI{45}{\degree} between the two exposures).
	These Figures also show the corresponding residuals of fit.
	The ideal curves (completely unpolarized light incident on the HWP, the diattenuations of the HWP, derotator and polarizers equal to 1, and no angle offsets) coincide with the x-axes of the graphs and are therefore not shown.
%
%
%
%
%

\section{Determination of normalized Stokes parameters and graphs of model fits of unpolarized star observations} 
\label{app:graphs_unpolstar}

\begin{figure}[htb!]
\centering 
\includegraphics[width=\hsize]{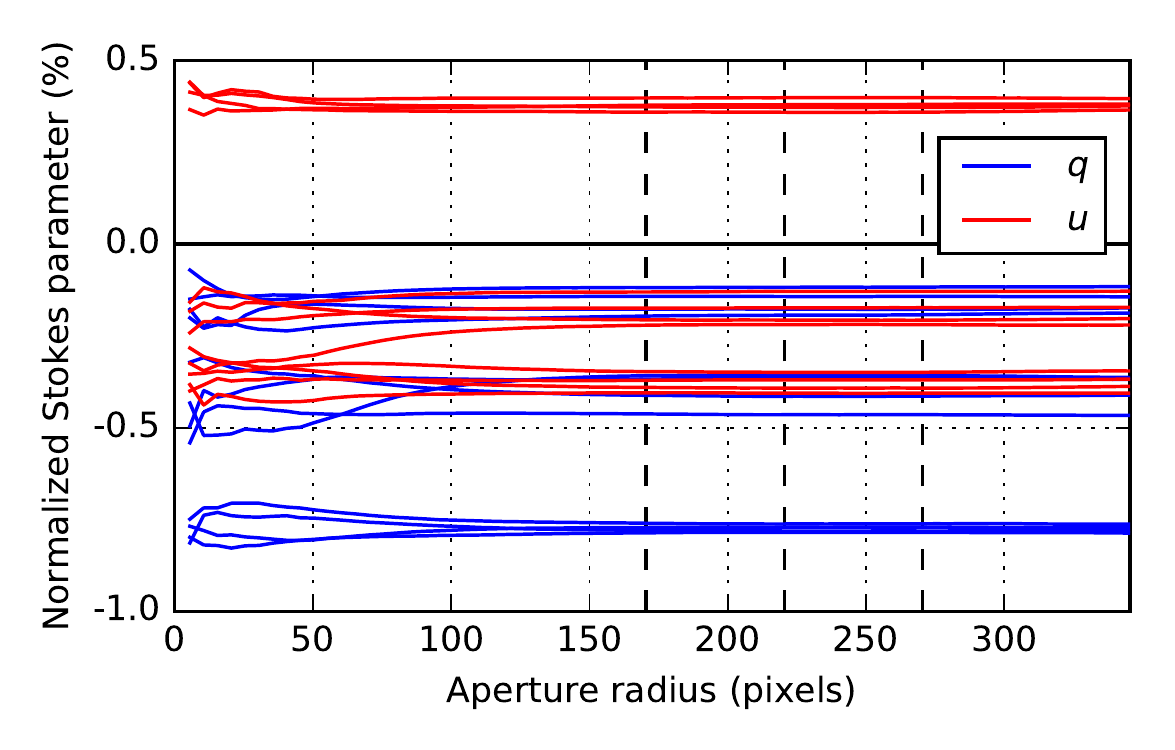} 
\caption{Normalized Stokes parameters $q$ and $u$ as a function of aperture radius for the observations of the unpolarized star HD~217343 (2018) in H-band. The central and outer dashed lines indicate the radii of the apertures from which the normalized Stokes parameters and their error bars (see Figs.~\ref{fig:M3_M4_DoLP_2018} and \ref{fig:M3_M4_STOKES_BB_H_2018}) have been determined, respectively.} 
\label{fig:STOKES_RADIUS_HD217343_BB_H} 
\end{figure}

The normalized Stokes parameters of the observations of the unpolarized stars are determined from apertures in the $Q$-, $U$-, $I_Q$- and $I_U$-images.
	For the data of HD~217343 (2018), we compute the signal in these images as the mean in an aperture minus the median of the background signal in a concentric annulus.
	We then calculate the normalized Stokes parameter $q$ or $u$ by dividing the signal from the $Q$- or $U$-image by that from the corresponding $I_Q$- or $I_U$-image according to Eq.~\ref{eq:model_normalized_stokes_parameter}. 
	The radii of the apertures used are determined from plots of the normalized Stokes parameters as a function of aperture radius (see Fig.~\ref{fig:STOKES_RADIUS_HD217343_BB_H}). 
	In all filters an aperture radius of 220 pixels is used, because at this radius the curves have approached a constant value. 
	The annulus to compute the background signal from starts at the outer radius of the aperture and has a width of 40 pixels.
	
For the data of HD~176425 (2016) we use the same method to compute the normalized Stokes parameters, but we do not subtract the background signal.
	This is because almost the complete image is filled with signal from the star and therefore there is no location to accurately determine the background signal from. 
	In Y-, J-, and H-band, where we use an aperture radius of 200 pixels, this is no problem because the background signal is very small.

\begin{figure}[htb!]
\centering 
\includegraphics[width=\hsize]{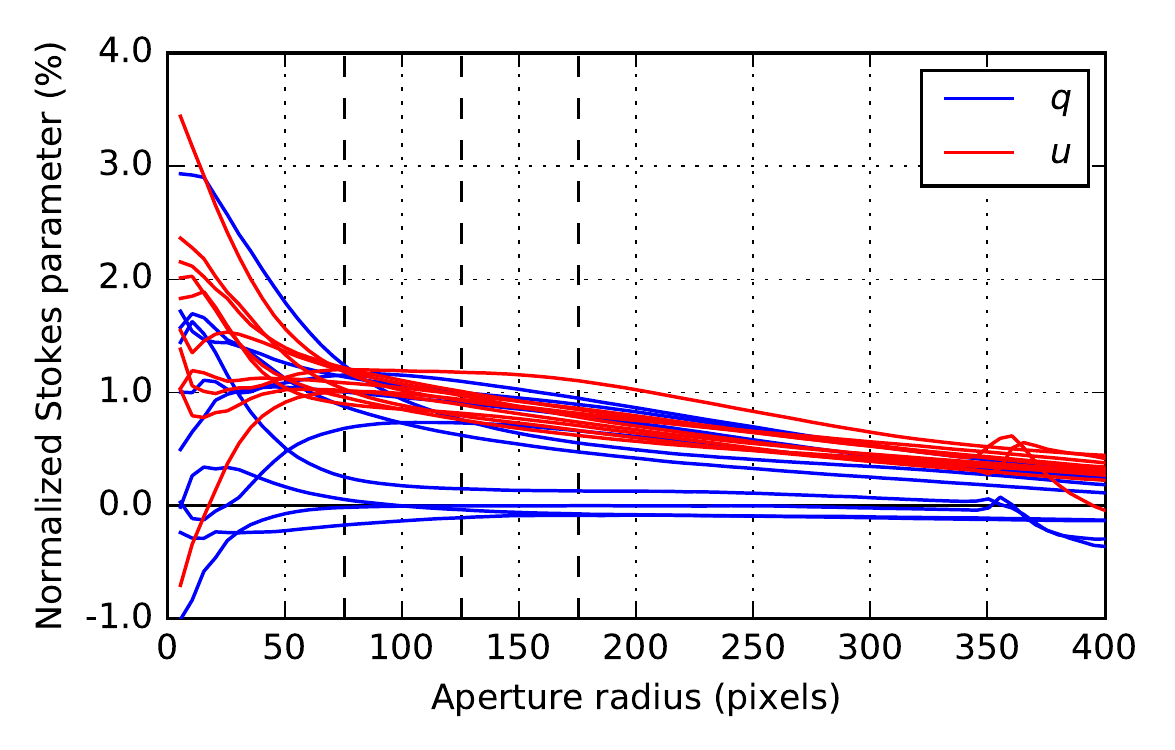} 
\caption{Normalized Stokes parameters $q$ and $u$ as a function of aperture radius for the observations of the unpolarized standard star HD~176425 (2016) in K$_\mathrm{s}$-band. The central and outer dashed lines indicate the radii of the apertures from which the normalized Stokes parameters and their error bars (see Figs.~\ref{fig:M3_M4_DoLP} and \ref{fig:M3_M4_STOKES_BB_Ks}) have been determined, respectively.} 
\label{fig:STOKES_RADIUS_HD176425_BB_Ks} 
\end{figure}

In K$_\mathrm{s}$-band however (see Fig.~\ref{fig:STOKES_RADIUS_HD176425_BB_Ks}), the curves of $q$ and $u$ versus aperture radius do not approach a constant value, but decrease with increasing aperture radius due to the much stronger background signal that most likely originates from thermal emission of the UT and SPHERE's uncooled optics upstream from IRDIS. 
	Since the intensity of the star's point spread function (PSF) decreases with increasing distance from the center, the thermal background becomes more prominent for larger aperture radii. 
	Although the thermal background is removed after computing the double difference ($Q$- and $U$-images), it is not removed after computing the double sum ($I_Q$- and $I_U$-images), and therefore the normalized Stokes parameters decrease with increasing aperture radius.
	An aperture radius of 125 pixels is selected for the measurements in K$_\mathrm{s}$-band, because at this radius: 1) the curves of the other filters start to approach a constant value, 2) the thermal background starts to become visible in the raw frames, and 3) the determined diattenuations of the UT and M4 are in line with expectations based on the determined diattenuations in the other filters and their deviation from the analytical values (see Fig.~\ref{fig:M3M4_IP}). \\
%
%
%

\begin{figure}[h!]
\centering 
\includegraphics[width=\hsize]{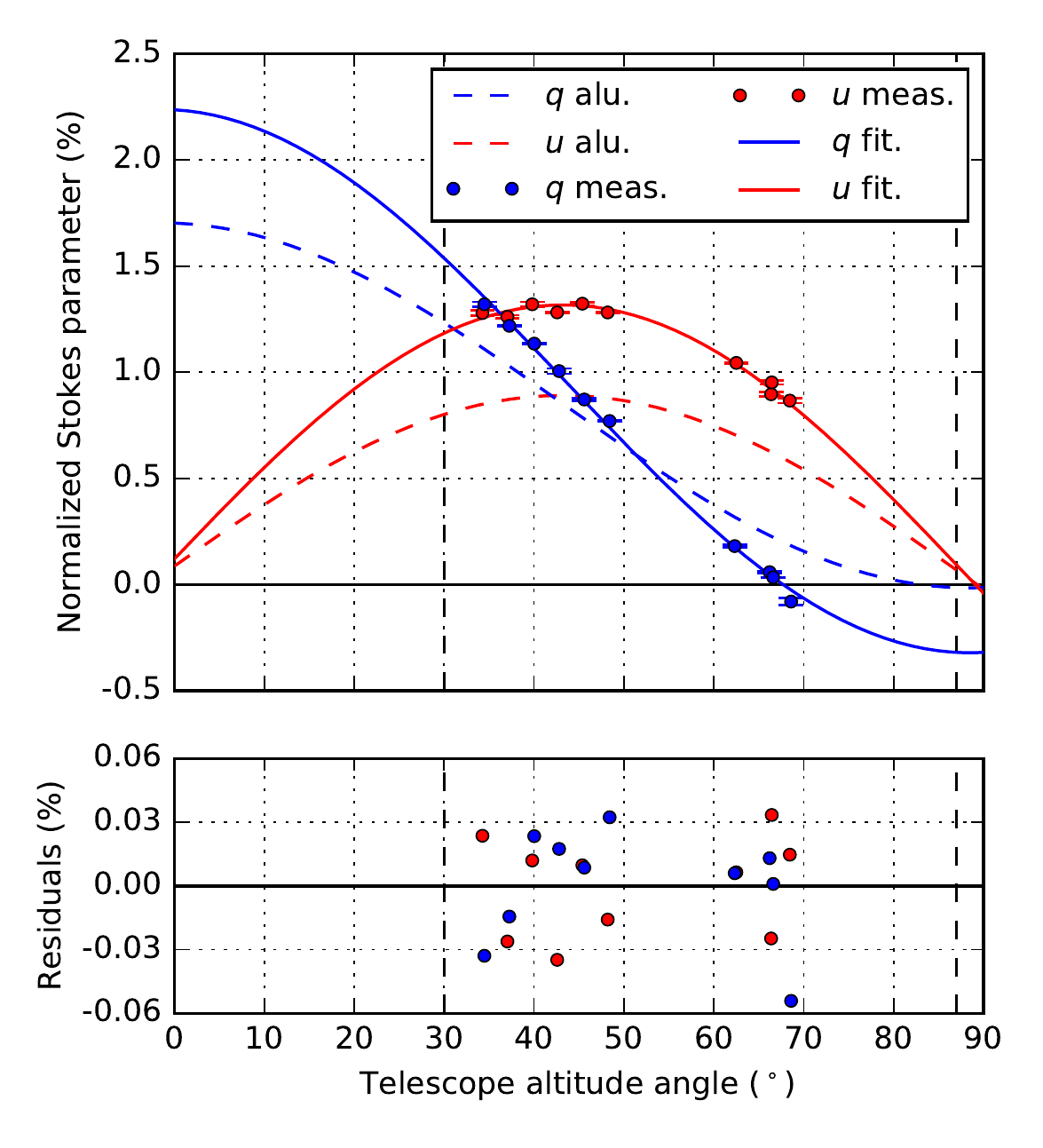} 
\caption{Analytical (aluminum), measured (including error bars) and fitted normalized Stokes parameters $q$ and $u$ as a function of telescope altitude angle for the observations of the unpolarized standard star HD~176425 (2016) in H-band. Note that for science observations the telescope altitude angle is restricted to $\SI{30}{\degree} \leq a \leq \SI{87}{\degree}$.} 
\label{fig:M3_M4_STOKES_BB_H} 
\end{figure}

\begin{figure}[h!]
\centering 
\includegraphics[width=\hsize]{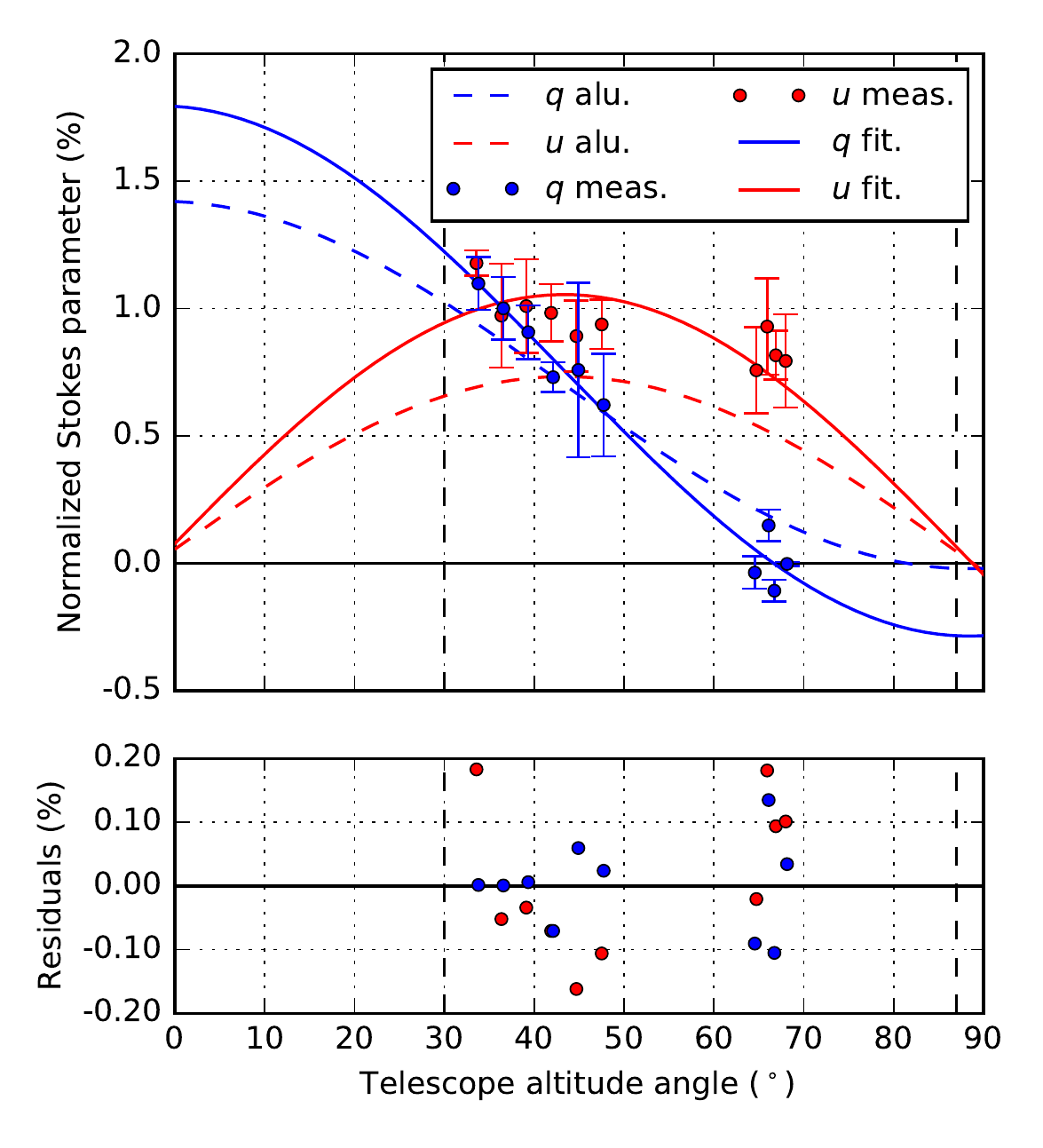} 
\caption{Analytical (aluminum), measured (including error bars) and fitted normalized Stokes parameters $q$ and $u$ as a function of telescope altitude angle for the observations of the unpolarized standard star HD~176425 (2016) in K$_\mathrm{s}$-band. Note that for science observations the telescope altitude angle is restricted to $\SI{30}{\degree} \leq a \leq \SI{87}{\degree}$.} 
\label{fig:M3_M4_STOKES_BB_Ks} 
\end{figure}

\begin{figure}[h!]
\centering 
\includegraphics[width=\hsize]{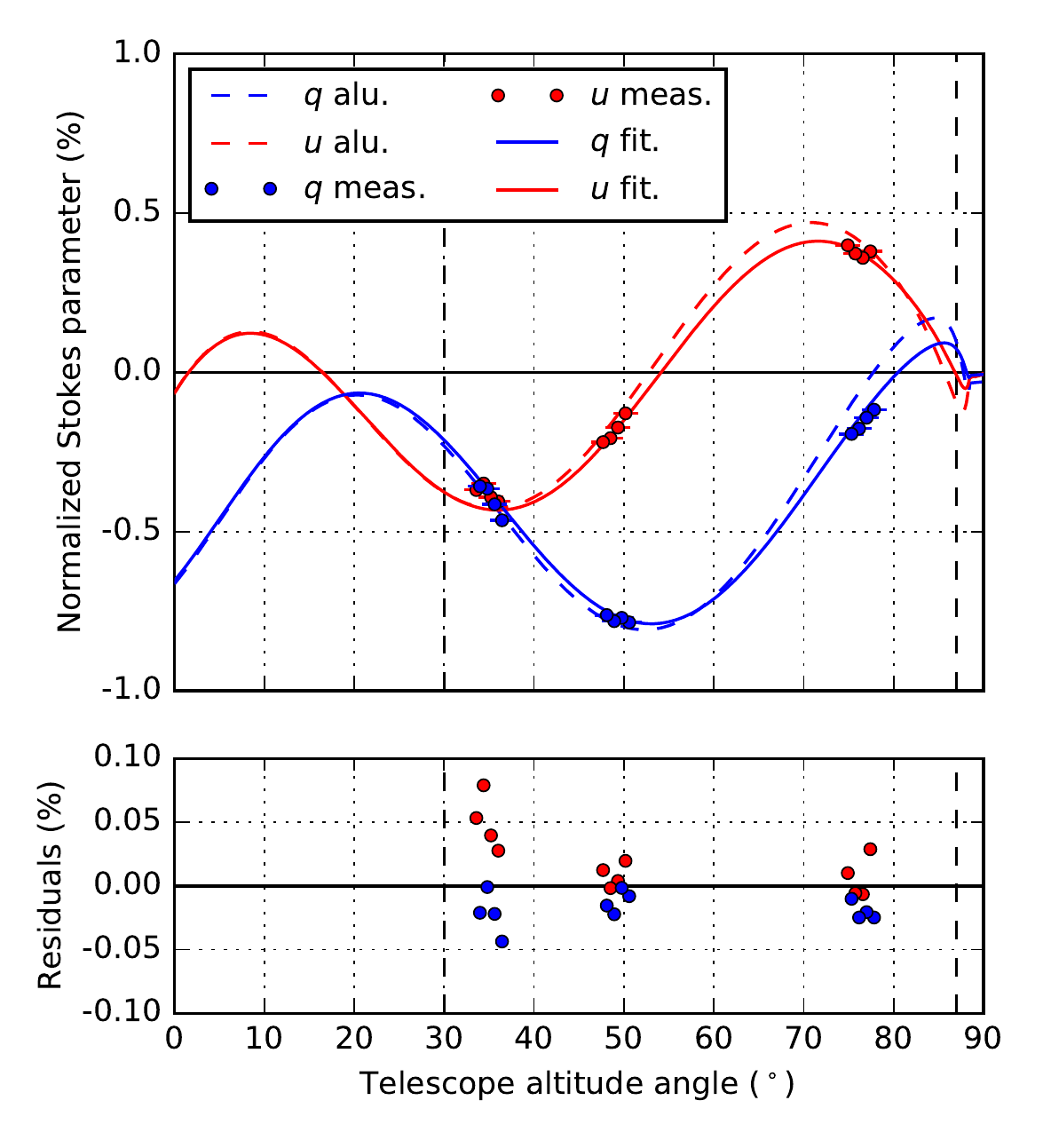} 
\caption{Analytical (aluminum), measured (including error bars) and fitted normalized Stokes parameters $q$ and $u$ as a function of telescope altitude angle for the observations of the unpolarized star HD~217343 (2018) in H-band. Note that for science observations the telescope altitude angle is restricted to $\SI{30}{\degree} \leq a \leq \SI{87}{\degree}$.} 
\label{fig:M3_M4_STOKES_BB_H_2018} 
\end{figure}

\noindent Figs.~\ref{fig:M3_M4_STOKES_BB_H} and \ref{fig:M3_M4_STOKES_BB_Ks} show the analytical, measured and fitted normalized Stokes parameters $q$ and $u$ of the observations of HD~176425 (2016) as a function of telescope altitude angle in H- and K$_\mathrm{s}$-band, respectively.
	Figure~\ref{fig:M3_M4_STOKES_BB_H_2018} shows the same graph for the observations of HD~217343 (2018) in H-band.
	The residuals of fit are also included in these Figures. 
	The analytical curves are computed from the Fresnel equations using the complex refractive index of aluminum.
	The error bars are calculated as half the difference between the normalized Stokes parameters determined from apertures with radii 50 pixels larger and smaller than the radius of the aperture used to calculate $q$ and $u$ used for determining the diattenuations (see Figs.~\ref{fig:STOKES_RADIUS_HD217343_BB_H} and ~\ref{fig:STOKES_RADIUS_HD176425_BB_Ks}). 
	The error bars show the uncertainty in the normalized Stokes parameters due to the dependency of the measured values on the chosen aperture radius.
	These uncertainties are small for all measurement except those of HD~176425 (2016) in K$_\mathrm{s}$-band because the thermal background could not be subtracted.
	Finally, note that because we did not keep the derotator fixed with its plane of incidence horizontal for the observations of HD~217343 (2018), crosstalk from the derotator causes the shape of the curves in Fig.~\ref{fig:M3_M4_STOKES_BB_H_2018} to be different from those of Figs.~\ref{fig:M3_M4_STOKES_BB_H} and \ref{fig:M3_M4_STOKES_BB_Ks}.
%
%
%
%
%

\section{Calculation of accuracies of fit and uncertainties in determined parameters}
\label{app:uncertainty_parameters}

To estimate the polarimetric accuracy of the instrument model, we calculate for each broadband filter the accuracies of fitting the model parameters to the calibration data.
	We compute these accuracies of fit as the corrected sample standard deviation of the residuals $s_\mathrm{res}$:
\begin{equation} 
	s_\mathrm{res} = \sqrt{\dfrac{\sum_{i=1}^n r_i^2}{n - k}},
	\label{eq:accuracy_fit_residuals} 
\end{equation} 
with $r_i$ the residuals of fit, $n$ the number data points and $k$ the number of parameters determined from the data set.
	The accuracies of fit are calculated separately for the polarized source measurements, the unpolarized source measurements and the two observations of unpolarized stars (denoted $s_\mathrm{rel}$, $s_\mathrm{unpol}$ and $s_\mathrm{star}$, respectively, in Sect.~\ref{sec:model_accuracy}).
	The results are shown in Table~\ref{tab:accuracies_of_fit}. 
%
\begin{table}[!hbtp]
\caption{Accuracies of fit of the polarized source measurements, the unpolarized source measurements and the observations of the unpolarized stars HD~176425 (2016) and HD~217343 (2018) in Y-, J-, H- and K$_\mathrm{s}$-band.}
\centering
\setlength{\tabcolsep}{1pt}
\begin{tabular}{l c c c c}
\hline\hline
Filter & \begin{tabular}{@{}c@{}} $s_\mathrm{res}$ (\%) \T \\ ~polarized~ \\ ~source~ \B \end{tabular} & \begin{tabular}{@{}c@{}} $s_\mathrm{res}$ (\%) \T \\ ~unpolarized~ \\ ~source~ \B \end{tabular} & \begin{tabular}{@{}c@{}} $s_\mathrm{res}$ (\%) \T \\ ~unpolarized~ \\ ~star 2016~ \B \end{tabular} & \begin{tabular}{@{}c@{}} $s_\mathrm{res}$ (\%) \T \\ ~unpolarized~ \\ ~star 2018~ \B \end{tabular} \T\B \\ 
\hline
BB\_Y & 0.73 & 0.023~~ & 0.058 & ~0.064 \T \\
BB\_J & 0.41 & 0.0070 & 0.047 & 0.072 \\
BB\_H & 0.58 & 0.0083 & 0.025 & 0.029 \\
BB\_K$_\mathrm{s}$ & 0.54 & 0.0085 & 0.10 & ~0.092 \B \\ 
\hline
\end{tabular}
\label{tab:accuracies_of_fit}
\end{table}

To compute the uncertainties of the determined model parameters, we approximate the covariance matrix of the model parameters $\varSigma$ as:
%
\begin{equation} 
	\varSigma = \tau (J^\mathrm{T} J)^{-1} \tau,
	\label{eq:covariance_parameters} 
\end{equation} 
where $J$ is the Jacobian matrix:
\begin{equation} 
	J = \begin{bmatrix} \dfrac{\partial x_1}{\partial\beta_1} & \cdots & \dfrac{\partial x_1}{\partial\beta_m} \\ \vdots & \ddots & \vdots \\ \dfrac{\partial x_n}{\partial\beta_1} & \cdots & \dfrac{\partial x_n}{\partial\beta_m} \end{bmatrix},
\end{equation}
with $\beta_1$ to $\beta_m$ the $m$ determined model parameters and $x_1$ to $x_n$ the model functions describing the $n$ measurements (Eq.~\ref{eq:model_normalized_stokes_parameter} with the model equations and the parallactic, altitude, derotator and HWP angles of the measurements substituted). 
	$\tau$ is an $m \times m$ matrix with on its diagonal for each model parameter the accuracy of fit ($s_\mathrm{res}$) of the measurements from which that parameter is determined (see Table~\ref{tab:accuracies_of_fit}).
	For example, the diagonal element of $\tau$ corresponding to the model parameter $\varDelta_\mathrm{der}$ in H-band is equal to $s_\mathrm{res}$ of the polarized source measurements in the same filter.
	Finally, we compute the $1\sigma$-errors (1 times the standard deviation) of the model parameters as the square root of the diagonal elements of $\varSigma$, and list them behind the $\pm$-signs in Tables~\ref{tab:parameters_instrument} and \ref{tab:parameters_telescope}.

By taking the diagonal values of $\varSigma$ as the uncertainties of the parameters, it is assumed that the parameter values are not correlated. 	
	However, in reality all the parameters are weakly correlated, in particular because the offset angles $\delta_\mathrm{HWP}$, $\delta_\mathrm{der}$ and $\delta_\mathrm{cal}$ are determined from the complete set of polarized source measurements. 
	In addition, the uncertainties of the parameters are computed using a linear approximation through the Jacobian.
	Therefore the uncertainties should be considered first order estimates only.

%

\end{appendix}

\end{document}